\documentclass{aa}  

\usepackage{mathrsfs}
\usepackage{url}
\usepackage{hyperref}
\hypersetup{
    colorlinks=true,
    linkcolor=blue,
    filecolor=black,      
    urlcolor=blue,
    citecolor=blue,
    pdftitle={Overleaf Example},
    pdfpagemode=FullScreen,
    }
\usepackage{comment}

\usepackage{graphicx}
\usepackage{txfonts}
\usepackage{lipsum}
\usepackage{subcaption}  
\usepackage{ulem}
\usepackage{xcolor}
                                
\usepackage{lscape}             
\usepackage{placeins}           

\begin{document}

   \title{ 
   Balmer decrements as a new diagnostic for period-bounce Cataclysmic Variable stars}

   \author{Santiago Hernández-Díaz\inst{1}
        \and Beate Stelzer\inst{1} \and Daniela Muñoz-Giraldo\inst{1}
        }

\institute{Institut für Astronomie und Astrophysik, Eberhard Karls Universität Tübingen, Sand 1, 72076 Tübingen, Germany\\
\email{hernandez@astro.uni-tuebingen.de}}
\date{Received 19 January 2026 / Accepted 07 June 2026}

 
\abstract
   {Cataclysmic variable stars (CVs) evolve toward shorter orbital periods ($P_{\rm orb}$) until they reach a minimum $P_{\rm orb}$ near $P_{\rm orb}\sim80$\,min. Beyond this point, the donor star becomes out of thermal equilibrium or increasingly degenerate, causing the system to “bounce back” to longer $P_{\rm orb}$ values. Such highly evolved systems are known as period-bouncers. Although 40–80\% of all CVs are expected to have reached this stage, period-bouncers come up for only 3–25\% of the observed CV population. This is likely a consequence of their intrinsic faintness associated with lower mass-transfer rates. Establishing new diagnostics to unveil this missing population of evolved CVs is therefore crucial.}
   {The aim of this study is to investigate whether period-bounce CVs can be distinguished from short-period pre-bounce CVs through their Balmer decrements, that is the Balmer line ratios, using optical spectra from the Sloan Digital Sky Survey (SDSS). Differences in their Balmer decrements are expected to trace the distinct physical conditions in the accretion discs of period-bouncers resulting from their lower mass-transfer rates.} 
   {Two samples of non-magnetic CVs with public SDSS optical spectra were constructed: one of short-period pre-bounce CVs and another of period-bounce CVs. For systems showing Balmer absorption from the white dwarf (WD), hydrogen-dominated atmosphere models were fitted and subtracted to correct for the WD component. H$\alpha$, H$\beta$, and H$\gamma$ fluxes were measured. We then investigated statistical relations between the Balmer decrements, the H$\beta$ line luminosity, and $P_{\rm orb}$, and compared the measured Balmer decrements with theoretical predictions from accretion disc models.}
   {Short-period pre-bounce CVs show flat Balmer decrements, that is Balmer line ratios close to unity, consistent with optically thick line emission and local thermodynamic equilibrium accretion disc models. In contrast, systems near and beyond the period minimum exhibit progressively steeper decrements ($\text{H}{\alpha}/\text{H}{\beta}>1$ and $\text{H}{\gamma}/\text{H}{\beta}<1$). This behaviour is attributed to their lower mass accretion rates, as inferred from the H$\beta$ line luminosity. We fitted a linear logistic regression model to the diagram of H$\gamma$/H$\beta$ versus H$\alpha$/H$\beta$. We establish that this diagnostic diagram effectively separates period-bouncers from pre-bounce CVs, with remaining contaminants including period-bouncer candidates and evolved pre-bounce CVs with low mass accretion rates.}
   {Our study shows that Balmer decrements are an effective discriminator between short-period pre-bouncers and period-bouncers, consistent with enlarged optically thin disc regions and a more pronounced radial stratification in the physical conditions of the accretion discs of period-bouncers. Our logistic-regression classifier based on the $\text{H}{\gamma}/\text{H}{\beta}$ and $\text{H}{\alpha}/\text{H}{\beta}$ ratios offers a robust and efficient tool for identifying period-bouncer candidates in optical spectroscopic surveys.}

   \maketitle
   \nolinenumbers

\section{Introduction}

Cataclysmic variable stars (CVs) are compact binary systems in which a white dwarf (WD) primary accretes material from a low-mass main-sequence secondary star (the donor) that fills its Roche lobe (see \citealt{Warner_Book} for a comprehensive review). The evolution of CVs is primarily driven by angular momentum loss through magnetic braking (\citealt{Mestel_1968}, \citealt{Mestel_1987}) and gravitational radiation (\citealt{Faulkner_1971}, \citealt{Paczynski_1981}), which causes the binary to evolve toward progressively shorter orbital periods ($P_{\rm orb}$) (\citealt{Rappaport_1982}, \citealt{Kolb_1993}, \citealt{Knigge_2011}, \citealt{Kalomeni_2016}). For short-period systems below the period gap at $2-3$\,h (\citealt{Kolb_1998}, \citealt{Howell_2001}) the angular momentum loss is believed to be mainly due to gravitational radiation. Eventually,  the system reaches the period minimum at $P_{\rm orb} \sim 80\,$min (\citealt{Rappaport_1982}, \citealt{Paczynski_1983}, \citealt{Knigge_2006}, \citealt{Knigge_2011}, \citealt{Patterson_2011}, \citealt{Goliasch_2015}). Once the binary reaches the minimum $P_{\rm orb}$, it evolves back toward longer $P_{\rm orb}$ values, as the donor starts expanding in response to continued mass loss. Two effects can cause the donor to expand. First, the donor is driven out of thermal equilibrium if mass is removed more rapidly than the star can thermally readjust, that is, when the mass-loss timescale becomes shorter than the thermal (Kelvin–Helmholtz) timescale. Second, if the donor instead manages to remain in thermal equilibrium while its mass steadily decreases, the donor’s mass will eventually drop below the threshold required for sustained hydrogen burning. The donor then becomes increasingly degenerate, causing its mass–radius relation to invert (\citealt{Knigge_2006}, \citealt{Knigge_2011}). The system accommodates this expanding donor by settling Roche-lobe overflow at larger orbital separations, and therefore evolving toward longer $P_{\rm orb}$ values (\citealt{Warner_Book}). CVs that have passed this evolutionary turning point are referred to as period-bouncers (\citealt{Patterson_1998}, \citealt{Daniela_2024}). 

While theoretical evolutionary models predict that the majority of the CV population is expected to be conformed by period-bouncers, with estimations between 40–80\% of the CV population (see e.g., \citealt{Kolb_1993}, \citealt{Goliasch_2015}, and \citealt{Belloni_2020}), their observed numbers remain very small. Only 39 systems have been confirmed to date (\citealt{Daniela_2026}), and current surveys suggest that period-bouncers represent just 3–25\% of the known CV population (\citealt{Inight_2023a}, \citealt{Pala_2020}, \citealt{Rodriguez_2025}).

Recent theoretical work has proposed new mechanisms that may explain the scarcity of observed PBs. Angular momentum loss during nova eruptions in systems with low-mass WDs may lead to mergers, preventing many systems from evolving into period-bouncers (\citealt{Schreiber_2016}, \citealt{Belloni_2018}). Additionally, \citet{Schreiber_2023} proposed that CVs may become magnetic after reaching the period minimum, with magnetic coupling between the WD and donor causing the transfer of the WD's spin angular momentum to the orbit and ultimately driving the systems to detachment.

CVs are commonly classified into magnetic systems, where the WD's magnetic field governs the accretion flow and either truncates or entirely suppresses the disc formation (see \citealt{Patterson_1994} and \citealt{Cropper_Polars}); and non-magnetic systems, where the accretion is mediated by the formation of an accretion disc around the WD.

Optical spectra of non-magnetic CVs are typically characterised by a blue continuum with superimposed emission features from H, He$_{\rm I}$, He$_{\rm II}$, Ca$_{\rm II}$, and Fe lines (see e.g. \citealt{Smith_2006}). Most of the optical line emission originates in the accretion disc, as indicated by the characteristic double-peaked line profiles and broad linewidths of thousands of kilometers per second, which trace the Keplerian velocity field of the rotating gas (\citealt{Smak_1969}, \citealt{Smak_1982}, \citealt{Horne_1986}). Low excitation transitions are generally attributed to the cooler, low-density regions in the outer parts of the accretion disc, while higher excitation transitions are thought to form in the inner parts of the disc, where the transition region between the inner edge of the accretion disc and the surface of the WD, the boundary layer (BL),  generates an intense radiation field (\citealt{Williams_1980}, \citealt{Tylenda_1981}, \citealt{Popham_1995}). Additional contributions can also arise from other structures of the binary system, including the gas stream flowing from the donor star (e.g. \citealt{Harlaftis_1996}), the hot spot formed by the collision of the stream of accreting material with the disc (e.g. \citealt{Neustroev_2016}), and the irradiated hemisphere of the donor (e.g. \citealt{Stephen_1992}).

A well-known characteristic of many CV spectra is that the Balmer emission lines exhibit nearly flat decrements, meaning Balmer line ratios close to unity (e.g. \citealt{{Williams_1983}}), indicating that the accretion disc is optically thick in these lines. Statistical studies of emission lines in CVs were first carried out by \citet{Echevarria_1988} and \citet{Gordon_2006}. \citet{Echevarria_1988} compiled measurements of the $\mathrm{H}\alpha$, $\mathrm{H}\beta$, $\mathrm{H}\gamma$, $\mathrm{H}\delta$, $\mathrm{He_{\rm I}}\,\lambda4471$, and $\mathrm{He_{\rm II}}\,\lambda4686$ emission lines available in the literature, and compared them with predictions from Local Thermodynamic Equilibrium (LTE) models (\citealt{Drake_1980}, \citealt{Williams_1980}, \citealt{Tylenda_1981}). \citet{Gordon_2006} applied multivariate statistical analyses—principal component analysis (PCA) and discriminant function analysis—to the spectroscopic Balmer emission line data set of \citet{Williams_1983}, finding that variations in Balmer line ratios correlate with $P_{\rm orb}$ and the $\mathrm{H}\beta$ equivalent width. \citet{Gordon_2006} also compared the observed Balmer decrements, that is the Balmer line ratios, with a wider variety of models, including LTE models of accretion discs (\citealt{Williams_1980}), non-LTE models (\citealt{Williams_1991}), and chromospheric models (\citealt{Williams_1995}).

In this work, we extend these earlier studies by focusing specifically on non-magnetic, short-period CVs. In short-period CVs, with lower mass accretion rates, the disc is thought to become cooler and increasingly optically thin (\citealt{Williams_1980}). The main focus of this study is to investigate whether the Balmer decrements can act as a diagnostic of the physical conditions in the accretion discs of period-bounce CVs, whose lower mass-transfer rates are expected to produce systematically different line ratios compared to pre-bounce CVs (\citealt{Williams_1980}, \citealt{Tylenda_1981}).

We describe the samples of short-period pre-bounce and period-bounce CVs in Sect.~\ref{sect.Sample}. Our methodology for measuring the line fluxes is detailed in Sect.~\ref{Line_Flux_sect}. In Sect.~\ref{subsect:results}, we present our results and establish a diagnostic diagram constructed using a linear logistic regression model trained on the measured Balmer decrements. This model allows to estimate the probability that a system is a period-bouncer based on its position in the Balmer decrement parameter space. We then compare our results with predictions from theoretical accretion disc models in Sect.~\ref{sec:TheoreticalModels}. Finally, in Sect.~\ref{Summary}, we provide a summary of our findings and present our conclusions.

\section{Sample}
\label{sect.Sample}

For this study we define two samples of CVs around the period minimum: one composed of short-period pre-bounce CVs and another of period-bounce CVs. The two samples are drawn entirely from systems observed in the Sloan Digital Sky Survey (SDSS), and make exclusive use of publicly available data collected during its first four phases, including Data Release 17 (hereafter, SDSS I-IV) (\citealt{York_2000}, \citealt{Blanton_2017}, \citealt{Abdurro_2022}). SDSS I-IV operated from 2000 to 2020, employing a multi-object, fibre-fed spectrograph. The spectral range covered $\simeq3800-9200\,\AA$ in the first two phases and was extended to $\simeq3600-10,000\,\AA$ in the later surveys (\citealt{Smee_2013}). 

Our sample of pre-bounce CVs is drawn from the catalogue of \citet{Inight_2023}, who compiled 507 CVs observed during SDSS I–IV. The catalogue provides \textit{Gaia} DR3 source identifiers \citep{Gaia_DR3} (hereafter, \textit{Gaia} DR3 IDs), $P_{\rm orb}$ values, and type classifications that we adopt here. To select systems approaching the minimum $P_{\rm orb}$ but not yet evolved past it, we required $P_{\text{orb}}< 130$\,min, and \textit{Gaia} magnitude $G< 18.5$\,mag. The latter condition is adopted to avoid contamination of the pre-bouncer sample with as yet unidentified period-bouncers, that might hide among the low-luminosity systems. Applying these criteria yields an initial sample of 65 pre-bounce CVs. 

However, several systems in this initial sample—V379\,Vir, PM\,J12507+1549, V406\,Vir, EZ\,Lyn, and SDSS\,J101421.55+063857.7—have been identified as period-bouncers (see \citealt{Daniela_2026}) and were therefore excluded from the sample of pre-bounce CVs. Two AM\,CVn systems, GP\,Com and SDSS\,J080449.49+161624.8, were also excluded, as these systems follow a distinct evolutionary path (see \citealt{Solheim_2010}). Three systems—V355\,UMa, LV\,Cnc, and SDSS\,J102905.24+485515.2—are classified as WZ\,Sge systems or WZ\,Sge candidates in \citet{Inight_2023}. These three systems were identified as period-bouncer candidates by \citet{Daniela_2024}. They were therefore retained in the analysis but treated as a separate group. A detailed discussion of these three systems is provided in Sect.~\ref{subsect:diagnostic_diagram}. Finally, the system OV\,Boo is likely a CV that formed directly from a detached WD–brown dwarf binary (see \citealt{Littlefair_2007}). According to the multiwavelength period-bouncer scorecard of \citet{Daniela_2024}—which consists of ten parameters derived from multiwavelength observational characteristics expected for an ideal period-bouncer—OV\,Boo has a 50\% probability of being a period-bouncer. Given this ambiguity, we excluded OV\,Boo from the analysis. After discarding these systems, the sample of pre-bounce CVs comprises 54 systems.

The period-bouncer sample is adopted from the catalogue of period-bounce CVs presented by \citet{Daniela_2026}, which builds upon the initial catalogue of \citet{Daniela_2024}. This catalogue comprises $39$ systems identified as period-bouncers based on their observational properties being consistent with expectations for highly evolved CVs. While only a subset of these systems have a direct detection of a very late-type donor, all exhibit multi-wavelength characteristics supporting their classification as period-bouncers. In this work, the period-bouncer sample initially comprised the 15 systems from \citet{Daniela_2026} that were observed during SDSS I–IV. This catalogue includes \textit{Gaia} DR3 IDs, $P_{\rm orb}$ values and magnetic/non-magnetic classifications (see \citealt{Daniela_2024}). 

\begin{figure}
    \centering
    \includegraphics[width=0.99\linewidth]{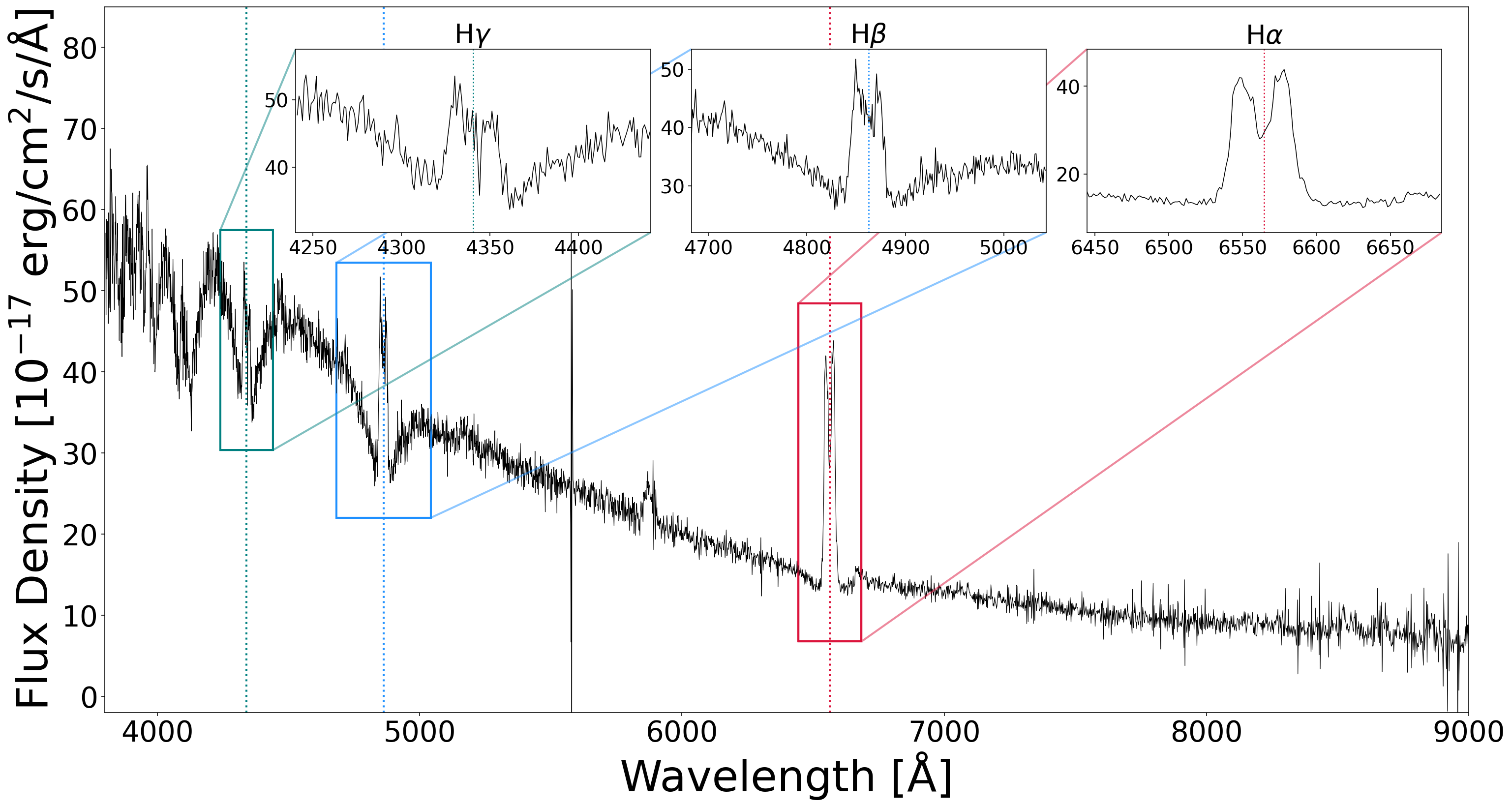}
    \caption{SDSS spectrum of V406\,Vir (plate 335, fibre 85, MJD 52000) showing double-peaked Balmer emission lines.}
    \label{Fig.V406_Vir}
\end{figure}

The number of confirmed magnetic period-bouncers is too small to allow a meaningful statistical analysis, therefore, we aim to focus on disc-dominated accretion. We discarded all magnetic CVs, that is, objects classified as polars or intermediate polars in \citet{Inight_2023}, or as magnetic in \citet{Daniela_2024}. We note that although V406\,Vir is known to host a magnetic WD, its field strength ($50\,\mathrm{kG} \lesssim \langle B \rangle \lesssim 100\,\mathrm{kG}$; \citealt{Pala_2018}) is too weak to significantly disrupt the accretion disc, as evidenced by the double-peaked Balmer emission lines (see Fig.~\ref{Fig.V406_Vir}) which trace the Keplerian velocity field of the rotating gas in the disc. This system was therefore retained in the sample. After discarding the magnetic CVs, the samples comprise 43 pre-bounce CVs, 12 period-bounce CVs, and 3 period-bounce candidates.

For both samples, we retrieved SDSS I-IV optical spectra using their J2000 coordinates and a search radius of 3\arcsec. Using the \textit{Gaia} DR3 IDs, the \textit{Gaia} DR3 photometry was retrieved through the \textit{Gaia} Archive\footnote{\citealt{gaia_archive}, \textit{Gaia} Archive} and distances were adopted from \citet{Bailer-Jones}.

\begin{figure}
    \centering
    \includegraphics[width=0.499\textwidth]{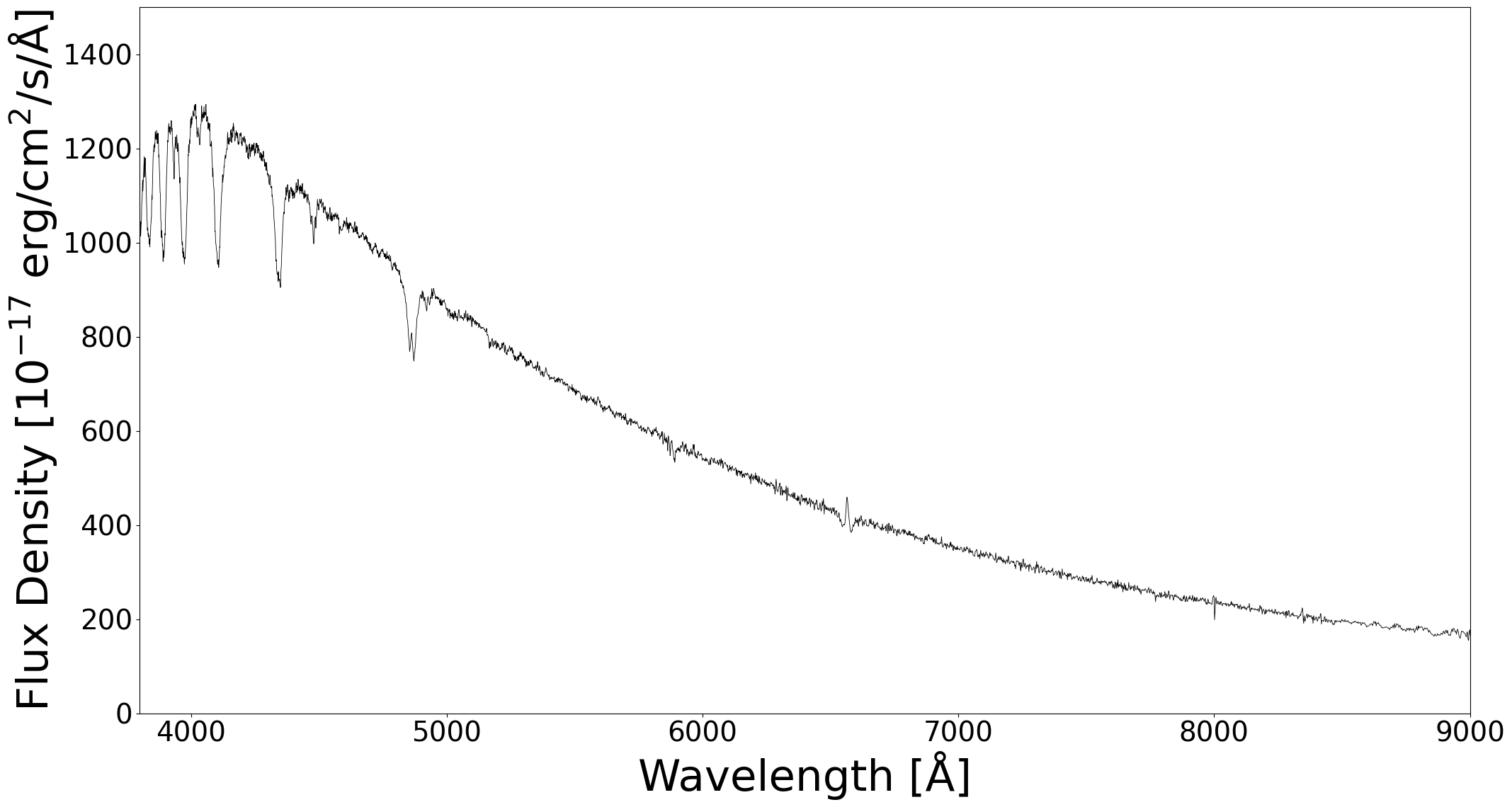}
    \vspace{0.6cm} 
    \includegraphics[width=0.499\textwidth]{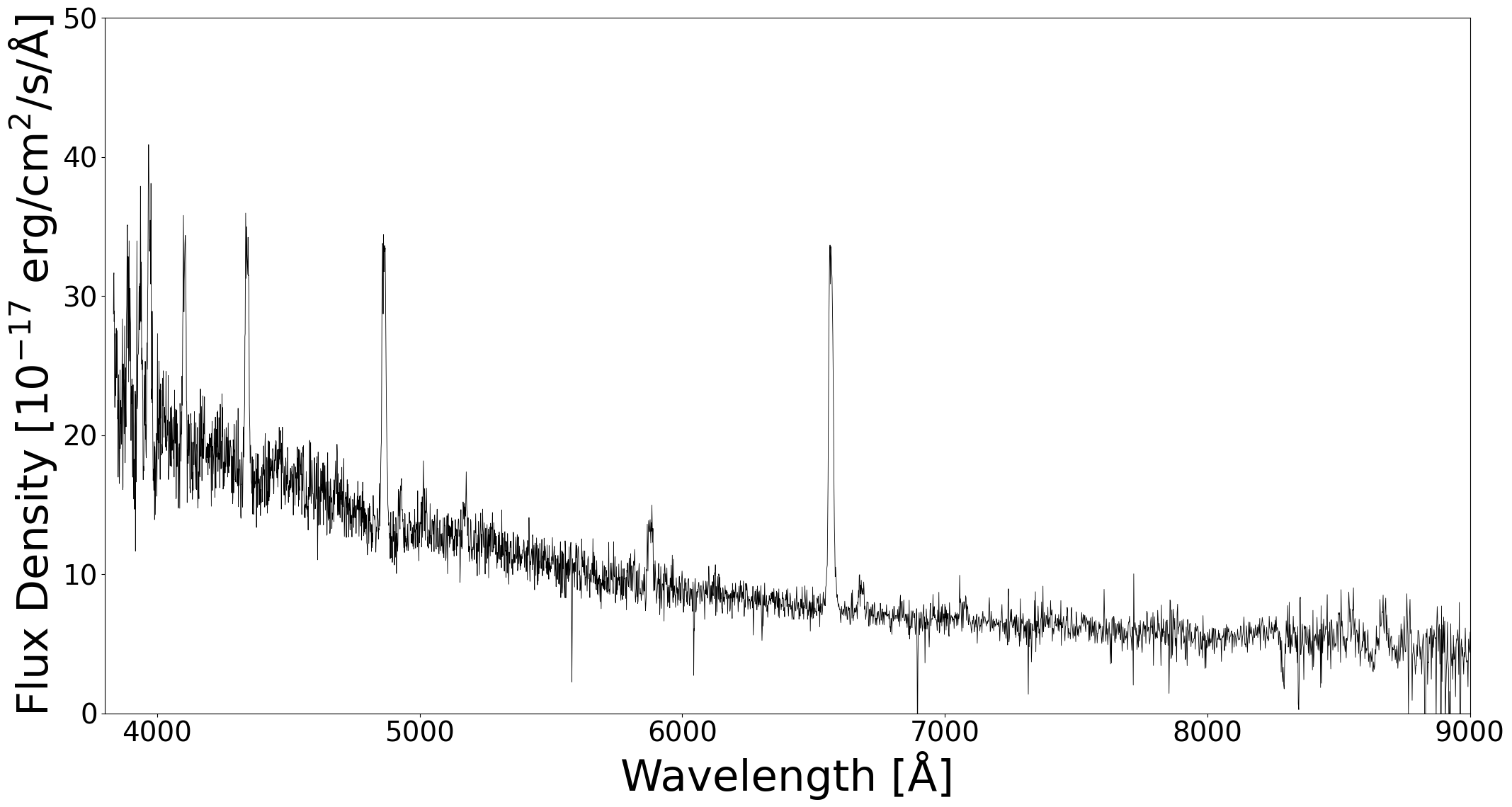}
    \caption{SDSS spectra of AK\,Cnc obtained at two different epochs. Top: AK\,Cnc observed during or near an outburst (plate 5293, fibre 322, MJD 55953). Bottom: AK\,Cnc observed in quiescence (plate 2575, fibre 318, MJD 54085).}
    \label{fig:AKCnc}
\end{figure}

\begin{figure}
    \centering
    \includegraphics[width=0.99\linewidth]{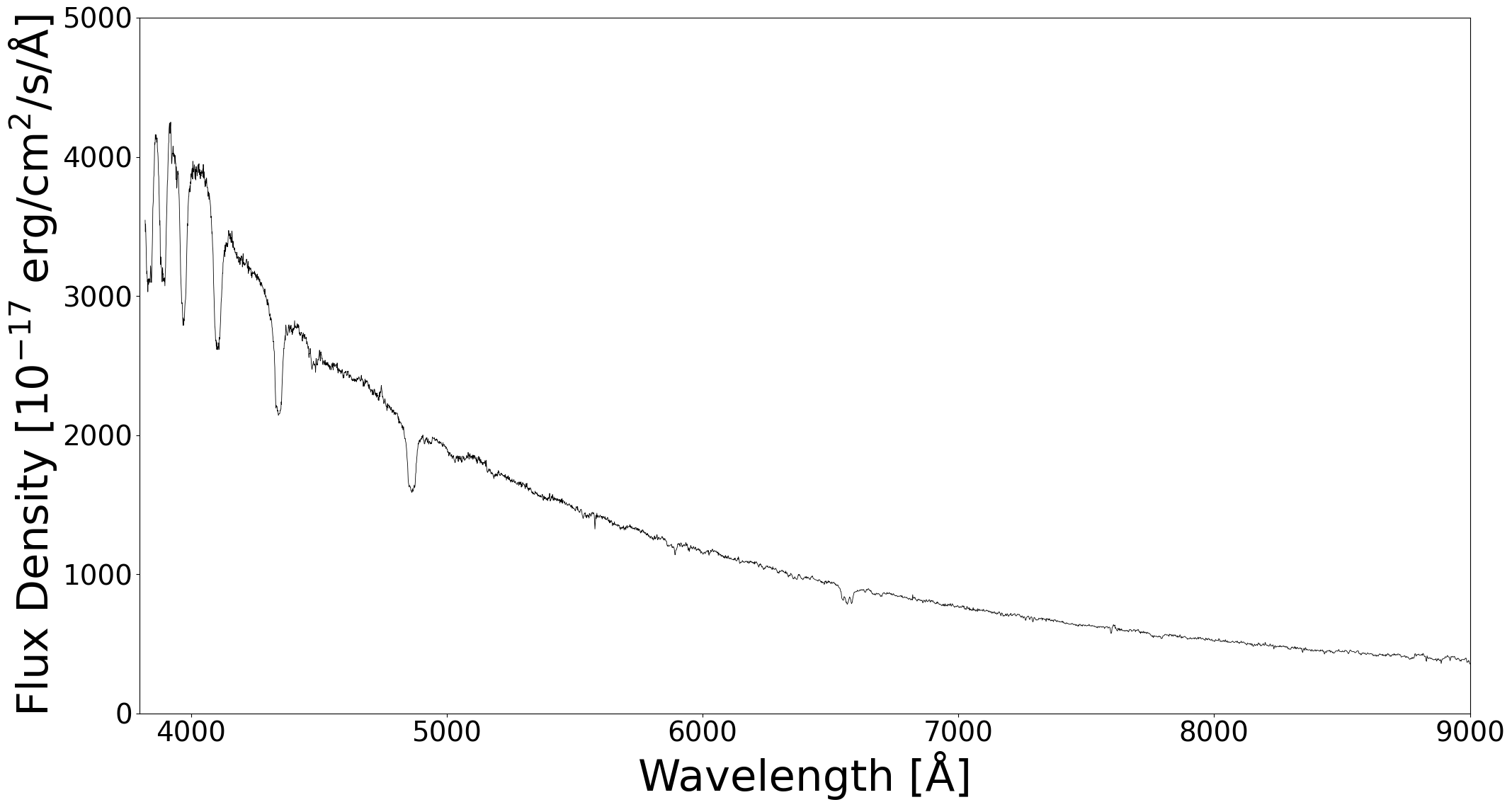}
    \caption{SDSS spectrum of QW\,Ser (plate 1721, fibre 21, MJD 53857) observed during or near an outburst.}
    \label{Fig.QWSer}
\end{figure}

Several systems have multiple SDSS observations obtained at different epochs. Some individual SDSS observations were discarded due to problems in the spectrum or because the spectrum shows evidence of being observed during, or close to, an outburst. In such cases, the continuum becomes significantly bluer with respect to the quiescent state (see e.g. \citealt{Bailey_1980}, \citealt{Cathy_1990}, and \citealt{Nogami_2004}), and the Balmer lines can appear in absorption as the optically thick accretion disc dominates the spectrum, producing broad absorption profiles (see e.g. \citealt{Hessman_1984} and \citealt{Kromer_2007}). Spectra observed near an outburst can lead to steeper Balmer decrements, that is $\text{H}{\alpha}/\text{H}{\beta}>1$ and $\text{H}{\gamma}/\text{H}{\beta}<1$, as the inner regions of the accretion disc become ionised or collapse onto the WD, causing the suppression of line emission from the densest and hottest disc regions that would otherwise produce a flatter decrement (see e.g. \citealt{Clarke_1984}). Including these observations would therefore bias the results. All spectra were visually inspected to identify signatures of an outburst and confirmed by comparison with spectra of the same objects obtained at different epochs (see Fig.~\ref{fig:AKCnc} for an example). For QW\,Ser there is only one available SDSS spectrum, which nevertheless shows clear evidence of having been obtained near an outburst (see Fig.~\ref{Fig.QWSer}).

The 12 discarded spectra, along with the reasons for their exclusion, are listed in Table~\ref{table:Discarded}.  Additionally, we note that the spectrum of IR\,Com (plate 2613, fibre 523, and MJD 54481) shows a gap in the blue region that affects the H$\gamma$ line (see Fig.~\ref{fig:IRCom} in Sect.~\ref{subsect:diagnostic_diagram}). Since the H$\alpha$ and H$\beta$ lines remain unaffected, the spectrum was retained.

\begin{figure}
    \centering
    \includegraphics[width=0.9\linewidth]{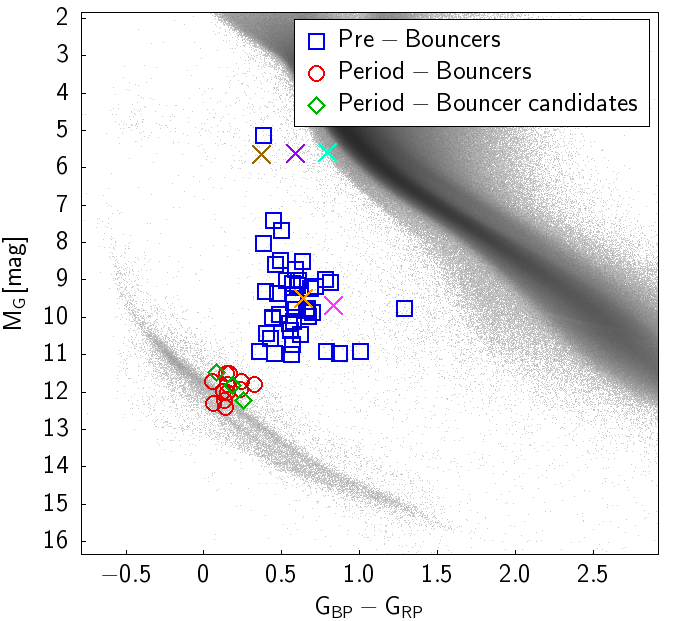}
    \caption{\textit{Gaia} colour–magnitude diagram for the samples of pre-bounce and period-bounce CVs. The grey points in the background represent the {\it Gaia}\,DR3 sources with distances from \citet{Bailer-Jones} and a parallax error of $<1\,\%$ of the parallax value, which are shown as a reference. The centroid positions of different CV types in the \textit{Gaia} colour-magnitude diagram derived by \citet{Abril_2020} are overlaid as crosses: brown for nova-like systems, cyan for old novae, orange for dwarf novae,  pink for polars, and purple for intermediate polars.}
    \label{Fig.CMD_Samples}
\end{figure}

The final samples comprise 12 period-bounce CVs with a total of 17 SDSS spectra, 3 period-bouncer candidates with 3 SDSS spectra, and 42 pre-bounce CVs with 62 SDSS spectra. A \textit{Gaia} colour–magnitude diagram for the final samples is shown in Fig.~\ref{Fig.CMD_Samples}. All pre-bounce CVs are located between the WD sequence and the main sequence, in the typical area for CVs (see the centroid position of different types of CVs from \citet{Abril_2020} overlaid as cross symbols in Fig.~\ref{Fig.CMD_Samples}). Period-bouncers and period-bouncer candidates lie within the WD locus, consistent with their evolutionary status, as the very low-mass donor in period-bouncers contributes primarily at infrared wavelengths.

The journal of observations and the results from the analysis described in Sect.~\ref{Line_Flux_sect} are available electronically at CDS (see Table~\ref{table:Master} in Appendix~\ref{sec:Table} for a description of contents).

\section{Emission line fluxes}
\label{Line_Flux_sect}

\subsection{Subtraction of the white dwarf contribution}
\label{WD_correction}

\begin{figure*}
    \centering
    \includegraphics[width=0.9\linewidth]{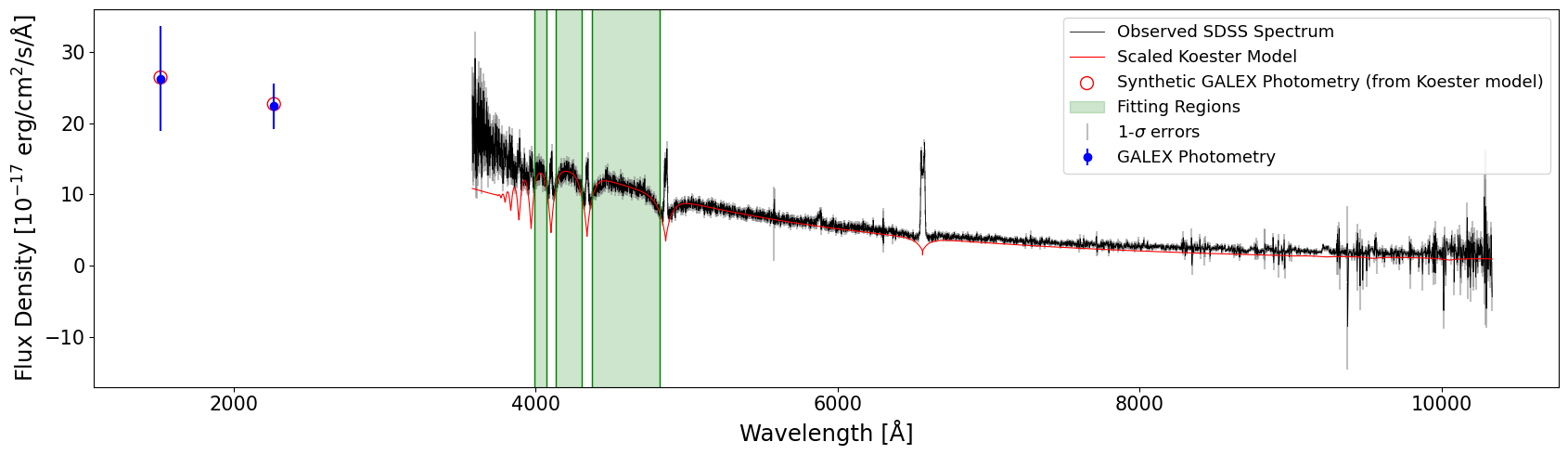}
    \caption{SDSS spectrum of PM\,J12192+2049 (plate 5978, fibre 185, and MJD 56073) fitted with a \citet{koester2010} DA WD photospheric model, as an example of a spectrum for which the WD contribution was subtracted. The best-fitting model corresponds to $\log(g)$=7.75 and $T_{\rm eff}$=11750\,K.}
    \label{Fig.WD_fitting}
\end{figure*}

The SDSS spectra of all systems were visually inspected to identify broad Balmer absorption from the WD photosphere superposed on the emission component from the accretion disc. Neglecting this WD contribution would systematically reduce the measured emission line fluxes, and because the strength of the WD Balmer line absorption varies depending on the WD effective temperature, $T_{\rm {eff}}$, and surface gravity, $\log(g)$ (see e.g. \citealt{Tremblay_2011}), failing to correct for this WD photospheric absorption would result in biased Balmer decrements. For typical WDs in CVs around the period minimum, with $T_{\rm {eff}}\sim10000-15000$\,K, correcting for the WD contribution generally produces flatter Balmer decrements as the absorption in H${\beta}$ is stronger than in H${\alpha}$. This is particularly relevant for the sample of period-bouncers, for which such corrections were required in all 17 observed spectra. 

For systems exhibiting WD absorption according to our visual inspection, we fitted a hydrogen-dominated (DA) WD atmosphere model to the spectrum and subtracted its contribution. The fitting procedure follows the same approach as in \citet{Daniela_2026}, which we summarise here. We employ pure-hydrogen, Local Thermodynamic Equilibrium (LTE) models from \citet{koester2010}, spanning effective temperatures $T_{\rm {eff}}=8000-30000$\,K. The grid spacing is 250\,K between 8000 and 20000\,K, and 1000 K above 20000\,K, with $\log(g)$=6.5 - 9.5 in steps of 0.25\,dex. Each spectral model is resampled to the same wavelength resolution as the observed spectra via cubic-spline interpolation.

To resolve possible degeneracies in the optical fitting, we incorporate near- and far-ultraviolet photometry from the \textit{Galaxy Evolution Explorer} \citep[GALEX;][]{Martin_2005, Morrissey_2007, Bianchi_2017}, as ultraviolet fluxes are highly sensitive to the WD temperature. Synthetic photometric fluxes for both GALEX bands are computed from the model spectra using the corresponding instrument filter response curves (see \citealt{Casagrande}). 

The model fitting is performed by minimising a chi-square function that includes both the SDSS spectroscopic and GALEX photometric contributions,

\begin{equation}
\begin{split}
& \chi^2 = \sum_{i} \left( \frac{f_{\text{obs},i}(\lambda) - K_{\text{WD}} f_{\text{model},i}(\lambda)}{\sigma_{\text{obs},i}(\lambda)}\right)^2\\
& +\sum_{j} \left( \frac{F_{\text{obs, GALEX},j} - K_{\text{WD}} F_{\text{model, GALEX},j}}{\sigma_{\text{obs, GALEX},j}}\right)^2,
\end{split}
\label{eq:chi_squared}
\end{equation}

\noindent Here, $f_{\text{obs},i}$ denotes the observed flux density values, with  $\sigma_{\text{obs},i}$ representing their associated uncertainties, while $f_{\text{model},i}$ are the surface flux density values from the WD model. The terms $F_{\text{obs, GALEX},j}$ and $\sigma_{\text{obs, GALEX},j}$ refer to the observed GALEX fluxes and their uncertainties, and $F_{\text{model, GALEX},j}$ are the synthetic GALEX surface fluxes derived from the WD model spectra. Finally, the scaling factor $K_{\text{WD}}$ is the free parameter to be optimised. Physically, $K_{\rm{WD}}$ corresponds to the dilution factor, $(R_{WD}/d)^2$, where $R_{WD}$ is the radius of the WD and $d$ is the distance to the object.

The fit is performed over specific wavelength regions selected to exclude the red optical range and to mask the Balmer emission cores as illustrated in Fig.~\ref{Fig.WD_fitting}. At wavelengths $\gtrsim5000$\,\AA, the accretion disc begins to contribute noticeably to the continuum flux, while in systems near the period minimum, the very low-mass donor star starts to contribute only at $\lambda\gtrsim7000\,\AA$ (see e.g. \citealt{Gansicke_2009}, \citealt{Zharikov_2013}, and \citealt{Amantayeva_2021}). The spectral regions selected for the fitting also encompass the broad Balmer absorption wings from the WD, which are important to constrain $\log(g)$.

Among the models yielding low total $\chi^2$ values (Eq.~\ref{eq:chi_squared}), the best-fit model is adopted as the one in best agreement with the GALEX photometry. This model is then subtracted from the observed spectrum to correct for the WD photospheric contribution. We note that for some systems GALEX photometry was unavailable. In these cases, the fit was performed using only the optical SDSS spectrum and the best-fit model was chosen as the one minimising the first component in the right-hand side of Eq.~\ref{eq:chi_squared}. Additionally, for a few systems where GALEX photometry was available but no model could satisfactorily reproduce the ultraviolet fluxes together with the SDSS spectrum, the GALEX photometry was found to be likely contaminated by nearby sources and was therefore excluded from the fitting process. These systems are 1RXS\,J101421.4+063855, SDSS\,J102905.24+485515.2, and IR\,Com. 

We note that we do not include an accretion disc component in our fitting procedure, implicitly assuming a small contribution in the optical blue and UV wavelengths. In studies where a power law is used as an approximation for the accretion disc, its UV contribution is typically $\sim$10–20\% for period-bouncers and negligible in some cases (see \citealt{Pala_2022}). As our approximation may bias our best-fitting models compared to methods that explicitly include a spectral component for the accretion disc, we assessed the impact of the WD subtraction for a subset of spectra by refitting models with $T_{\mathrm{eff}}$ offset by $\pm500$\,K relative to the best-fit value and remeasuring the line fluxes (see Sect.~\ref{Line_Flux_Measurement}). We found that the changes are within the uncertainties of the flux values, thus supporting the robustness of our analysis.

Table~\ref{table:WD_fittings} in Appendix~\ref{sec:WD_fittings} lists the 34 SDSS observations to which a WD model was fitted to correct for the WD absorption in the spectral lines. For each spectrum, the table reports the $T_{\rm eff}$ and $\log(g)$ of the best-fitting DA WD atmosphere model, together with the corresponding scaling factor $K_{\rm WD}$. A flag indicates whether GALEX photometry was included in the fitting procedure. These WD parameters should be interpreted as approximate physical parameters, particularly for pre-bounce CVs, where the accretion disc is expected to contribute to the optical continuum.

\subsection{Emission line flux measurements}
\label{Line_Flux_Measurement}

The fluxes of the emission lines were measured using an interactive routine that allows to define the local continuum and the integration window of the emission line. Two continuum regions on either side of the line are selected and fitted with a weighted ordinary least squares linear regression (WOLS(Y|X); see \citealt{weighted_linear_regression}), where the flux density uncertainties are used as weights $1/\sigma_{\text{f}}^2$. This linear model serves as a local continuum estimate across the line profile. The line flux is then measured by subtracting this local continuum and integrating the residual line profile within the chosen wavelength window. Integration is performed numerically by summing the residual flux density in each spectral bin multiplied by its corresponding wavelength width across the selected spectral window. This ensures that the full flux of the line is recovered regardless of its detailed profile shape. The calculation of the uncertainties is explained in Appendix~\ref{sec:uncertainties}.

\section{Results}
\label{subsect:results}

\begin{figure*}
    \centering
    \includegraphics[width=0.45\linewidth]{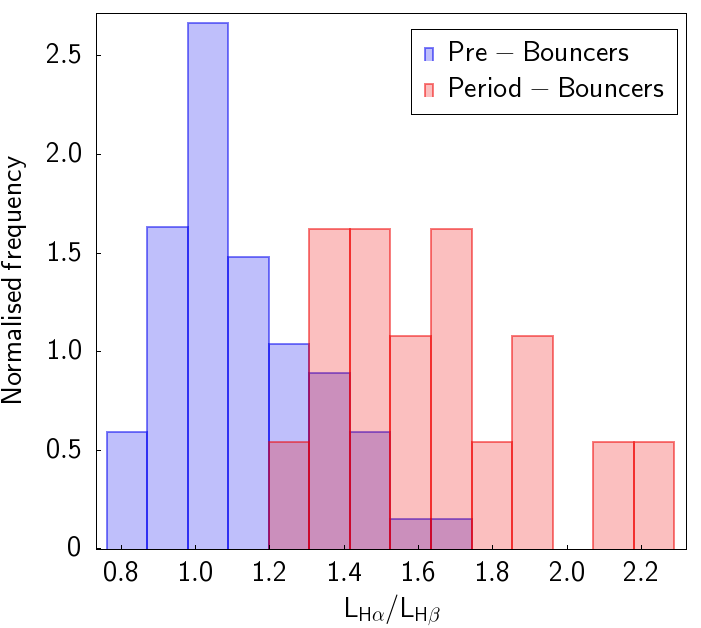}
    \includegraphics[width=0.45\linewidth]{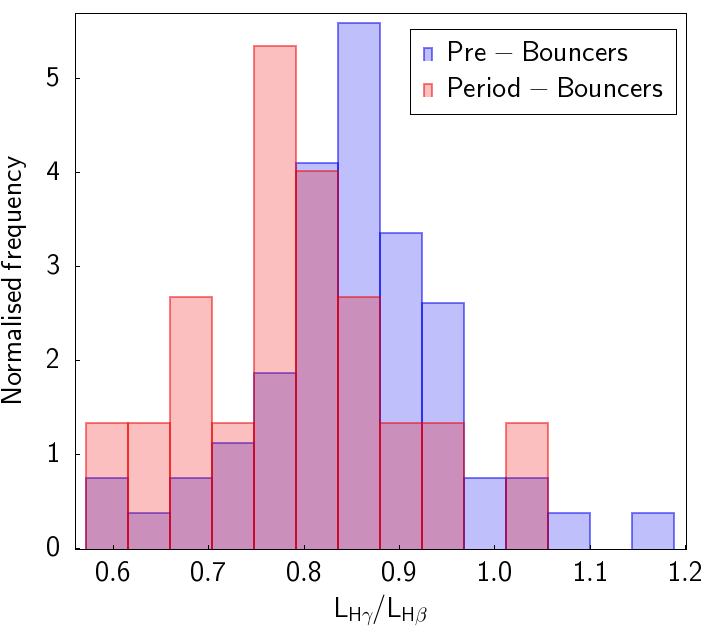}
    \caption{Area normalised distributions of the observed Balmer decrements for the samples of pre-bouncers and period-bouncers. The distributions are normalised so that the total area under each distribution equals one, making the distributions equivalent to empirical probability density functions. Left: $L_{\mathrm{H}{\alpha}}/L_{\mathrm{H}{\beta}}$ ratios. Right: $L_{\mathrm{H}{\gamma}}/L_{\mathrm{H}{\beta}}$ ratios.}
    \label{Fig.Ratios_Distributions}
\end{figure*}

We measured the H${\alpha}$, H${\beta}$, and H${\gamma}$ line fluxes for 82 SDSS spectra, comprising 17 spectra of period-bounce CVs, 3 spectra of period-bouncer candidates, and 62 spectra of short-period pre-bounce CVs (see Sect.~\ref{sect.Sample}). Using the distances from \citet{Bailer-Jones}, we converted these fluxes into line luminosities (hereafter $L_{\mathrm{H}{\alpha}}$, $L_{\mathrm{H}{\beta}}$, and $L_{\mathrm{H}{\gamma}}$) (see Table~\ref{table:Master} and Appendix~\ref{sec:uncertainties}).

For most pre-bouncer systems, we find nearly flat Balmer decrements, with $L_{\mathrm{H}{\alpha}}/L_{\mathrm{H}{\beta}}\sim 1$ and $L_{\mathrm{H}{\gamma}}/L_{\mathrm{H}{\beta}}\sim 0.85$ (see Fig.~\ref{Fig.Ratios_Distributions}), in agreement with previous studies of Balmer decrements in CVs (see \citealt{Williams_1983}). 
In contrast, the distributions of $L_{\mathrm{H}{\alpha}}/L_{\mathrm{H}{\beta}}$ and $L_{\mathrm{H}{\gamma}}/L_{\mathrm{H}{\beta}}$ ratios for period-bouncers are steeper, that is they exhibit systematically higher $L_{\mathrm{H}{\alpha}}/L_{\mathrm{H}{\beta}}$ and lower $L_{\mathrm{H}{\gamma}}/L_{\mathrm{H}{\beta}}$ values than pre-bouncers (see Fig.~\ref{Fig.Ratios_Distributions}). These results are consistent with previous individual studies of period-bouncers (see e.g. \citealt{Pala_2018_QZLib}, \citealt{Amantayeva_2021} and \citealt{Neustroev_2023}). Our findings on a larger sample suggests that we can statistically separate the population of period-bouncers from pre-bouncer systems based on their Balmer decrements, which reflect differences in the physical conditions under which the Balmer lines are formed in each class.

\begin{figure*}
    \centering
    \includegraphics[width=0.45\linewidth]{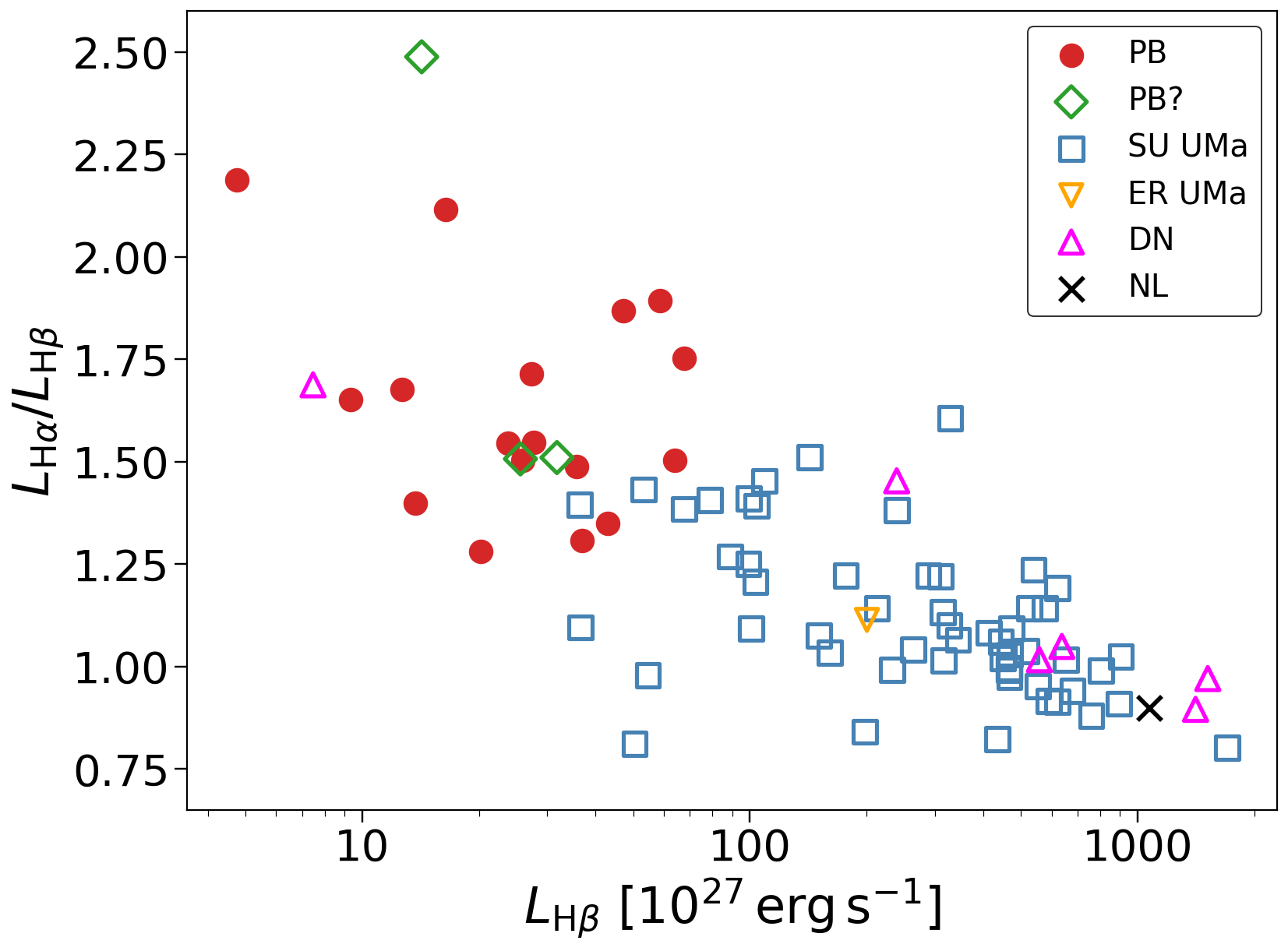}
    \includegraphics[width=0.45\linewidth]{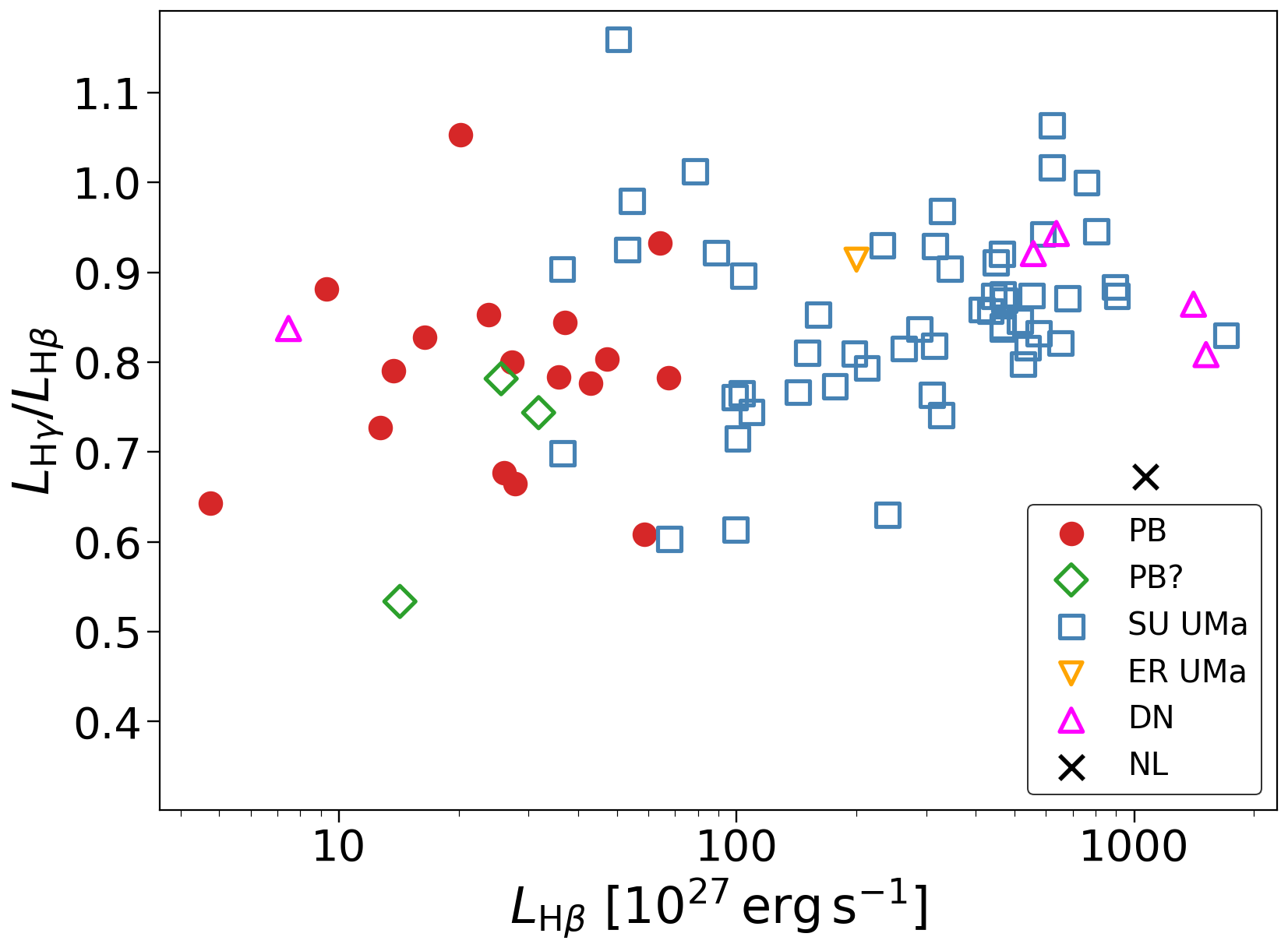}
    \caption{Observed Balmer decrements as a function of the $\mathrm{H}{\beta}$ line luminosity. Left: $L_{\mathrm{H}{\alpha}}/L_{\mathrm{H}{\beta}}$ ratio versus $L_{\mathrm{H}{\beta}}$. Right: $L_{\mathrm{H}{\gamma}}/L_{\mathrm{H}{\beta}}$ ratio versus $L_{\mathrm{H}{\beta}}$. Symbol shapes and colours identify different CV subclasses as follows: PB = period-bouncer, PB? = period-bouncer candidate, SU\,UMa = SU\,UMa-type dwarf nova, ER\,UMa = ER\,UMa-type dwarf nova, DN = dwarf nova, and NL = nova-like variable.}
    \label{Fig.BDvsLHbeta}
\end{figure*}

Figure~\ref{Fig.BDvsLHbeta} presents the $L_{\mathrm{H}{\alpha}}/L_{\mathrm{H}{\beta}}$ and $L_{\mathrm{H}{\gamma}}/L_{\mathrm{H}{\beta}}$ ratios as a function of the H$\beta$ line luminosity, which we chose as a proxy for the mass accretion rate. As can be seen, H$\beta$ line luminosities are systematically lower for period-bouncers and are associated with steeper Balmer decrements. The observed correlations are quantified for the full data set of pre-bounce and period-bounce CVs with the Spearman correlation coefficient (\citealt{Spearman_1904}), which gives $\rho_{\rm s}=-0.77$ for the $L_{\mathrm{H}{\alpha}}/L_{\mathrm{H}{\beta}}$ ratios and $\rho_{\rm s}=0.35$ for the $L_{\mathrm{H}{\gamma}}/L_{\mathrm{H}{\beta}}$ ratios, each against $L_{\mathrm{H}{\beta}}$. This continuously steeper observed Balmer decrements towards lower $L_{\mathrm{H}{\beta}}$ suggests that as CVs evolve toward and beyond the minimum $P_{\rm orb}$, their mass-transfer rates decline steadily, leading to cooler and more optically thin emission regions (see Sect.~\ref{sec:TheoreticalModels}). We note that no correction was applied for the effect of binary inclination, since the orientation of most systems is unknown. The spread due to different inclinations likely contributes to the observed scatter in Fig.~\ref{Fig.BDvsLHbeta}.

\begin{figure}
    \centering
    \includegraphics[width=0.9\linewidth]{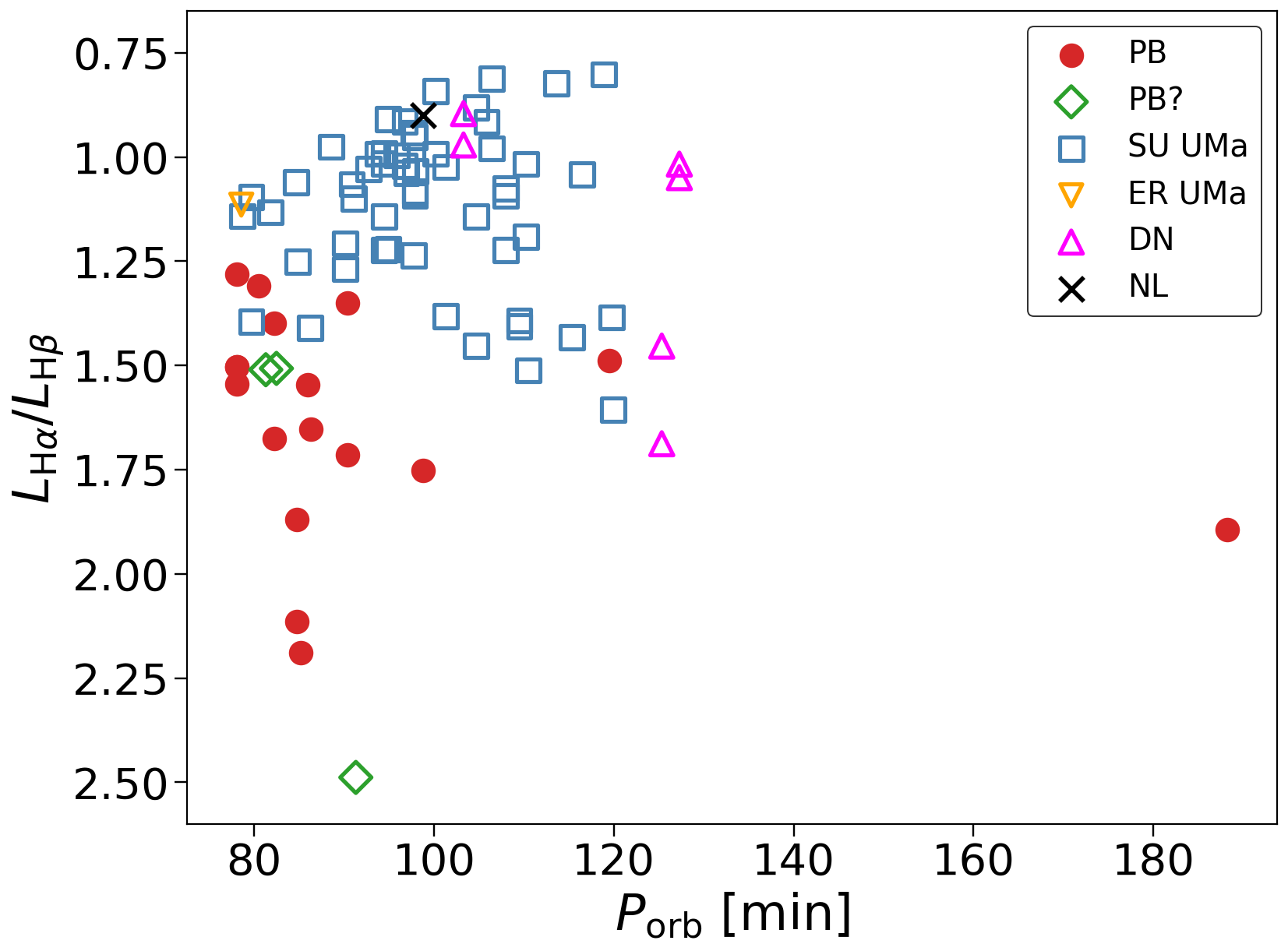}
    \caption{Observed $L_{\mathrm{H}{\alpha}}/L_{\mathrm{H}{\beta}}$ ratio as a function of $P_{\rm orb}$. Symbol shapes and colours identify different CV subclasses with the same notation as in Fig.~\ref{Fig.BDvsLHbeta}.}
    \label{Fig.BD_vs_Porb}
\end{figure}

The relation between the $L_{\mathrm{H}{\alpha}}/L_{\mathrm{H}{\beta}}$ ratio and $P_{\rm orb}$ (see Fig.~\ref{Fig.BD_vs_Porb}) further supports this evolutionary interpretation. In particular, for $L_{\mathrm{H}{\alpha}}/L_{\mathrm{H}{\beta}}\gtrsim1.5$, the sample is dominated almost exclusively by period-bouncers and period-bouncer candidates.

The left panel of Figure~\ref{Fig.Diagnostic_Diagram} shows the relationship between the $L_{\mathrm{H}{\gamma}}/L_{\mathrm{H}{\beta}}$ and $L_{\mathrm{H}{\alpha}}/L_{\mathrm{H}{\beta}}$ ratios for different CV subclasses. An anti-correlation is observed across the full data set, with a Spearman coefficient $\rho_{\rm s}=-0.44$, consistent with the behaviour expected from theoretical models of accretion discs (see Sect.~\ref{sec:TheoreticalModels}).

\subsection{Diagnostic diagram}
\label{subsect:diagnostic_diagram}

To statistically separate period-bouncers from short-period pre-bounce CVs, we trained a linear logistic regression model (\citealt{Berkson_1944}, \citealt{Cox_1958}) to the $L_{\mathrm{H}{\gamma}}/L_{\mathrm{H}{\beta}}$ versus $L_{\mathrm{H}{\alpha}}/L_{\mathrm{H}{\beta}}$ diagram. The model has the form,

\begin{equation}
P_{\text{PB}}=\dfrac{1}{1+\exp\left[-\left(\beta_{0} + \beta_{1}\cdot \left(\dfrac{L_{\mathrm{H}\alpha}}{L_{\mathrm{H}\beta}}\right) + \beta_{2}\cdot \left(\dfrac{L_{\mathrm{H}\gamma}}{L_{\mathrm{H}\beta}}\right)\right)\right]}
\label{Eq.LRModel}
\end{equation}

\noindent where $P_{\text{PB}}$ denotes the probability that a system is a period-bouncer and $\beta_{i}$ ($i = 0,1,2$) are the model coefficients. We note that the three period-bouncer candidates are not used for the training and evaluation of the model. 

\begin{figure*}
    \centering
    \includegraphics[width=0.45\linewidth]{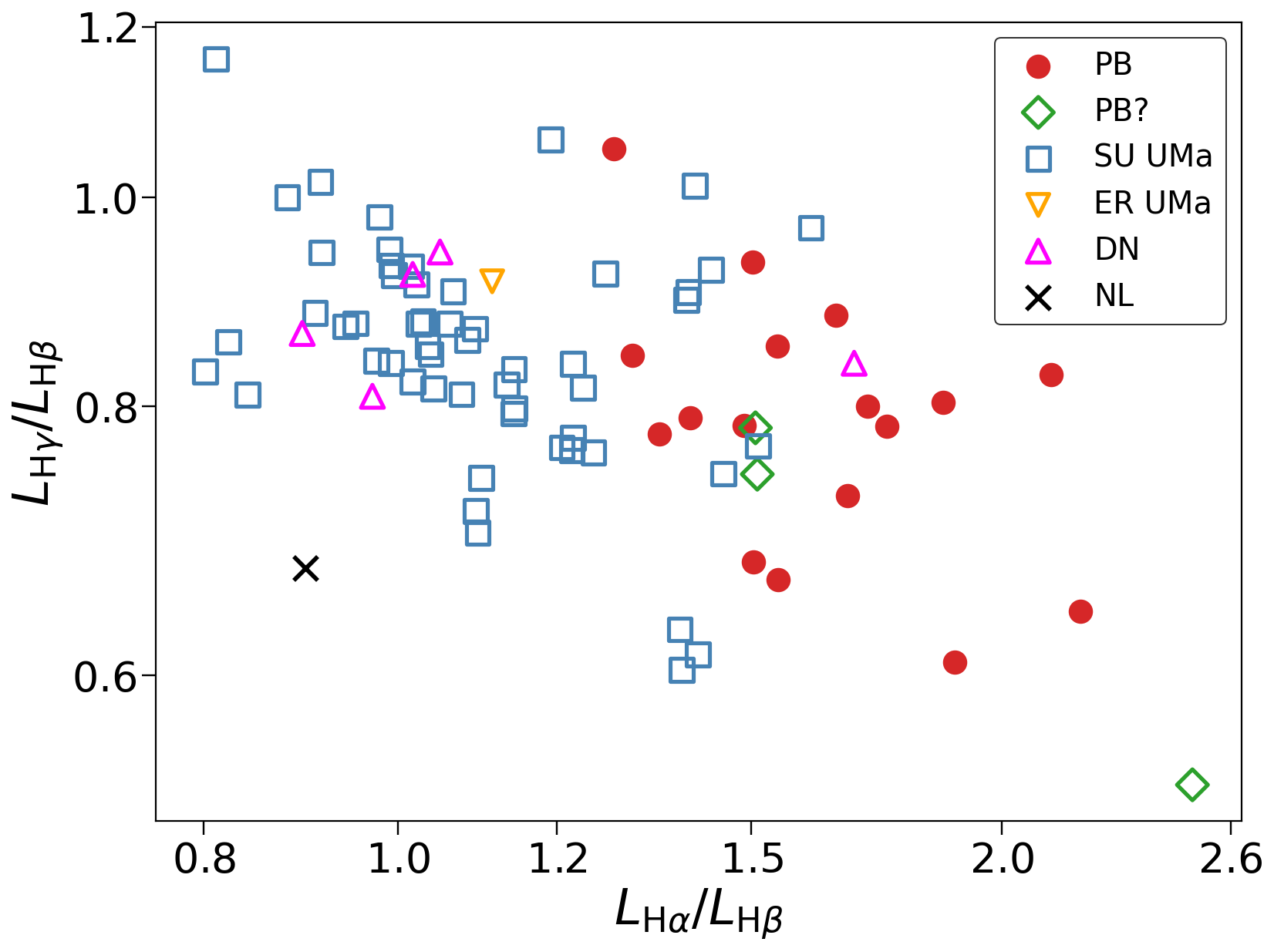}
    \includegraphics[width=0.50\linewidth]{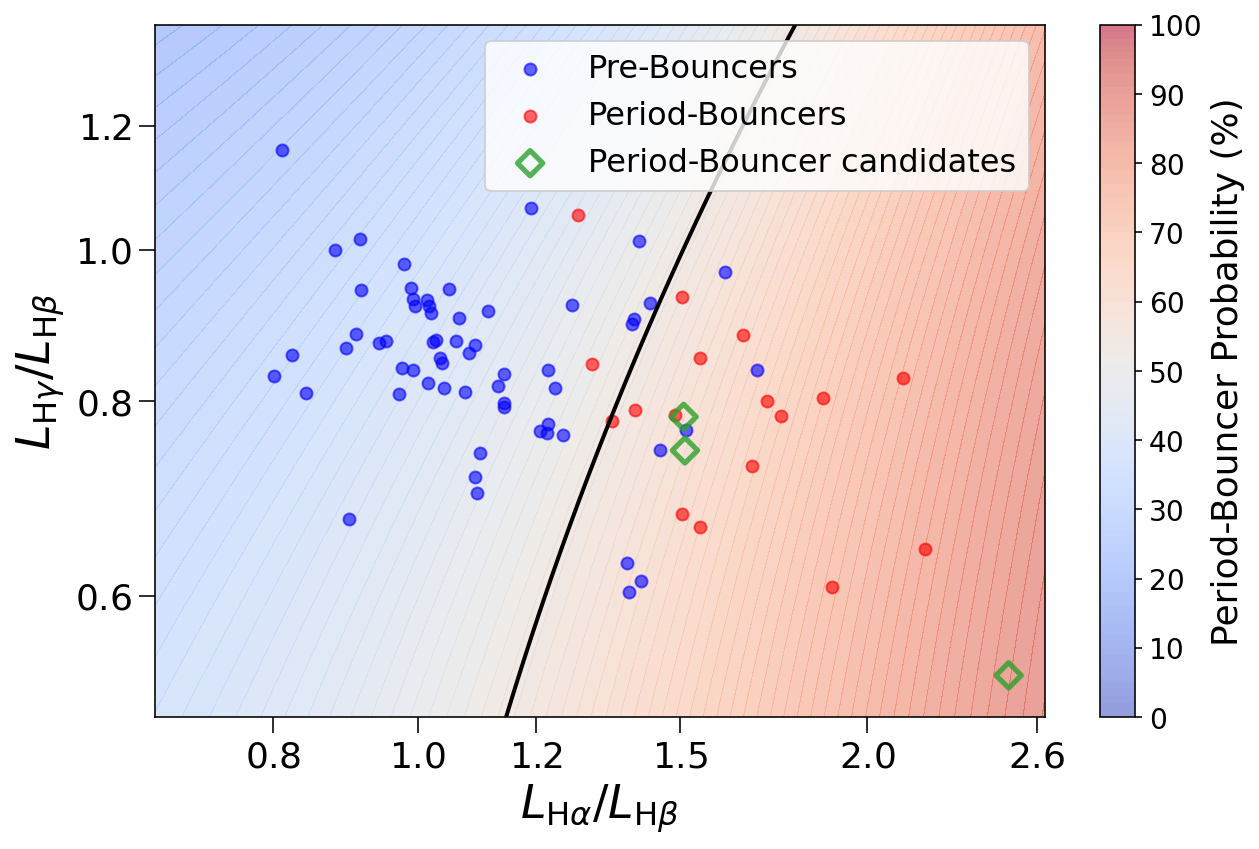}
    \caption{Observed $L_{\mathrm{H}{\gamma}}/L_{\mathrm{H}{\beta}}$ ratio versus $L_{\mathrm{H}{\alpha}}/L_{\mathrm{H}{\beta}}$ ratio. Left panel: Symbol shapes and colours identify different CV subclasses with the same notation as in Fig.~\ref{Fig.BDvsLHbeta}. Right panel: Same data as in the left panel. The solid black line indicates the decision boundary at $P_{\text{PB}}=0.535$ that discriminates between pre-bounce and period-bounce CVs according to a linear logistic regression model (see Appendix~\ref{sec:Final_Model}). The colour gradient in the background represents the predicted probability of a system being a period-bouncer, with redder (bluer) tones corresponding to lower (higher) probability.}
    \label{Fig.Diagnostic_Diagram}
\end{figure*}

This approach allows us to estimate the probability of a system being a period-bouncer as a function of its position in the Balmer decrement parameter space, and it provides a quantitative boundary between the two samples, that we find to be located at $P_{\text{PB}}=0.535$ (see Appendix~\ref{sec:Final_Model}). In the right panel of Fig.~\ref{Fig.Diagnostic_Diagram}, we illustrate this decision boundary and the predicted probability field across the observed Balmer decrement parameter space, together with the observed decrements of our targets.

\begin{figure}
	\centering
	\includegraphics[width=0.95\linewidth]{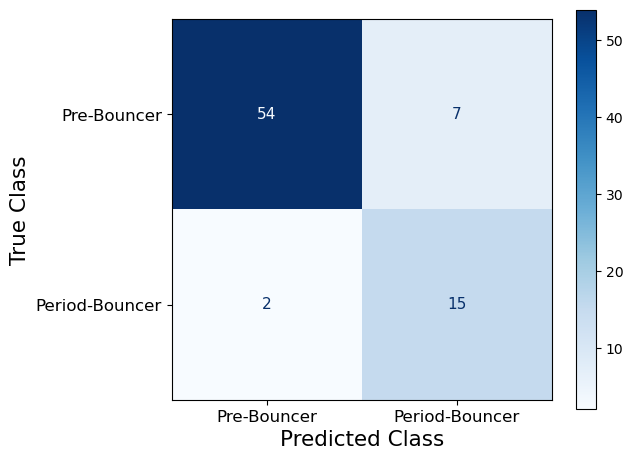}
	\caption{Confusion matrix for the logistic regression model. The matrix shows the number of true positives, true negatives, false positives, and false negatives for each class on the complete data set.}
	\label{Fig.Confusion_Matrix}
\end{figure}

The performance of the model was evaluated using standard classification metrics, including accuracy, precision, recall, and F1-score (see Eqs.~\ref{eq:precision}–\ref{eq:f1} in Appendix~\ref{sec:Evaluation_Model}). The results demonstrate a high discriminative capability with an accuracy of $0.885 \pm 0.031$, a global precision of $0.833 \pm 0.047$, a global recall of $0.888 \pm 0.016$, and a global F1-score of $0.847 \pm 0.036$. Details on the model training and evaluation procedures, and additional class-specific metrics are provided in Appendix~\ref{sec:Logistic_Regression}. The final model coefficients and their uncertainties are $\beta_{0}=-0.953\pm 0.085$,
$\beta_{1}=1.40\pm 0.11$, and 
$\beta_{2}=-1.01\pm 0.16$. In Fig.~\ref{Fig.Confusion_Matrix}, the confusion matrix corresponding to the model's classification is presented. The model correctly classifies 54 out of 61 pre-bouncers and 15 out of 17 period-bouncers. We note that the pre-bouncer sample used for the training and evaluation of the model contains 61 systems, rather than 62, as in one spectrum of IR\,Com the H$\gamma$ line could not be measured.

\subsubsection{Incorrectly classified period-bouncers}
\label{subsect:diagnostic_diagram_PBs}

\begin{figure}
    \centering
    \includegraphics[width=0.499\textwidth]{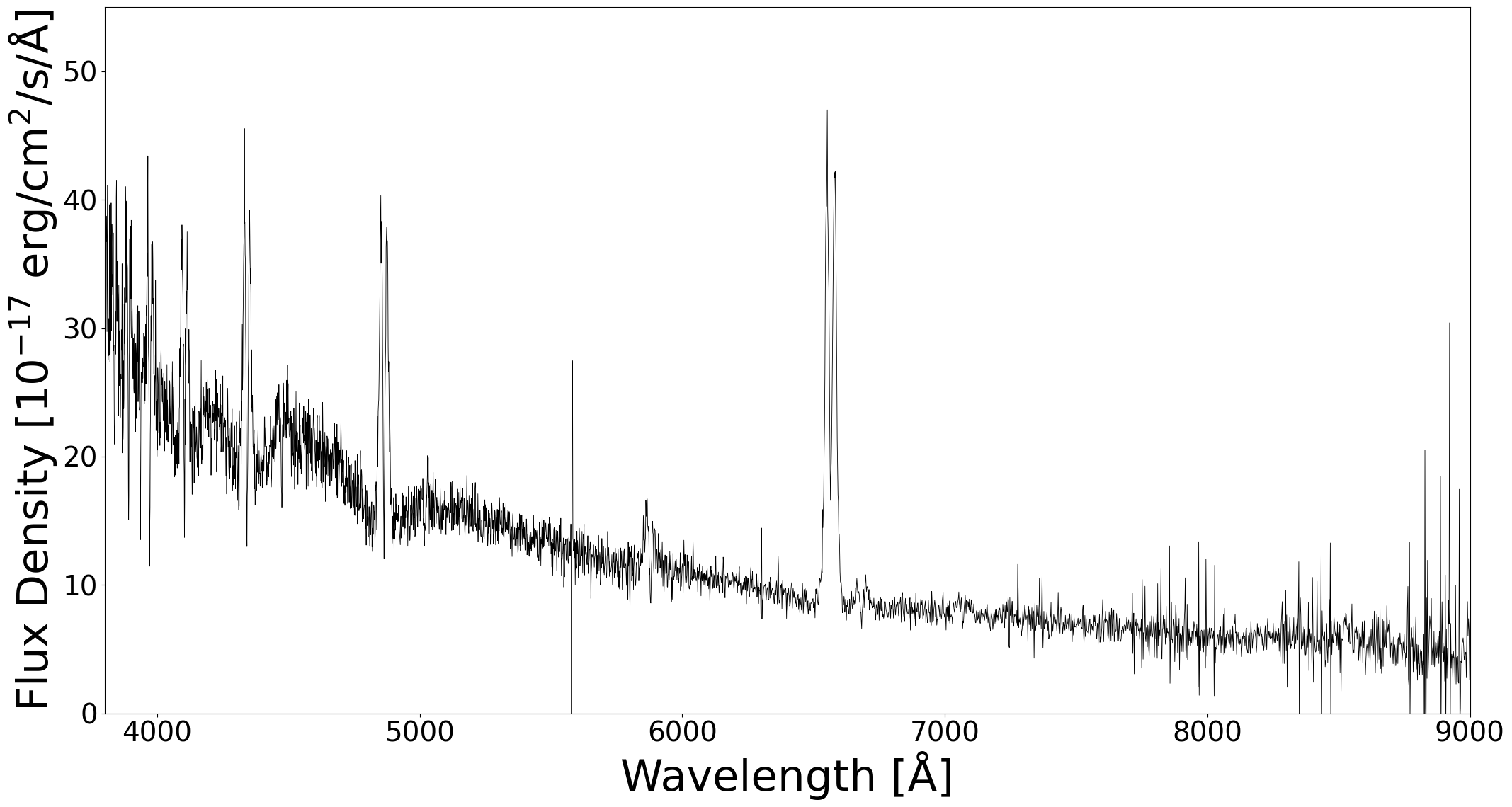}
    \vspace{0.6cm} 
    \includegraphics[width=0.499\textwidth]{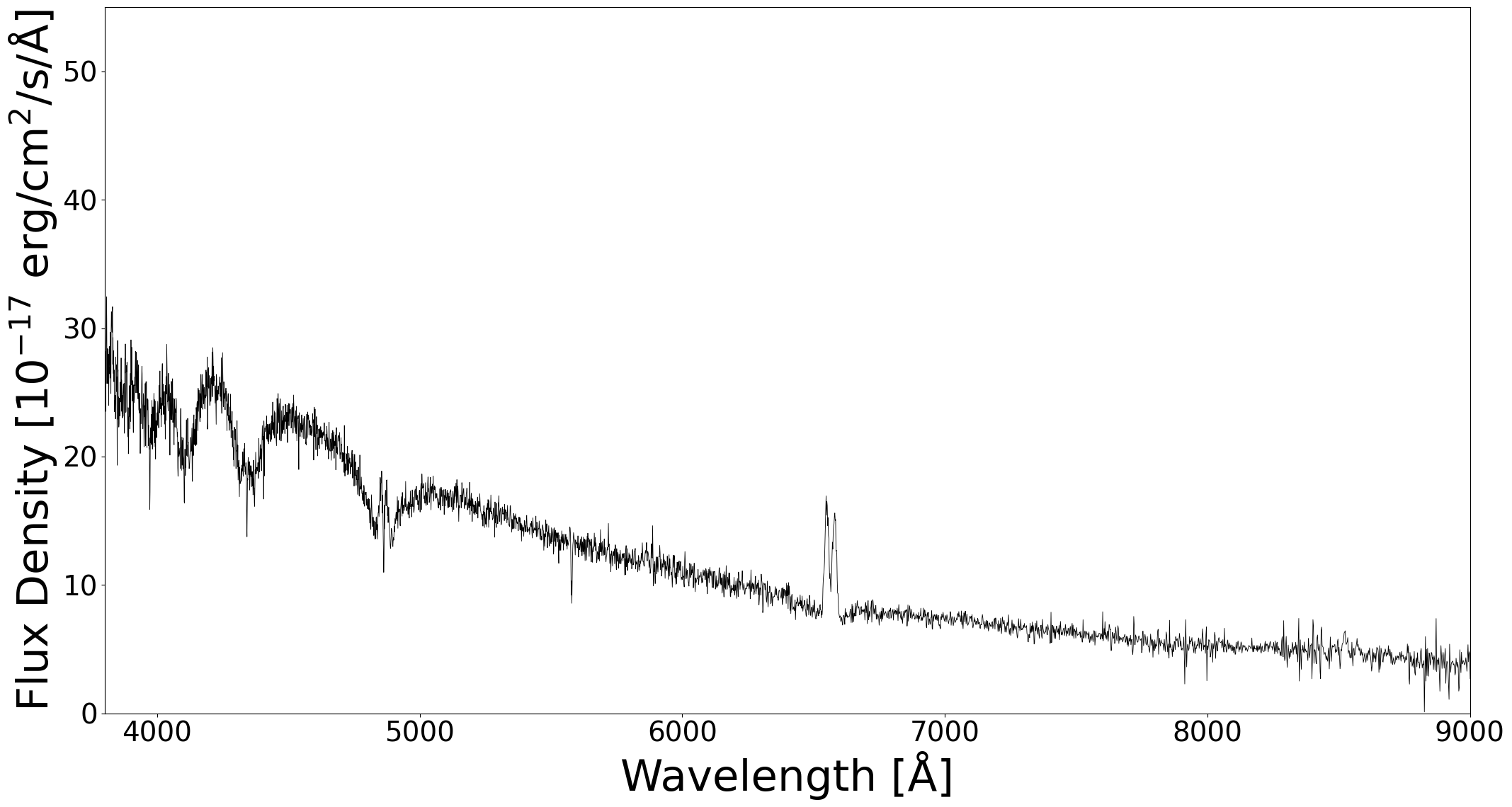}
    \caption{SDSS spectra of SDSS\,J143317.78+101122.8 obtained at two different epochs. Top: plate 1709, fibre 153, MJD 53533. Bottom: plate 5465, fibre 138, MJD 55988.}
    \label{fig:SDSSJ143317.78+101122.8}
\end{figure}

As indicated in Fig.~\ref{Fig.Confusion_Matrix}, the trained logistic regression model fails for only two spectra to identify the period-bouncer classification. The first case is V406\,Vir (plate 335, fibre 85, MJD 52000), which lies close to the model’s decision boundary with $P_{\text{PB}}=51\%$. The second misclassified spectrum belongs to SDSS\,J143317.78+101122.8 (plate 5465, fibre 138, MJD 55988), which shows a more pronounced deviation with $P_{\text{PB}}=44\%$. For this system, a second SDSS spectrum is available (plate 1709, fibre 153, MJD 53533), which is correctly classified as a period-bouncer with $P_{\text{PB}}=55\%$, as it exhibits noticeably steeper Balmer decrements. As shown in Fig.~\ref{fig:SDSSJ143317.78+101122.8}, the two spectra of SDSS\,J143317.78+101122.8 differ significantly: the spectrum observed during MJD 53533 shows much stronger Balmer emission lines, whereas the continuum slope remains mostly unchanged between the two epochs. Possibly, the misclassified spectrum was obtained near an outburst, when additional absorption from the accretion disc could have weakened the emission lines.  

\subsubsection{Classification of period-bouncer candidates}
\label{subsect:diagnostic_diagram_PBs_Candidates}

The three period-bouncer candidates (green rhombus in Fig.~\ref{Fig.Diagnostic_Diagram})—V355\,UMa, LV\,Cnc, and SDSS\,J102905.24+485515.2—are classified as period-bouncers by the logistic regression model, with predicted probabilities of 59\%, 60\%, and 88\%, respectively. As mentioned in Sect.~\ref{sect.Sample}, \citet{Inight_2023} classifies them as WZ\,Sge systems (V355\,UMa) or WZ\,Sge candidates (LV\,Cnc and SDSS\,J102905.24+485515.2). WZ\,Sge-type dwarf novae are characterised by rare, large-amplitude outbursts, typically $\sim$8\,mag, with recurrence times of about a decade. They generally exhibit low mass accretion rates and have a short orbital period ($P_{\rm orb}\lesssim 86$\,min) (see \citealt{Kato_2015} for a comprehensive review). As evolved CVs near the period minimum, WZ\,Sge systems are classical period-bouncer candidates. Thus, it is not surprising that the physical conditions in their accretion discs, as indicated by their Balmer decrements, closely resemble those of known period-bouncers.

Based on the multiwavelength period-bouncer scorecard of \citet{Daniela_2024}, V355\,UMa and SDSS\,J102905.24+485515.2 have scores of 80\% and 71\%, respectively. Both sources lie outside the western half of the sky (Galactic longitude $l \geq 180^\circ$) with German data rights on the extended ROentgen Survey with an Imaging Telescope Array (eROSITA)\footnote{extended ROentgen Survey with an Imaging Telescope Array \citep[eROSITA;][]{predehl2021} on board the Spektrum-Roentgen-Gamma mission \citep[SRG;][]{sunyaev2021}.}. Consequently, \citet{Daniela_2024} could not estimate an instantaneous mass accretion rate based on their X-ray luminosities, which is essential for their confirmation as period-bouncers. LV\,Cnc has a period-bouncer score of 79\% and, although detected with eROSITA (within the German sky area), it did not satisfy the period-bouncer selection criteria defined by \citet{Daniela_2024} (see also \citealt{Daniela_2024b}). The position of these systems in the Balmer decrement parameter space further supports their status as period-bouncer candidates, and future measurements of their mass ratios or secondary spectral types will enable a definitive confirmation.

\subsubsection{Incorrectly classified pre-bouncers}
\label{subsect:diagnostic_diagram_nPBs}

Seven spectra corresponding to pre-bouncer systems are misclassified as period-bouncers by the logistic regression model (Fig.~\ref{Fig.Confusion_Matrix} and Table~\ref{table:misclassified}). In the cases of IR\,Com and QZ\,Ser, the steeper Balmer decrements could arise from a transient state or atypical characteristics of the system, as we explain in the following.

\begin{table}[h!]
    \caption{SDSS spectra of pre-bounce CVs misclassified by the logistic regression model.}
    \label{table:misclassified}
    \centering
    \begin{tabular}{c c c c}
        \hline\hline
        Name & Plate & Fibre & MJD \\
        \hline
        IR\,Com & 5985 & 232 & 56089 \\ 
        QZ\,Ser & 2171 & 7 & 53557 \\
        V521\,Peg & 7578 & 83 & 56956 \\
        HY\,Psc & 380 & 575 & 51792 \\
        QU\,Aqr & 984 & 533 & 52442 \\
        OU\,Vir & 306 & 4 & 51637 \\
        SDSS\,J083845.23+491055.5 & 445 & 89 & 51873 \\
        \hline
    \end{tabular}
\end{table}

\begin{figure}
    \centering
    \includegraphics[width=0.499\textwidth]{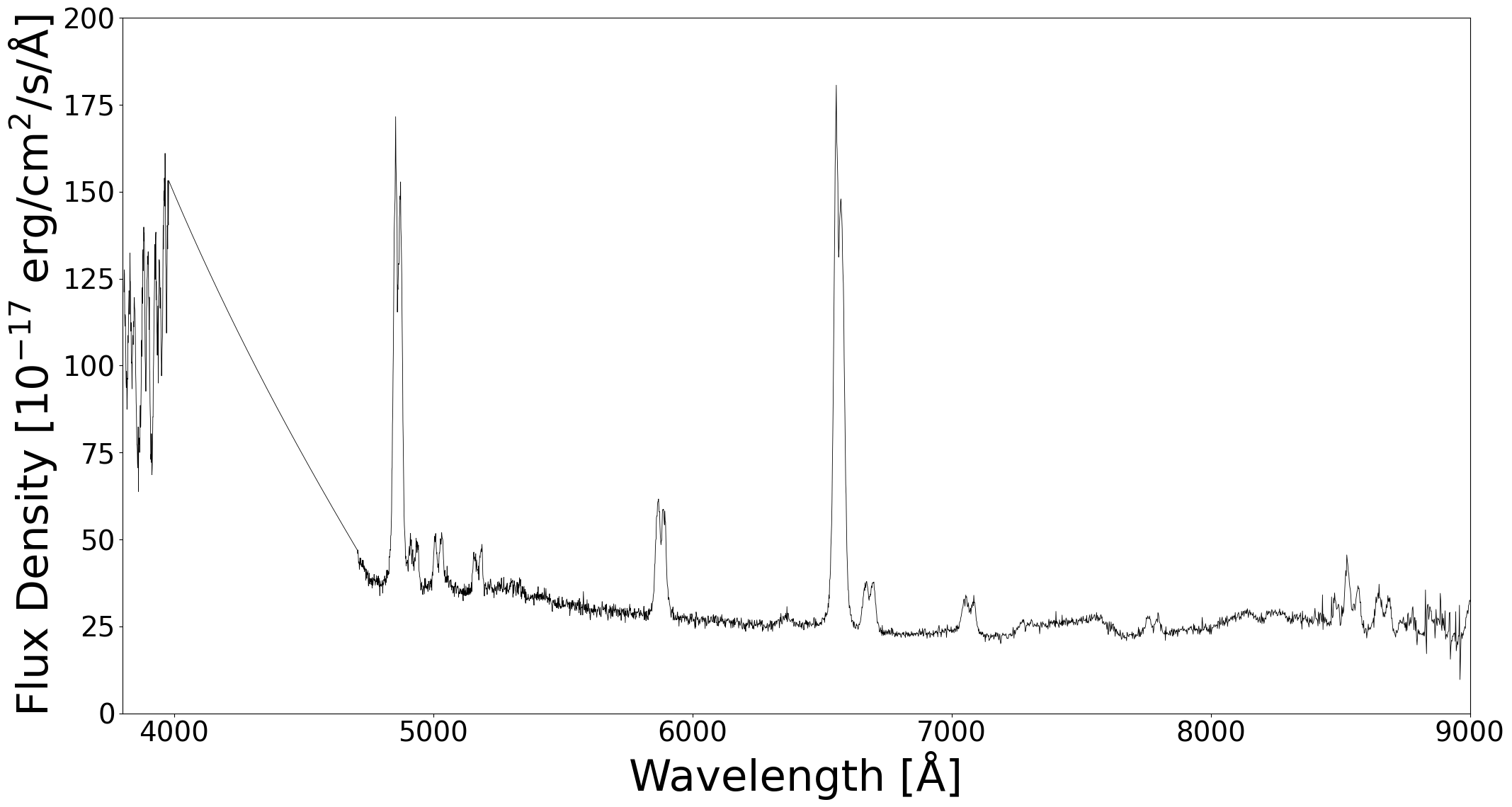}
    \vspace{0.6cm} 
    \includegraphics[width=0.499\textwidth]{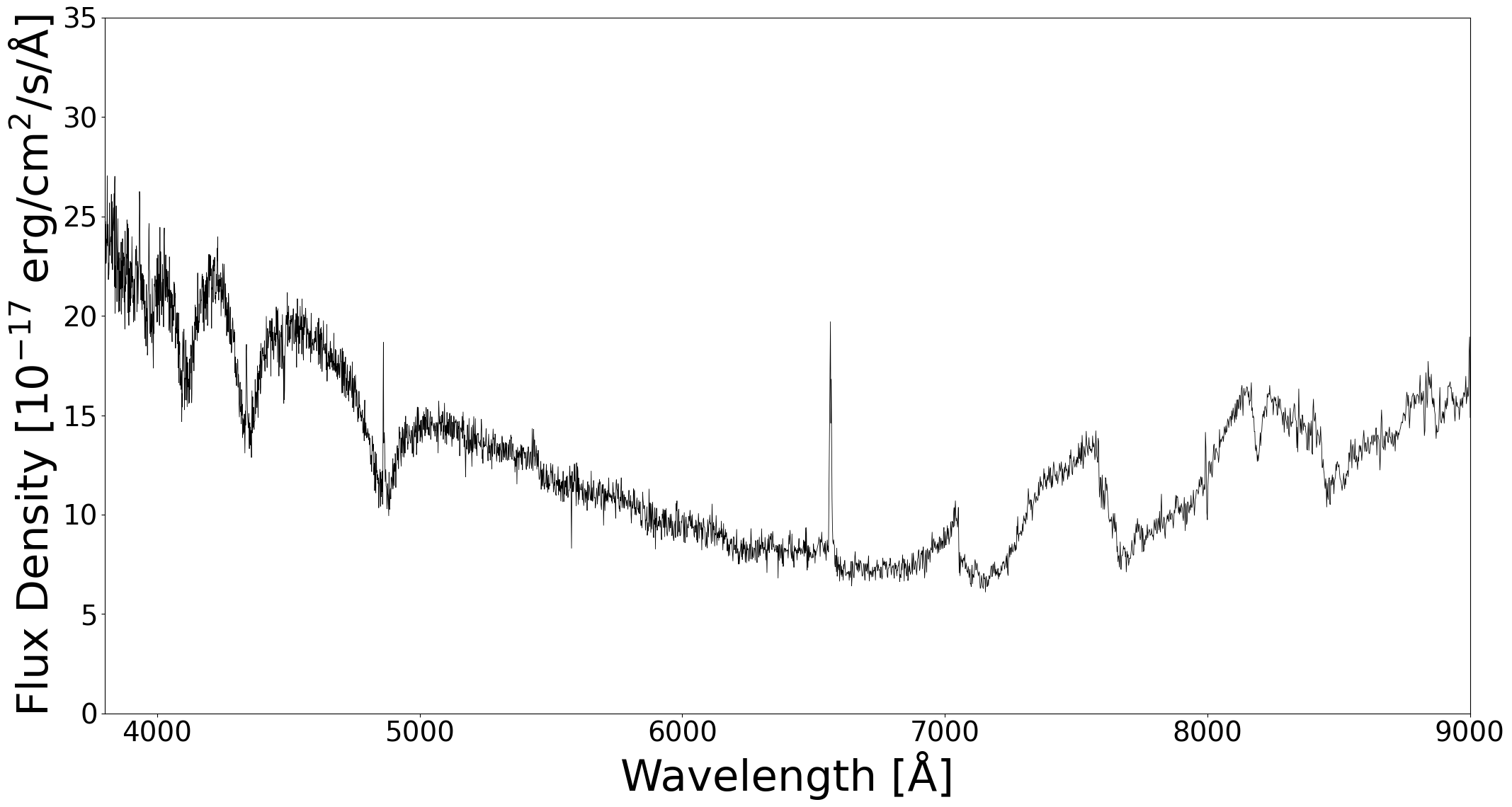}
    \caption{SDSS spectra of IR\,Com obtained at two different epochs. Top: plate 2613, fibre 523, MJD 54481. Bottom: plate 5985, fibre 232, MJD 56089, shows the system in a low-state of reduced Balmer emission.}
    \label{fig:IRCom}
\end{figure}

IR\,Com was studied in greater detail by \citet{Gansicke_2014}, who reported that the SDSS spectrum obtained at MJD 56089 corresponds to an unusually long low-state lasting more than two years, as indicated by its long-term light curve. During this phase, accretion nearly ceased, and the WD photosphere and its M dwarf companion dominated the optical emission. During this state, the Balmer emission lines were strongly diminished and narrow compared to those in the other available spectrum of IR\,Com. The two SDSS spectra of IR\,Com are shown in Fig.~\ref{fig:IRCom}. This extended low state, rather than the system’s true evolutionary status, is the main reason why the spectrum of IR\,Com obtained at MJD 56089 is misclassified by our model as a period-bouncer. As suggested by \citet{Gansicke_2014}, the WD in IR\,Com may possess a weak magnetic field (of a few hundred kG to a few MG), which could account for the occurrence of this extended low state. Low states are generally associated with a reduction in the mass transfer rate from the donor, and in systems with a magnetic WD, where the accretion flow is directly channelled onto the WD, a decrease in the mass-loss rate leads to a rapid decline in the system’s brightness (\citealt{Kafka_2005}).

QZ\,Ser is a dwarf nova that exhibits several unusual characteristics. \citet{Thorstensen_2002} identified it as a short-period CV with $P_{\rm orb}=119.75$\,min and a K-type secondary star, an uncommon combination for systems in this evolutionary regime, where late-M dwarf donors are typically observed. The secondary’s mass is significantly lower than that of a typical K-type star, suggesting that it had experienced substantial nuclear evolution prior to the onset of mass transfer. \citet{Chavez_2022} proposed that the system may host a third body in a close, nearly circular, and coplanar orbit to account for the observed very long photometric period of 277.72 days. Such a tertiary companion could introduce perturbations in the eccentricity of the central binary, leading to periodic shifts in the position of the inner Lagrangian point (L$_1$), and consequently driving cyclic changes in the system’s brightness and mass-transfer rate. Periods of reduced mass transfer may thus lead QZ\,Ser to exhibit Balmer decrements that are more similar to those observed in period-bouncers.

For the remaining misclassified pre-bounce CVs, there is no additional observational information that could be linked to their misclassification as period-bouncers. We note that partial overlap between short-period pre-bounce systems and period-bouncers is expected, as no sharp evolutionary transition is predicted. For example, V521\,Peg has been suggested to lie on the evolutionary boundary between SU\,UMa and WZ\,Sge-type systems (see \citealt{Szegedi_2022}). With an orbital period of $P_{\rm orb}=86.2$\,min \citep{Rodriguez_2005}, the system is located near the period minimum . Although a donor star of spectral type M5 or later has been identified spectroscopically \citep{Rodriguez_2005}, indicating that the system still hosts a non-degenerate secondary, this system can exhibit Balmer decrements consistent with low mass accretion rates.

In order to assess whether any of the misclassified pre-bounce CVs should instead be considered as new period-bouncer candidates, we applied the period-bouncer scorecard introduced by \citet{Daniela_2024}. All systems yield final scores inconsistent with a period-bouncer classification, with V521\,Peg attaining the highest score (42$\%$), and the remaining systems having scores $\leq 17\%$, well below the lowest score obtained by a bona fide confirmed period-bouncer (64$\%$ for V406\,Vir).

\section{Comparison with theoretical models}
\label{sec:TheoreticalModels}

The systematically enhanced Balmer decrements that are observed in period-bounce CVs compared to those seen in short-period pre-bounce CVs indicate a distinct line formation regime in period-bouncers. This motivates a comparison with predictions from theoretical models.

While a fraction of the Balmer line emission may arise from the hotspot, the accretion stream, or from radiation reflected by the secondary star, the typically observed double-peaked line profiles and broad linewidths—often reaching several thousand kilometres per second—(see \citealt{Smak_1969}, \citealt{Smak_1982}, and \citealt{Horne_1986}), are indicative of emission originating primarily in the accretion disc. This is further supported by observations in high-inclination systems, where the blue side of the emission line profile is eclipsed by the donor star before the red side (see e.g. \citealt{Greenstein_1959} and \citealt{Young_1981}), strongly indicating that the dominant emission originates in extended regions of the accretion disc rotating in the prograde direction around the WD. 

Photoionisation-recombination models predict $L_{\mathrm{H}{\alpha}}/L_{\mathrm{H}{\beta}}$ ratios of $2.5-3.5$ (\citealt{Osterbrock_1989}), significantly higher than the ratios observed in our sample. While emission from the photoionisation-recombination processes may be present in the accretion disc or accretion stream, the observed flatter decrements—including some inverted $L_{\mathrm{H}{\alpha}}/L_{\mathrm{H}{\beta}}$ ratios in pre-bouncer systems—indicate that pure photoionisation-recombination can not account for the line formation processes in these systems.

The flat Balmer decrements measured in pre-bounce CVs can be explained with the LTE accretion disc models of \citet{Williams_1980}. In these models, the continuum originates from the hot, optically thick inner disc, which radiates approximately as a blackbody, while the outer disc is optically thin to the continuum but optically thick in the Balmer lines. In this framework, the line source function is given approximately by the Planck function at the local disc temperature, that is, the temperature at the specific radius where the line is formed, as determined by the balance between dissipative heating and radiative losses. Accounting for Doppler broadening, which scales with the rest wavelength of the transition of the emitted line, these models produce nearly flat Balmer decrements. 

The models of \citet{Williams_1980} showed that lower mass accretion rates lead to lower temperatures throughout the accretion disc, with larger fractions of the outer disc becoming optically thin to the continuum. Observational evidence suggesting the presence of optically thin outer disc regions has been reported for some systems (see e.g. \citealt{Littlefair_2001} and \citealt{Neustroev_2016}). This results in steeper Balmer decrements, as H${\alpha}$ approaches the maximum intensity of the local Planck function more closely than H${\beta}$ at lower local temperatures.

The calculations of \citet{Williams_1980} extend down to mass accretion rates of $\dot{M}=10^{-12}\,M_\odot\,\mathrm{yr}^{-1}$, for which they predict $L_{\mathrm{H}{\alpha}}/L_{\mathrm{H}{\beta}}\sim 1.2$. This value is lower than the $L_{\mathrm{H}{\alpha}}/L_{\mathrm{H}{\beta}}$ ratios measured in period-bouncers. These models were not computed for lower mass accretion rates, and it is therefore unclear whether extrapolation to $\dot{M}\lesssim10^{-13}\,M_\odot\,\mathrm{yr}^{-1}$ could reproduce the values observed in period-bouncers. 

The LTE models from \citet{Tylenda_1981}, which also consider accretion discs with low mass accretion rates and an optically thin continuum in the outer regions, predict significantly steeper Balmer decrements than those from \citet{Williams_1980}. In particular, these models predict $L_{\mathrm{H}{\alpha}}/L_{\mathrm{H}{\beta}}\sim 1.8$–$1.3$ and $L_{\mathrm{H}{\gamma}}/L_{\mathrm{H}{\beta}}\sim 0.6$–$0.7$ for $\dot{M}\sim10^{-12}$–$10^{-9}\,M_\odot\,\text{yr}^{-1}$. Overall, the predictions for the $\mathrm{H}\alpha/\mathrm{H}\beta$ ratios from \citet{Tylenda_1981} are more consistent with the steeper $\mathrm{H}\alpha/\mathrm{H}\beta$ Balmer decrements observed in period-bouncers, but they predict significantly lower $\mathrm{H}\gamma/\mathrm{H}\beta$ ratios than those observed in period-bouncers. Moreover, they remain significantly steep for a wide range of mass accretion rates in contrast to what is observed in pre-bounce CVs.

The primary difference between the two sets of models lies in their treatment of viscosity. A larger value of the dimensionless coefficient $\alpha$ \citep{Shakura_Sunyaev_1973} implies more efficient angular momentum transport, leading to a lower surface density, reduced optical thickness, and a higher temperature at a given radius. The treatment of the viscosity in \citet{Williams_1980} is equivalent to values of $\alpha>1$, while the models of \citet{Tylenda_1981} adopt the standard $\alpha$-disc formalism with $\alpha=1$. This difference in viscous efficiency directly affects the predicted Balmer decrements. We note that both these values of $\alpha$ are unrealistically high, as observational estimates constrain $\alpha$ to lie in the range $0.1-0.4$ (see \citealt{King_2007}).

At low temperatures ($\leq 10000$\,K) and in low-density accretion discs—conditions expected in period-bouncers—the higher energy levels of hydrogen are expected to no longer follow LTE (see \citealt{Williams_1988} and \citealt{Dumont_1991}). This departure from LTE is strongest for all energy levels in the cooler and less dense surface layers. Thus, a more accurate treatment of the physics in discs with low mass accretion rates requires non-LTE modelling, where both collisional and radiative processes are included in determining the hydrogen level populations and resulting line emission. Such models were developed by \citet{Williams_1988} and \citet{Williams_1991}. 

\begin{figure}
    \centering
    \includegraphics[width=0.49\textwidth]{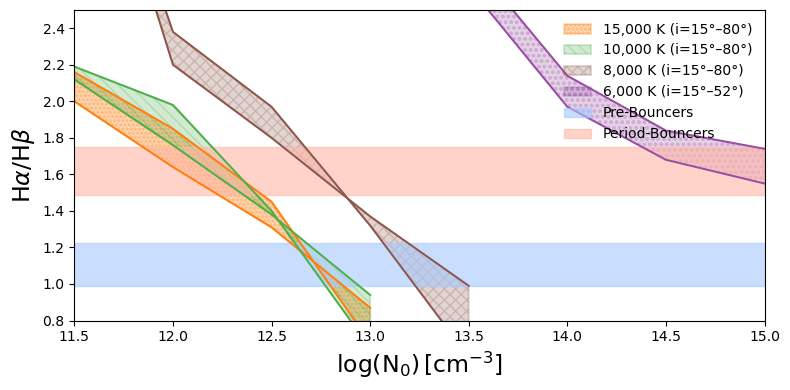}
    \vspace{0.6cm} 
    \includegraphics[width=0.49\textwidth]{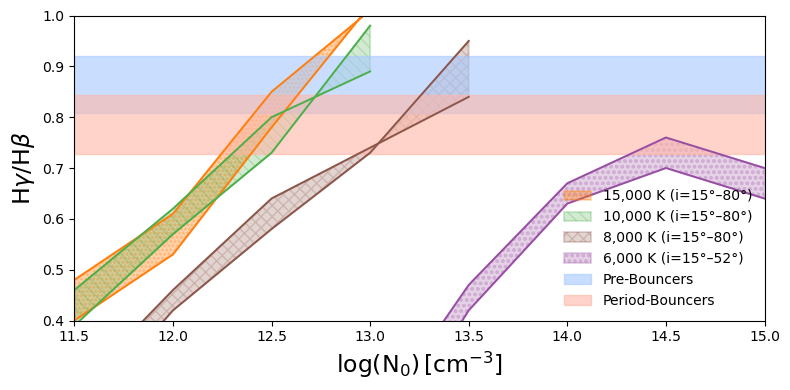}
    \caption{Predicted Balmer decrements from the non-LTE radiative transfer calculations of \citet{Williams_1991} for different disc temperatures and midplane densities. The density at the disc midplane is represented by the logarithm of the neutral plus ionised hydrogen number density ($\log (N_0)$). The horizontal shaded bands represent the 25th–75th percentile ranges of the observed Balmer decrements in pre-bouncer (blue) and period-bouncer (red) systems.}
    \label{fig:nLTE_models}
\end{figure}

In Fig.~\ref{fig:nLTE_models}, we adopt the \citet{Williams_1991} non-LTE radiative transfer calculations, which predict Balmer decrements as a function of temperature, midplane density, and disc inclinations, and we compare them with our measurements. In these models, the density is represented by the logarithm of the neutral plus ionised hydrogen number density at the disc midplane, $\log (N_0)$. The horizontal, shaded colour bands represent the ranges spanned by the 25th to 75th percentiles of the observed Balmer decrements in our period-bouncer (red) and pre-bouncer (blue) samples. We note that the observations represent values integrated over the entire accretion disc, while the theoretical Balmer decrements correspond to emission from a single point in the disc. 

The comparison with the non-LTE models suggests that, in period-bouncers the Balmer lines originate in cooler or less dense regions of the accretion disc (see Fig.~\ref{fig:nLTE_models}). The difference—also observed in the distributions of observed Balmer decrements (see Fig.~\ref{Fig.Ratios_Distributions} in Sect.~\ref{subsect:results})—is more pronounced in the $\mathrm{H}\alpha/\mathrm{H}\beta$ ratios than for the $\mathrm{H}\gamma/\mathrm{H}\beta$ ratios, indicating that the physical conditions of the regions where the higher-order Balmer lines form are relatively similar in both CV populations, while the regions producing $\mathrm{H}\alpha$ differ more strongly. Furthermore, while in pre-bouncers the observed $\mathrm{H}\alpha/\mathrm{H}\beta$ and $\mathrm{H}\gamma/\mathrm{H}\beta$ ratios are consistent with similar temperatures and midplane densities, this is not true for period-bouncers. Assuming comparable temperatures, the $\mathrm{H}\alpha/\mathrm{H}\beta$ ratios in period-bouncers imply lower midplane densities than those inferred from $\mathrm{H}\gamma/\mathrm{H}\beta$ ratios. This reinforces the view that, in the accretion discs of period-bouncers, the $\mathrm{H}\alpha$ emission arises in regions of the accretion disc with markedly different physical conditions relative to those in which the higher-order Balmer lines form, indicating that period-bouncers could have a stronger radial disc stratification than pre-bounce CVs.

These results suggest that the viscosity parameter $\alpha$ may not be constant throughout the disc but instead decreases with radius. The radial dependence of the $\alpha$ parameter was already anticipated by \citet{Shakura_Sunyaev_1973}. At large radii, the accretion discs of period-bouncers are expected to be sufficiently cool that the gas becomes less ionised (see e.g. \citealt{Williams_1980}). Under such conditions, the magnetorotational instability, which is widely thought to drive angular-momentum transport in accretion discs (\citealt{Balbus_1991}), can be significantly suppressed (see \citealt{Gammie_1996}). This would lead to a reduced effective viscosity in the outer disc regions. Theoretical models including a radially varying $\alpha$ parameter may therefore be capable of reproducing the full range of Balmer decrements observed in short-period CVs.

The radial distribution of the Balmer line–forming regions can be probed directly with Doppler tomography, which provides a two-dimensional velocity map of CV accretion discs from emission-line spectra observed at multiple orbital phases (\citealt{Marsh_1988}, \citealt{Marsh_2005}). Such studies showed that in CVs with low mass accretion rates, higher-order Balmer lines appear to extend to comparatively higher velocities in the Doppler maps relative to $\mathrm{H}\alpha$, while $\mathrm{H}\alpha$ typically shows stronger emission at lower velocities. Assuming a Keplerian velocity field, this behaviour indicates that the higher-order Balmer lines originate preferentially from regions closer to the WD, whereas $\mathrm{H}\alpha$ traces emission from larger disc radii (see e.g. \citealt{Pala_2018_QZLib} and \citealt{Mennickent_2006}). This supports our findings from the Balmer decrements. However, Doppler tomography in long-period CVs might suggest a similar stratification (see e.g. \citealt{Neustroev_2002}). This appears to question our interpretation that the radial stratification is systematically stronger in the accretion discs of period-bouncers compared to those of less evolved CVs. So far, only a small number of systems has been analysed with Doppler tomography including at least $\mathrm{H}\alpha$ and $\mathrm{H}\beta$ simultaneously, and - given its diagnostic power - this topic deserves  more attention.

An additional possible contributor to the observed Balmer emission lines is a chromosphere or corona above the disc. Hot coronae can emerge naturally through a thermal instability (see \citealt{Shaviv_1986}, \citealt{Adam_1988}, \citealt{Czerny_1989}, and \citealt{Kool_1999}) or sound waves propagating vertically in the accretion disc (see \citealt{Murray_1992}). In addition, the BL can release a substantial fraction of the accretion energy, producing ultraviolet (UV) and soft and hard X-ray emission in CVs (see e.g. \citealt{Hertfelder_2013} and references therein). This radiation can be reprocessed by the disc's surface layers, giving rise to a hot chromosphere. 

Accretion disc models including chromosphere heating by reprocessed radiation from the BL were presented by \citet{Williams_1992} and \citet{Williams_1995}. \citet{Williams_1992} showed that the presence of a chromosphere above a steady-state $\alpha$-disc mainly increases the intensity of the hydrogen emission lines in the inner, optically thick regions of the disc. In \citet{Williams_1995}, these calculations were extended to a range of mass accretion rates. For the lowest mass accretion rate considered $\dot{M}\sim10^{-13}\,M_\odot,\text{yr}^{-1}$, the resulting Balmer decrement was $\sim 1.1$. To investigate whether other processes could produce a more extended chromosphere, such as magnetohydrodynamic waves (see \citealt{Tout_1992}, \citealt{Stone_1996}, and \citealt{Srivastava_2021}), \citet{Williams_1995} presented models in which the BL luminosity was increased by a factor of 10 relative to that expected from the mass accretion rate. The resulting effect was modest, yielding about a 14\% increase in line strength and only a minor change in the Balmer decrements toward ratios near unity. Thus, although the presence of a chromosphere in the accretion discs of our sample is likely, its influence can not reproduce the steeper Balmer decrements observed in period-bouncers.

\section{Summary and conclusions}
\label{Summary}

We have shown that the Balmer decrements observed in CVs constitute key diagnostics of the physical conditions and emission mechanisms operating within these systems. We determined H${\alpha}$, H${\beta}$, and H${\gamma}$ line luminosities for a sample of 82 SDSS spectra, including 17 spectra of period-bounce CVs, 3 spectra of period-bouncer candidates, and 62 spectra of short-period pre-bounce CVs. Our results show that the Balmer decrements observed in period-bouncers are significantly steeper, especially in the $\mathrm{H}\alpha/\mathrm{H}\beta$ ratios (see Fig.~\ref{Fig.Ratios_Distributions}), indicating different physical conditions in their accretion discs compared to those of pre-bounce CVs with higher mass accretion rates.

We find that lower $\mathrm{H}\beta$ luminosities correlate with steeper Balmer decrements (see Fig.~\ref{Fig.BDvsLHbeta}), directly linking the steeper Balmer decrements observed in period-bouncers to their intrinsically lower mass-accretion rates. This interpretation is further supported by the relation between the $\mathrm{H}\alpha/\mathrm{H}\beta$ ratio and $P_{\rm orb}$ (see Fig.~\ref{Fig.BD_vs_Porb}). As additional period-bounce CVs are discovered and confirmed, larger samples of optical spectroscopy will allow these relationships to be constrained with greater precision.

We trained a linear logistic regression model on the H$\gamma$/H$\beta$ versus H$\alpha$/H$\beta$ diagram (see Fig.~\ref{Fig.Diagnostic_Diagram}). This diagnostic diagram shows a high capability to separate period-bounce CVs from pre-bounce CVs, providing an efficient and robust means of identifying new period-bouncer candidates in large spectroscopic surveys such as SDSS.

We note, however, that steep Balmer decrements do not uniquely identify period-bounce CVs, as systems observed near an outburst and other low accretion rate CVs, such as evolved SU\,UMa or WZ\,Sge systems, can also show steep Balmer decrements without necessarily being located beyond the period minimum. To maximise its efficiency, this diagnostic must be combined with other indicators characteristic of period-bounce CVs, such as $P_{\rm orb}$ near the $\sim80$\,min minimum or ultraviolet and infrared colours (see \citealt{Daniela_2024} for a detailed description on period-bouncer diagnostics).

Finally, the observed Balmer decrements in short-period pre-bounce and period-bounce CVs were compared with calculations from LTE, non-LTE, and chromospheric accretion-disc models (see Sect.~\ref{sec:TheoreticalModels}). The steep Balmer decrements observed in period-bouncers reflect cool, extended, and predominantly optically thin discs resulting from very low mass accretion rates. Together with a more pronounced radial stratification in the physical conditions of the disc, this provides a consistent explanation for their observed systematically steeper Balmer decrements.

\section*{Data availability}
\label{sec:Data_availability}
The table described by Table~\ref{table:Master} is available at the CDS.

\begin{acknowledgements}
We wish to thank the anonymous referee for constructive comments and suggestions that helped improve the clarity and quality of this article. We are also grateful to Axel Schwope and Valery Suleimanov for helpful discussions and valuable comments on this work.

SHD acknowledges financial support from Deutsche Forschungsgemeinschaft (DFG) under grant number STE\,1068/6-2. This work has made use of data from the European Space Agency (ESA) mission \textit{Gaia} (\href{https://www.cosmos.esa.int/gaia}{https://www.cosmos.esa.int/gaia}), processed by the \textit{Gaia} Data Processing and Analysis Consortium (DPAC, \href{https://www.cosmos.esa.int/web/gaia/dpac/consortium}{https://www.cosmos.esa.int/web/gaia/dpac/consortium}). Funding for the DPAC has been provided by national institutions, in particular the institutions participating in the \textit{Gaia} Multilateral Agreement. Funding for the Sloan Digital Sky Survey IV has been provided by the Alfred P. Sloan Foundation, the U.S. Department of Energy Office of Science, and the Participating Institutions. 

SDSS-IV acknowledges support and resources from the Center for High Performance Computing  at the University of Utah. The SDSS website is \url{www.sdss4.org}.

SDSS-IV is managed by the Astrophysical Research Consortium for the Participating Institutions of the SDSS Collaboration including the Brazilian Participation Group, the Carnegie Institution for Science, Carnegie Mellon University, Center for Astrophysics | Harvard \& Smithsonian, the Chilean Participation Group, the French Participation Group, Instituto de Astrof\'isica de Canarias, The Johns Hopkins University, Kavli Institute for the Physics and Mathematics of the Universe (IPMU) / University of Tokyo, the Korean Participation Group, Lawrence Berkeley National Laboratory, Leibniz Institut f\"ur Astrophysik Potsdam (AIP),  Max-Planck-Institut f\"ur Astronomie (MPIA Heidelberg), Max-Planck-Institut f\"ur Astrophysik (MPA Garching), Max-Planck-Institut f\"ur Extraterrestrische Physik (MPE), National Astronomical Observatories of China, New Mexico State University, New York University, University of Notre Dame, Observat\'ario Nacional / MCTI, The Ohio State University, Pennsylvania State University, Shanghai Astronomical Observatory, United Kingdom Participation Group, Universidad Nacional Aut\'onoma de M\'exico, University of Arizona, University of Colorado Boulder, University of Oxford, University of Portsmouth, University of Utah, University of Virginia, University of Washington, University of Wisconsin, Vanderbilt University, and Yale University. 

Based on observations made with the NASA Galaxy Evolution Explorer. GALEX is operated for NASA by the California Institute of Technology under NASA contract NAS5-98034.

\end{acknowledgements}

%

\bibliography{BIB.bib}

@ARTICLE{weighted_linear_regression,
       author = {{York}, Derek},
        title = "{Least-Squares Fitting of a Straight Line}",
      journal = {Can. J. Phys.},
         year = 1966,
        month = may,
       volume = {44},
       number = {5},
        pages = {1079-1086},
          doi = {10.1139/p66-090},
       adsurl = {https://ui.adsabs.harvard.edu/abs/1966CaJPh..44.1079Y}
}

@ARTICLE{Williams_1980,
       author = {{Williams}, R.~E.},
        title = "{Emission lines from the accretion disks of cataclysmic variables.}",
      journal = {\apj},
         year = 1980,
        month = feb,
       volume = {235},
        pages = {939-944},
          doi = {10.1086/157698},
       adsurl = {https://ui.adsabs.harvard.edu/abs/1980ApJ...235..939W}
}

@ARTICLE{Gordon_2006,
       author = {{Sarty}, Gordon E. and {Wu}, Kinwah},
        title = "{Multivariate Characterization of Hydrogen Balmer Emission in Cataclysmic Variables}",
      journal = {\pasa},
         year = 2006,
        month = nov,
       volume = {23},
       number = {3},
        pages = {106-118},
          doi = {10.1071/AS06011},
archivePrefix = {arXiv},
       eprint = {astro-ph/0608392},
 primaryClass = {astro-ph},
       adsurl = {https://ui.adsabs.harvard.edu/abs/2006PASA...23..106S}
}

@ARTICLE{Knigge_2011,
       author = {{Knigge}, Christian and {Baraffe}, Isabelle and {Patterson}, Joseph},
        title = "{The Evolution of Cataclysmic Variables as Revealed by Their Donor Stars}",
      journal = {\apjs},
         year = 2011,
        month = jun,
       volume = {194},
       number = {2},
          eid = {28},
        pages = {28},
          doi = {10.1088/0067-0049/194/2/28},
archivePrefix = {arXiv},
       eprint = {1102.2440},
 primaryClass = {astro-ph.SR},
       adsurl = {https://ui.adsabs.harvard.edu/abs/2011ApJS..194...28K}
}

@article{Amantayeva_2021,
doi = {10.3847/1538-4357/ac0e36},
url = {https://dx.doi.org/10.3847/1538-4357/ac0e36},
year = {2021},
month = {sep},
publisher = {The American Astronomical Society},
volume = {918},
number = {2},
pages = {58},
author = {Amantayeva, A. and Zharikov, S. and Page, K. L. and Pavlenko, E. and Sosnovskij, A. and Khokhlov, S. and Ibraimov, M.},
title = {Period Bouncer Cataclysmic Variable EZ Lyn in Quiescence},
journal = {ApJ}
}

@article{Pala_2018_QZLib,
    author = {Pala, A F and Schmidtobreick, L and Tappert, C and Gänsicke, B T and Mehner, A},
    title = {The cataclysmic variable QZ Lib: a period bouncer},
    journal = {MNRAS},
    volume = {481},
    number = {2},
    pages = {2523-2535},
    year = {2018},
    month = {09},
    issn = {0035-8711},
    doi = {10.1093/mnras/sty2434},
    url = {https://doi.org/10.1093/mnras/sty2434},
    eprint = {https://academic.oup.com/mnras/article-pdf/481/2/2523/25805495/sty2434.pdf},
}

@ARTICLE{Williams_1983,
       author = {{Williams}, G.},
        title = "{Spectroscopy of cataclysmic variables. I. Observations.}",
      journal = {\apjs},
         year = 1983,
        month = nov,
       volume = {53},
        pages = {523-552},
          doi = {10.1086/190900},
       adsurl = {https://ui.adsabs.harvard.edu/abs/1983ApJS...53..523W}
}

@article{Neustroev_2016,
	author = {{Neustroev}, V. V. and {Zharikov}, S. V. and {Borisov}, N. V.},
	title = {Voracious vortexes in cataclysmic variables - A multi-epoch tomographic study of HT Cassiopeia},
	DOI= "10.1051/0004-6361/201526363",
	url= "https://doi.org/10.1051/0004-6361/201526363",
	journal = {A\&A},
	year = 2016,
	volume = 586,
	pages = "A10",
	month = "",
}

@ARTICLE{Williams_1991,
       author = {{Williams}, Glen A.},
        title = "{Hydrogen line emission from accretion disks. II - A grid of models with applications to the optically thin regions of disks.}",
      journal = {\aj},
         year = 1991,
        month = may,
       volume = {101},
        pages = {1929-1941},
          doi = {10.1086/115818},
       adsurl = {https://ui.adsabs.harvard.edu/abs/1991AJ....101.1929W}
}

@ARTICLE{Drake_1980,
       author = {{Drake}, S.~A. and {Ulrich}, R.~K.},
        title = "{The emission-line spectrum from a slab of hydrogen at moderate to high densities.}",
      journal = {\apjs},
         year = 1980,
        month = feb,
       volume = {42},
        pages = {351-383},
          doi = {10.1086/190654},
       adsurl = {https://ui.adsabs.harvard.edu/abs/1980ApJS...42..351D}
}

@ARTICLE{Tylenda_1981,
       author = {{Tylenda}, R.},
        title = "{Radiation from Optically Thin Accretion Discs}",
      journal = {\actaa},
         year = 1981,
        month = jan,
       volume = {31},
        pages = {127},
       adsurl = {https://ui.adsabs.harvard.edu/abs/1981AcA....31..127T}
}

@article{Neustroev_2023,
    author = {Neustroev, Vitaly V and Mäntynen, Iikka},
    title = {A brown dwarf donor and an optically thin accretion disc with a complex stream impact region in the period-bouncer candidate BW Sculptoris},
    journal = {MNRAS},
    volume = {523},
    number = {4},
    pages = {6114-6137},
    year = {2023},
    month = {06},
    issn = {0035-8711},
    doi = {10.1093/mnras/stad1730},
    url = {https://doi.org/10.1093/mnras/stad1730},
    eprint = {https://academic.oup.com/mnras/article-pdf/523/4/6114/50836290/stad1730.pdf},
}

@ARTICLE{Williams_1988,
       author = {{Williams}, Glen A. and {Shipman}, Harry L.},
        title = "{Hydrogen Line Emission from Optically Thin Accretion Disks}",
      journal = {\apj},
         year = 1988,
        month = mar,
       volume = {326},
        pages = {738},
          doi = {10.1086/166132},
       adsurl = {https://ui.adsabs.harvard.edu/abs/1988ApJ...326..738W}
}

@ARTICLE{Dumont_1991,
       author = {{Dumont}, A.~M. and {Lasota}, J.~P. and {Collin-Souffrin}, S. and {King}, A.~R.},
        title = "{Optically thin accretion discs}",
      journal = {A\&A},
         year = 1991,
        month = feb,
       volume = {242},
       number = {2},
        pages = {503-509},
       adsurl = {https://ui.adsabs.harvard.edu/abs/1991A&A...242..503D}
}

@article{Zharikov_2013,
	author = {{Zharikov, S.} and {Tovmassian, G.} and {Aviles, A.} and {Michel, R.} and {Gonzalez-Buitrago, D.} and {García-Díaz, Ma. T.}},
	title = {The accretion disk in the post period-minimum   cataclysmic variable SDSS J080434.20 + 510349.2⋆},
	DOI= "10.1051/0004-6361/201220099",
	url= "https://doi.org/10.1051/0004-6361/201220099",
	journal = {A\&A},
	year = 2013,
	volume = 549,
	pages = "A77",
	month = "",
}

@BOOK{Osterbrock_1989,
  author    = {Osterbrock, Donald E.},
  title     = {Astrophysics of Gaseous Nebulae and Active Galactic Nuclei},
  year      = {1989},
  publisher = {University Science Books},
  address   = {Mill Valley, CA},
  adsurl    = {https://ui.adsabs.harvard.edu/abs/1989agna.book.....O}
}

@article{Littlefair_2001,
    author = {Littlefair, S.P. and Dhillon, V.S. and Marsh, T.R. and Harlaftis, E.T.},
    title = {Mirror eclipses in the cataclysmic variable IP Peg},
    journal = {MNRAS},
    volume = {327},
    number = {2},
    pages = {475-482},
    year = {2001},
    month = {10},
    issn = {0035-8711},
    doi = {10.1046/j.1365-8711.2001.04744.x},
    url = {https://doi.org/10.1046/j.1365-8711.2001.04744.x},
    eprint = {https://academic.oup.com/mnras/article-pdf/327/2/475/3501538/327-2-475.pdf},
}

@ARTICLE{Williams_1995,
       author = {{Williams}, Glen A.},
        title = "{Hydrogen Line Emission From Various Accretion Disk Models}",
      journal = {\aj},
         year = 1995,
        month = jan,
       volume = {109},
        pages = {319},
          doi = {10.1086/117275},
       adsurl = {https://ui.adsabs.harvard.edu/abs/1995AJ....109..319W}
}

@ARTICLE{Shaviv_1986,
       author = {{Shaviv}, G. and {Wehrse}, R.},
        title = "{The vertical temperature stratification and corona formation of accretion disc atmospheres}",
      journal = {A\&A},
         year = 1986,
        month = apr,
       volume = {159},
       number = {1-2},
        pages = {L5-L7},
       adsurl = {https://ui.adsabs.harvard.edu/abs/1986A&A...159L...5S}
}

@INPROCEEDINGS{Williams_1992,
       author = {{Williams}, G.~A.},
        title = "{Accretion Disk Chromospheres due to Boundary Layer Radiation}",
    booktitle = {Cataclysmic Variable Stars},
         year = 1992,
       editor = {{Vogt}, Nikolaus},
       series = {ASP Conference Series},
       volume = {29},
        month = jan,
        pages = {90},
       adsurl = {https://ui.adsabs.harvard.edu/abs/1992ASPC...29...90W}
}

@ARTICLE{Adam_1988,
       author = {{Adam}, J. and {Stoerzer}, H. and {Wehrse}, R. and {Shaviv}, G.},
        title = "{Radiation from accretion discs}",
      journal = {A\&A},
         year = 1988,
        month = mar,
       volume = {193},
       number = {1-2},
        pages = {L1-L3},
       adsurl = {https://ui.adsabs.harvard.edu/abs/1988A&A...193L...1A}
}

@ARTICLE{Clarke_1984,
       author = {{Clarke}, J.~T. and {Bowyer}, S.},
        title = "{Spectrophotometry of the outburst and quiescent states of the dwarf novae RX And and KT Per.}",
      journal = {A\&A},
         year = 1984,
        month = nov,
       volume = {140},
        pages = {345-351},
       adsurl = {https://ui.adsabs.harvard.edu/abs/1984A&A...140..345C}
}

@ARTICLE{Horne_1986,
       author = {{Horne}, K. and {Marsh}, T.~R.},
        title = "{Emission line formation in accretion discs}",
      journal = {\mnras},
         year = 1986,
        month = feb,
       volume = {218},
        pages = {761-773},
          doi = {10.1093/mnras/218.4.761},
       adsurl = {https://ui.adsabs.harvard.edu/abs/1986MNRAS.218..761H}
}

@ARTICLE{Marsh_1988,
       author = {{Marsh}, T.~R.},
        title = "{A spectroscopic study of the deeply eclipsing dwarf nova IP Peg.}",
      journal = {\mnras},
         year = 1988,
        month = apr,
       volume = {231},
        pages = {1117-1138},
          doi = {10.1093/mnras/231.4.1117},
       adsurl = {https://ui.adsabs.harvard.edu/abs/1988MNRAS.231.1117M}
}

@article{Hertfelder_2013,
	author = {{Hertfelder, Marius} and {Kley, Wilhelm} and {Suleimanov, Valery} and {Werner, Klaus}},
	title = {The boundary layer in compact binaries},
	DOI= "10.1051/0004-6361/201322542",
	url= "https://doi.org/10.1051/0004-6361/201322542",
	journal = {A\&A},
	year = 2013,
	volume = 560,
	pages = "A56",
	month = "",
}

@article{Srivastava_2021,
author = {Srivastava, A. K. and Ballester, J. L. and Cally, P. S. and Carlsson, M. and Goossens, M. and Jess, D. B. and Khomenko, E. and Mathioudakis, M. and Murawski, K. and Zaqarashvili, T. V.},
title = {Chromospheric Heating by Magnetohydrodynamic Waves and Instabilities},
journal = {Journal of Geophysical Research: Space Physics},
volume = {126},
number = {6},
pages = {e2020JA029097},
doi = {https://doi.org/10.1029/2020JA029097},
url = {https://agupubs.onlinelibrary.wiley.com/doi/abs/10.1029/2020JA029097},
eprint = {https://agupubs.onlinelibrary.wiley.com/doi/pdf/10.1029/2020JA029097},
note = {e2020JA029097 2020JA029097},
year = {2021}
}

@article{koester2010,
       author = {{Koester}, D.},
        title = {White dwarf spectra and atmosphere models},
      journal = {Memorie della Societa Astronomica Italiana},
         year = 2010,
        month = jan,
       volume = {81},
        pages = {921-931},
       adsurl = {https://ui.adsabs.harvard.edu/abs/2010MmSAI..81..921K}
}

@ARTICLE{Tremblay_2011,
       author = {{Tremblay}, P. -E. and {Bergeron}, P. and {Gianninas}, A.},
        title = "{An Improved Spectroscopic Analysis of DA White Dwarfs from the Sloan Digital Sky Survey Data Release 4}",
      journal = {\apj},
         year = 2011,
        month = apr,
       volume = {730},
       number = {2},
          eid = {128},
        pages = {128},
          doi = {10.1088/0004-637X/730/2/128},
archivePrefix = {arXiv},
       eprint = {1102.0056},
 primaryClass = {astro-ph.SR},
       adsurl = {https://ui.adsabs.harvard.edu/abs/2011ApJ...730..128T}
}

@ARTICLE{Greenstein_1959,
       author = {{Greenstein}, Jesse L. and {Kraft}, Robert P.},
        title = "{The Binary System Nova DQ Herculis. I. The Spectrum and Radial Velocity during the Eclipse Cycle.}",
      journal = {\apj},
         year = 1959,
        month = jul,
       volume = {130},
        pages = {99},
          doi = {10.1086/146700},
       adsurl = {https://ui.adsabs.harvard.edu/abs/1959ApJ...130...99G}
}

@ARTICLE{Young_1981,
       author = {{Young}, P. and {Schneider}, D.~P. and {Shectman}, S.~A.},
        title = "{The voracious vortex in HT Cassiopeiae.}",
      journal = {\apj},
         year = 1981,
        month = may,
       volume = {245},
        pages = {1035-1042},
          doi = {10.1086/158880},
       adsurl = {https://ui.adsabs.harvard.edu/abs/1981ApJ...245.1035Y}
}

@ARTICLE{Abril_2020,
       author = {{Abril}, Javier and {Schmidtobreick}, Linda and {Ederoclite}, Alessandro and {L{\'o}pez-Sanjuan}, Carlos},
        title = "{Disentangling cataclysmic variables in Gaia's HR diagram}",
      journal = {\mnras},
         year = 2020,
        month = feb,
       volume = {492},
       number = {1},
        pages = {L40-L44},
          doi = {10.1093/mnrasl/slz181},
archivePrefix = {arXiv},
       eprint = {1912.01531},
 primaryClass = {astro-ph.SR},
       adsurl = {https://ui.adsabs.harvard.edu/abs/2020MNRAS.492L..40A}
}

@article{Inight_2023,
    author = {Inight, Keith and Gänsicke, Boris T and Breedt, Elmé and Israel, Henry T and Littlefair, Stuart P and Manser, Christopher J and Marsh, Tom R and Mulvany, Tim and Pala, Anna Francesca and Thorstensen, John R},
    title = {A catalogue of cataclysmic variables from 20 yr of the Sloan Digital Sky Survey with new classifications, periods, trends, and oddities},
    journal = {MNRAS},
    volume = {524},
    number = {4},
    pages = {4867-4898},
    year = {2023},
    month = {07},
    issn = {0035-8711},
    doi = {10.1093/mnras/stad2018},
    url = {https://doi.org/10.1093/mnras/stad2018},
    eprint = {https://academic.oup.com/mnras/article-pdf/524/4/4867/56261577/stad2018.pdf},
}

@ARTICLE{Gaia_DR3,
       author = {{Gaia Collaboration} and {Vallenari}, A. and {Brown}, A.~G.~A. and {Prusti}, T. and {de Bruijne}, J.~H.~J. and {Arenou}, F. and {Babusiaux}, C. and {Biermann}, M. and {Creevey}, O.~L. and {Ducourant}, C. and {Evans}, D.~W. and {Eyer}, L. and {Guerra}, R. and {Hutton}, A. and {Jordi}, C. and {Klioner}, S.~A. and {Lammers}, U.~L. and {Lindegren}, L. and {Luri}, X. and {Mignard}, F. and {Panem}, C. and {Pourbaix}, D. and {Randich}, S. and {Sartoretti}, P. and {Soubiran}, C. and {Tanga}, P. and {Walton}, N.~A. and {Bailer-Jones}, C.~A.~L. and {Bastian}, U. and {Drimmel}, R. and {Jansen}, F. and {Katz}, D. and {Lattanzi}, M.~G. and {van Leeuwen}, F. and {Bakker}, J. and {Cacciari}, C. and {Casta{\~n}eda}, J. and {De Angeli}, F. and {Fabricius}, C. and {Fouesneau}, M. and {Fr{\'e}mat}, Y. and {Galluccio}, L. and {Guerrier}, A. and {Heiter}, U. and {Masana}, E. and {Messineo}, R. and {Mowlavi}, N. and {Nicolas}, C. and {Nienartowicz}, K. and {Pailler}, F. and {Panuzzo}, P. and {Riclet}, F. and {Roux}, W. and {Seabroke}, G.~M. and {Sordo}, R. and {Th{\'e}venin}, F. and {Gracia-Abril}, G. and {Portell}, J. and {Teyssier}, D. and {Altmann}, M. and {Andrae}, R. and {Audard}, M. and {Bellas-Velidis}, I. and {Benson}, K. and {Berthier}, J. and {Blomme}, R. and {Burgess}, P.~W. and {Busonero}, D. and {Busso}, G. and {C{\'a}novas}, H. and {Carry}, B. and {Cellino}, A. and {Cheek}, N. and {Clementini}, G. and {Damerdji}, Y. and {Davidson}, M. and {de Teodoro}, P. and {Nu{\~n}ez Campos}, M. and {Delchambre}, L. and {Dell'Oro}, A. and {Esquej}, P. and {Fern{\'a}ndez-Hern{\'a}ndez}, J. and {Fraile}, E. and {Garabato}, D. and {Garc{\'\i}a-Lario}, P. and {Gosset}, E. and {Haigron}, R. and {Halbwachs}, J. -L. and {Hambly}, N.~C. and {Harrison}, D.~L. and {Hern{\'a}ndez}, J. and {Hestroffer}, D. and {Hodgkin}, S.~T. and {Holl}, B. and {Jan{\ss}en}, K. and {Jevardat de Fombelle}, G. and {Jordan}, S. and {Krone-Martins}, A. and {Lanzafame}, A.~C. and {L{\"o}ffler}, W. and {Marchal}, O. and {Marrese}, P.~M. and {Moitinho}, A. and {Muinonen}, K. and {Osborne}, P. and {Pancino}, E. and {Pauwels}, T. and {Recio-Blanco}, A. and {Reyl{\'e}}, C. and {Riello}, M. and {Rimoldini}, L. and {Roegiers}, T. and {Rybizki}, J. and {Sarro}, L.~M. and {Siopis}, C. and {Smith}, M. and {Sozzetti}, A. and {Utrilla}, E. and {van Leeuwen}, M. and {Abbas}, U. and {{\'A}brah{\'a}m}, P. and {Abreu Aramburu}, A. and {Aerts}, C. and {Aguado}, J.~J. and {Ajaj}, M. and {Aldea-Montero}, F. and {Altavilla}, G. and {{\'A}lvarez}, M.~A. and {Alves}, J. and {Anders}, F. and {Anderson}, R.~I. and {Anglada Varela}, E. and {Antoja}, T. and {Baines}, D. and {Baker}, S.~G. and {Balaguer-N{\'u}{\~n}ez}, L. and {Balbinot}, E. and {Balog}, Z. and {Barache}, C. and {Barbato}, D. and {Barros}, M. and {Barstow}, M.~A. and {Bartolom{\'e}}, S. and {Bassilana}, J. -L. and {Bauchet}, N. and {Becciani}, U. and {Bellazzini}, M. and {Berihuete}, A. and {Bernet}, M. and {Bertone}, S. and {Bianchi}, L. and {Binnenfeld}, A. and {Blanco-Cuaresma}, S. and {Blazere}, A. and {Boch}, T. and {Bombrun}, A. and {Bossini}, D. and {Bouquillon}, S. and {Bragaglia}, A. and {Bramante}, L. and {Breedt}, E. and {Bressan}, A. and {Brouillet}, N. and {Brugaletta}, E. and {Bucciarelli}, B. and {Burlacu}, A. and {Butkevich}, A.~G. and {Buzzi}, R. and {Caffau}, E. and {Cancelliere}, R. and {Cantat-Gaudin}, T. and {Carballo}, R. and {Carlucci}, T. and {Carnerero}, M.~I. and {Carrasco}, J.~M. and {Casamiquela}, L. and {Castellani}, M. and {Castro-Ginard}, A. and {Chaoul}, L. and {Charlot}, P. and {Chemin}, L. and {Chiaramida}, V. and {Chiavassa}, A. and {Chornay}, N. and {Comoretto}, G. and {Contursi}, G. and {Cooper}, W.~J. and {Cornez}, T. and {Cowell}, S. and {Crifo}, F. and {Cropper}, M. and {Crosta}, M. and {Crowley}, C. and {Dafonte}, C. and {Dapergolas}, A. and {David}, M. and {David}, P. and {de Laverny}, P. and {De Luise}, F. and {De March}, R. and {De Ridder}, J. and {de Souza}, R. and {de Torres}, A. and {del Peloso}, E.~F. and {del Pozo}, E. and {Delbo}, M. and {Delgado}, A. and {Delisle}, J. -B. and {Demouchy}, C. and {Dharmawardena}, T.~E. and {Di Matteo}, P. and {Diakite}, S. and {Diener}, C. and {Distefano}, E. and {Dolding}, C. and {Edvardsson}, B. and {Enke}, H. and {Fabre}, C. and {Fabrizio}, M. and {Faigler}, S. and {Fedorets}, G. and {Fernique}, P. and {Fienga}, A. and {Figueras}, F. and {Fournier}, Y. and {Fouron}, C. and {Fragkoudi}, F. and {Gai}, M. and {Garcia-Gutierrez}, A. and {Garcia-Reinaldos}, M. and {Garc{\'\i}a-Torres}, M. and {Garofalo}, A. and {Gavel}, A. and {Gavras}, P. and {Gerlach}, E. and {Geyer}, R. and {Giacobbe}, P. and {Gilmore}, G. and {Girona}, S. and {Giuffrida}, G. and {Gomel}, R. and {Gomez}, A. and {Gonz{\'a}lez-N{\'u}{\~n}ez}, J. and {Gonz{\'a}lez-Santamar{\'\i}a}, I. and {Gonz{\'a}lez-Vidal}, J.~J. and {Granvik}, M. and {Guillout}, P. and {Guiraud}, J. and {Guti{\'e}rrez-S{\'a}nchez}, R. and {Guy}, L.~P. and {Hatzidimitriou}, D. and {Hauser}, M. and {Haywood}, M. and {Helmer}, A. and {Helmi}, A. and {Sarmiento}, M.~H. and {Hidalgo}, S.~L. and {Hilger}, T. and {H{\l}adczuk}, N. and {Hobbs}, D. and {Holland}, G. and {Huckle}, H.~E. and {Jardine}, K. and {Jasniewicz}, G. and {Jean-Antoine Piccolo}, A. and {Jim{\'e}nez-Arranz}, {\'O}. and {Jorissen}, A. and {Juaristi Campillo}, J. and {Julbe}, F. and {Karbevska}, L. and {Kervella}, P. and {Khanna}, S. and {Kontizas}, M. and {Kordopatis}, G. and {Korn}, A.~J. and {K{\'o}sp{\'a}l}, {\'A}. and {Kostrzewa-Rutkowska}, Z. and {Kruszy{\'n}ska}, K. and {Kun}, M. and {Laizeau}, P. and {Lambert}, S. and {Lanza}, A.~F. and {Lasne}, Y. and {Le Campion}, J. -F. and {Lebreton}, Y. and {Lebzelter}, T. and {Leccia}, S. and {Leclerc}, N. and {Lecoeur-Taibi}, I. and {Liao}, S. and {Licata}, E.~L. and {Lindstr{\o}m}, H.~E.~P. and {Lister}, T.~A. and {Livanou}, E. and {Lobel}, A. and {Lorca}, A. and {Loup}, C. and {Madrero Pardo}, P. and {Magdaleno Romeo}, A. and {Managau}, S. and {Mann}, R.~G. and {Manteiga}, M. and {Marchant}, J.~M. and {Marconi}, M. and {Marcos}, J. and {Marcos Santos}, M.~M.~S. and {Mar{\'\i}n Pina}, D. and {Marinoni}, S. and {Marocco}, F. and {Marshall}, D.~J. and {Martin Polo}, L. and {Mart{\'\i}n-Fleitas}, J.~M. and {Marton}, G. and {Mary}, N. and {Masip}, A. and {Massari}, D. and {Mastrobuono-Battisti}, A. and {Mazeh}, T. and {McMillan}, P.~J. and {Messina}, S. and {Michalik}, D. and {Millar}, N.~R. and {Mints}, A. and {Molina}, D. and {Molinaro}, R. and {Moln{\'a}r}, L. and {Monari}, G. and {Mongui{\'o}}, M. and {Montegriffo}, P. and {Montero}, A. and {Mor}, R. and {Mora}, A. and {Morbidelli}, R. and {Morel}, T. and {Morris}, D. and {Muraveva}, T. and {Murphy}, C.~P. and {Musella}, I. and {Nagy}, Z. and {Noval}, L. and {Oca{\~n}a}, F. and {Ogden}, A. and {Ordenovic}, C. and {Osinde}, J.~O. and {Pagani}, C. and {Pagano}, I. and {Palaversa}, L. and {Palicio}, P.~A. and {Pallas-Quintela}, L. and {Panahi}, A. and {Payne-Wardenaar}, S. and {Pe{\~n}alosa Esteller}, X. and {Penttil{\"a}}, A. and {Pichon}, B. and {Piersimoni}, A.~M. and {Pineau}, F. -X. and {Plachy}, E. and {Plum}, G. and {Poggio}, E. and {Pr{\v{s}}a}, A. and {Pulone}, L. and {Racero}, E. and {Ragaini}, S. and {Rainer}, M. and {Raiteri}, C.~M. and {Rambaux}, N. and {Ramos}, P. and {Ramos-Lerate}, M. and {Re Fiorentin}, P. and {Regibo}, S. and {Richards}, P.~J. and {Rios Diaz}, C. and {Ripepi}, V. and {Riva}, A. and {Rix}, H. -W. and {Rixon}, G. and {Robichon}, N. and {Robin}, A.~C. and {Robin}, C. and {Roelens}, M. and {Rogues}, H.~R.~O. and {Rohrbasser}, L. and {Romero-G{\'o}mez}, M. and {Rowell}, N. and {Royer}, F. and {Ruz Mieres}, D. and {Rybicki}, K.~A. and {Sadowski}, G. and {S{\'a}ez N{\'u}{\~n}ez}, A. and {Sagrist{\`a} Sell{\'e}s}, A. and {Sahlmann}, J. and {Salguero}, E. and {Samaras}, N. and {Sanchez Gimenez}, V. and {Sanna}, N. and {Santove{\~n}a}, R. and {Sarasso}, M. and {Schultheis}, M. and {Sciacca}, E. and {Segol}, M. and {Segovia}, J.~C. and {S{\'e}gransan}, D. and {Semeux}, D. and {Shahaf}, S. and {Siddiqui}, H.~I. and {Siebert}, A. and {Siltala}, L. and {Silvelo}, A. and {Slezak}, E. and {Slezak}, I. and {Smart}, R.~L. and {Snaith}, O.~N. and {Solano}, E. and {Solitro}, F. and {Souami}, D. and {Souchay}, J. and {Spagna}, A. and {Spina}, L. and {Spoto}, F. and {Steele}, I.~A. and {Steidelm{\"u}ller}, H. and {Stephenson}, C.~A. and {S{\"u}veges}, M. and {Surdej}, J. and {Szabados}, L. and {Szegedi-Elek}, E. and {Taris}, F. and {Taylor}, M.~B. and {Teixeira}, R. and {Tolomei}, L. and {Tonello}, N. and {Torra}, F. and {Torra}, J. and {Torralba Elipe}, G. and {Trabucchi}, M. and {Tsounis}, A.~T. and {Turon}, C. and {Ulla}, A. and {Unger}, N. and {Vaillant}, M.~V. and {van Dillen}, E. and {van Reeven}, W. and {Vanel}, O. and {Vecchiato}, A. and {Viala}, Y. and {Vicente}, D. and {Voutsinas}, S. and {Weiler}, M. and {Wevers}, T. and {Wyrzykowski}, {\L}. and {Yoldas}, A. and {Yvard}, P. and {Zhao}, H. and {Zorec}, J. and {Zucker}, S. and {Zwitter}, T.},
        title = "{Gaia Data Release 3. Summary of the content and survey properties}",
      journal = {A\&A},
         year = 2023,
        month = jun,
       volume = {674},
          eid = {A1},
        pages = {A1},
          doi = {10.1051/0004-6361/202243940},
archivePrefix = {arXiv},
       eprint = {2208.00211},
 primaryClass = {astro-ph.GA},
       adsurl = {https://ui.adsabs.harvard.edu/abs/2023A&A...674A...1G}
}

@online{gaia_archive,
  author    = {{European Space Agency}},
  title     = {Gaia Archive},
  year      = {2024},
  url       = {https://gea.esac.esa.int/archive/},
  note      = {\url{https://gea.esac.esa.int/archive/} Accessed: 2025-06-15},
}

@misc{Bailer-Jones,
       author = {{Bailer-Jones}, C.~A.~L. and {Rybizki}, J. and {Fouesneau}, M. and {Demleitner}, M. and {Andrae}, R.},
        title = "{VizieR Online Data Catalog: Distances to 1.47 billion stars in Gaia EDR3 (Bailer-Jones+, 2021)}",
 howpublished = {VizieR On-line Data Catalog: I/352.  Originally published in: 2021AJ....161..147B},
         year = 2021,
        month = feb,
          eid = {I/352},
       adsurl = {https://ui.adsabs.harvard.edu/abs/2021yCat.1352....0B}
}

@article{York_2000,
doi = {10.1086/301513},
url = {https://doi.org/10.1086/301513},
year = {2000},
month = {sep},
publisher = {},
volume = {120},
number = {3},
pages = {1579},
author = {York, Donald G. and Adelman, J. and Anderson, Jr., John E. and Anderson, Scott F. and Annis, James and Bahcall, Neta A. and Bakken, J. A. and Barkhouser, Robert and Bastian, Steven and Berman, Eileen and Boroski, William N. and Bracker, Steve and Briegel, Charlie and Briggs, John W. and Brinkmann, J. and Brunner, Robert and Burles, Scott and Carey, Larry and Carr, Michael A. and Castander, Francisco J. and Chen, Bing and Colestock, Patrick L. and Connolly, A. J. and Crocker, J. H. and Csabai, István and Czarapata, Paul C. and Davis, John Eric and Doi, Mamoru and Dombeck, Tom and Eisenstein, Daniel and Ellman, Nancy and Elms, Brian R. and Evans, Michael L. and Fan, Xiaohui and Federwitz, Glenn R. and Fiscelli, Larry and Friedman, Scott and Frieman, Joshua A. and Fukugita, Masataka and Gillespie, Bruce and Gunn, James E. and Gurbani, Vijay K. and de Haas, Ernst and Haldeman, Merle and Harris, Frederick H. and Hayes, J. and Heckman, Timothy M. and Hennessy, G. S. and Hindsley, Robert B. and Holm, Scott and Holmgren, Donald J. and Huang, Chi-hao and Hull, Charles and Husby, Don and Ichikawa, Shin-Ichi and Ichikawa, Takashi and Ivezić, Željko and Kent, Stephen and Kim, Rita S. J. and Kinney, E. and Klaene, Mark and Kleinman, A. N. and Kleinman, S. and Knapp, G. R. and Korienek, John and Kron, Richard G. and Kunszt, Peter Z. and Lamb, D. Q. and Lee, B. and Leger, R. French and Limmongkol, Siriluk and Lindenmeyer, Carl and Long, Daniel C. and Loomis, Craig and Loveday, Jon and Lucinio, Rich and Lupton, Robert H. and MacKinnon, Bryan and Mannery, Edward J. and Mantsch, P. M. and Margon, Bruce and McGehee, Peregrine and McKay, Timothy A. and Meiksin, Avery and Merelli, Aronne and Monet, David G. and Munn, Jeffrey A. and Narayanan, Vijay K. and Nash, Thomas and Neilsen, Eric and Neswold, Rich and Newberg, Heidi Jo and Nichol, R. C. and Nicinski, Tom and Nonino, Mario and Okada, Norio and Okamura, Sadanori and Ostriker, Jeremiah P. and Owen, Russell and Pauls, A. George and Peoples, John and Peterson, R. L. and Petravick, Donald and Pier, Jeffrey R. and Pope, Adrian and Pordes, Ruth and Prosapio, Angela and Rechenmacher, Ron and Quinn, Thomas R. and Richards, Gordon T. and Richmond, Michael W. and Rivetta, Claudio H. and Rockosi, Constance M. and Ruthmansdorfer, Kurt and Sandford, Dale and Schlegel, David J. and Schneider, Donald P. and Sekiguchi, Maki and Sergey, Gary and Shimasaku, Kazuhiro and Siegmund, Walter A. and Smee, Stephen and Smith, J. Allyn and Snedden, S. and Stone, R. and Stoughton, Chris and Strauss, Michael A. and Stubbs, Christopher and SubbaRao, Mark and Szalay, Alexander S. and Szapudi, Istvan and Szokoly, Gyula P. and Thakar, Anirudda R. and Tremonti, Christy and Tucker, Douglas L. and Uomoto, Alan and Vanden Berk, Dan and Vogeley, Michael S. and Waddell, Patrick and Wang, Shu-i and Watanabe, Masaru and Weinberg, David H. and Yanny, Brian and Yasuda, Naoki},
title = {The Sloan Digital Sky Survey: Technical Summary},
journal = {AJ}
}

@ARTICLE{Daniela_2024,
       author = {{Mu{\~n}oz-Giraldo}, Daniela and {Stelzer}, Beate and {Schwope}, Axel},
        title = "{Cataclysmic variables around the period-bounce: An eROSITA-enhanced multiwavelength catalog}",
      journal = {A\&A},
         year = {2024a},
        month = jul,
       volume = {687},
          eid = {A305},
        pages = {A305},
          doi = {10.1051/0004-6361/202449358},
archivePrefix = {arXiv},
       eprint = {2401.17298},
 primaryClass = {astro-ph.SR},
       adsurl = {https://ui.adsabs.harvard.edu/abs/2024A&A...687A.305M}
}

@article{Daniela_2024b,
  title={Catalog of Cataclysmic Variables Around the Period-bounce: New Systems},
  author={Mu{\~n}oz-Giraldo, Daniela and Stelzer, Beate and Schwope, Axel},
  journal={Research Notes of the AAS},
  volume={8},
  number={11},
  pages={279},
  year={2024b},
  publisher={The American Astronomical Society}
}

@ARTICLE{Hessman_1984,
       author = {{Hessman}, F.~V. and {Robinson}, E.~L. and {Nather}, R.~E. and {Zhang}, E. -H.},
        title = "{Time-resolved spectroscopy of SS Cygni at minimum and maximum light.}",
      journal = {\apj},
         year = 1984,
        month = nov,
       volume = {286},
        pages = {747-759},
          doi = {10.1086/162651},
       adsurl = {https://ui.adsabs.harvard.edu/abs/1984ApJ...286..747H}
}

@ARTICLE{Littlefair_2007,
       author = {{Littlefair}, S.~P. and {Dhillon}, V.~S. and {Marsh}, T.~R. and {G{\"a}nsicke}, B.~T. and {Baraffe}, I. and {Watson}, C.~A.},
        title = "{SDSS J150722.30+523039.8: a cataclysmic variable formed directly from a detached white dwarf/brown dwarf binary?}",
      journal = {\mnras},
         year = 2007,
        month = oct,
       volume = {381},
       number = {2},
        pages = {827-834},
          doi = {10.1111/j.1365-2966.2007.12285.x},
archivePrefix = {arXiv},
       eprint = {0708.0097},
 primaryClass = {astro-ph},
       adsurl = {https://ui.adsabs.harvard.edu/abs/2007MNRAS.381..827L}
}

@ARTICLE{Solheim_2010,
       author = {{Solheim}, J. -E.},
        title = "{AM CVn Stars: Status and Challenges}",
      journal = {\pasp},
         year = 2010,
        month = oct,
       volume = {122},
       number = {896},
        pages = {1133},
          doi = {10.1086/656680},
       adsurl = {https://ui.adsabs.harvard.edu/abs/2010PASP..122.1133S}
}

@article{Pala_2018,
    author = {Pala, A F and Gänsicke, B T and Marsh, T R and Breedt, E and Hermes, J J and Landstreet, J D and Schreiber, M R and Townsley, D M and Wang, L and Aungwerojwit, A and Hambsch, F–J and Monard, B and Myers, G and Nelson, P and Pickard, R and Poyner, G and Reichart, D E and Stubbings, R and Godon, P and Szkody, P and De Martino, D and Dhillon, V S and Knigge, C and Parsons, S G},
    title = {Evidence for mass accretion driven by spiral shocks onto the white dwarf in SDSS J123813.73–033933.0},
    journal = {MNRAS},
    volume = {483},
    number = {1},
    pages = {1080-1103},
    year = {2018},
    month = {11},
    issn = {0035-8711},
    doi = {10.1093/mnras/sty3174},
    url = {https://doi.org/10.1093/mnras/sty3174},
    eprint = {https://academic.oup.com/mnras/article-pdf/483/1/1080/27053271/sty3174.pdf},
}

@article{Bailey_1980,
    author = {Bailey, Jeremy},
    title = {Colour variations during dwarf nova outbursts},
    journal = {MNRAS},
    volume = {190},
    number = {2},
    pages = {119-123},
    year = {1980},
    month = {02},
    issn = {0035-8711},
    doi = {10.1093/mnras/190.2.119},
    url = {https://doi.org/10.1093/mnras/190.2.119},
    eprint = {https://academic.oup.com/mnras/article-pdf/190/2/119/2983170/mnras190-0119.pdf},
}

@ARTICLE{Cathy_1990,
       author = {{Mansperger}, Cathy S. and {Kaitchuck}, Ronald H.},
        title = "{Spectroscopy of the Dwarf Nova TW Virginis Caught on the Rise to Outburst}",
      journal = {\apj},
         year = 1990,
        month = jul,
       volume = {358},
        pages = {268},
          doi = {10.1086/168983},
       adsurl = {https://ui.adsabs.harvard.edu/abs/1990ApJ...358..268M}
}

@article{Nogami_2004,
    author = {Nogami, Daisaku and Iijima, Takashi},
    title = {Dramatic Spectral Evolution of WZ Sagittae during the 2001 Superoutburst},
    journal = {PASJ},
    volume = {56},
    number = {sp1},
    pages = {S163-S182},
    year = {2004},
    month = {03},
    issn = {0004-6264},
    doi = {10.1093/pasj/56.sp1.S163},
    url = {https://doi.org/10.1093/pasj/56.sp1.S163},
    eprint = {https://academic.oup.com/pasj/article-pdf/56/sp1/S163/54709364/pasj_56_sp1_s163.pdf},
}

@ARTICLE{Blanton_2017,
       author = {{Blanton}, Michael R. and {Bershady}, Matthew A. and {Abolfathi}, Bela and {Albareti}, Franco D. and {Allende Prieto}, Carlos and {Almeida}, Andres and {Alonso-Garc{\'\i}a}, Javier and {Anders}, Friedrich and {Anderson}, Scott F. and {Andrews}, Brett and {Aquino-Ort{\'\i}z}, Erik and {Arag{\'o}n-Salamanca}, Alfonso and {Argudo-Fern{\'a}ndez}, Maria and {Armengaud}, Eric and {Aubourg}, Eric and {Avila-Reese}, Vladimir and {Badenes}, Carles and {Bailey}, Stephen and {Barger}, Kathleen A. and {Barrera-Ballesteros}, Jorge and {Bartosz}, Curtis and {Bates}, Dominic and {Baumgarten}, Falk and {Bautista}, Julian and {Beaton}, Rachael and {Beers}, Timothy C. and {Belfiore}, Francesco and {Bender}, Chad F. and {Berlind}, Andreas A. and {Bernardi}, Mariangela and {Beutler}, Florian and {Bird}, Jonathan C. and {Bizyaev}, Dmitry and {Blanc}, Guillermo A. and {Blomqvist}, Michael and {Bolton}, Adam S. and {Boquien}, M{\'e}d{\'e}ric and {Borissova}, Jura and {van den Bosch}, Remco and {Bovy}, Jo and {Brandt}, William N. and {Brinkmann}, Jonathan and {Brownstein}, Joel R. and {Bundy}, Kevin and {Burgasser}, Adam J. and {Burtin}, Etienne and {Busca}, Nicol{\'a}s G. and {Cappellari}, Michele and {Delgado Carigi}, Maria Leticia and {Carlberg}, Joleen K. and {Carnero Rosell}, Aurelio and {Carrera}, Ricardo and {Chanover}, Nancy J. and {Cherinka}, Brian and {Cheung}, Edmond and {G{\'o}mez Maqueo Chew}, Yilen and {Chiappini}, Cristina and {Choi}, Peter Doohyun and {Chojnowski}, Drew and {Chuang}, Chia-Hsun and {Chung}, Haeun and {Cirolini}, Rafael Fernando and {Clerc}, Nicolas and {Cohen}, Roger E. and {Comparat}, Johan and {da Costa}, Luiz and {Cousinou}, Marie-Claude and {Covey}, Kevin and {Crane}, Jeffrey D. and {Croft}, Rupert A.~C. and {Cruz-Gonzalez}, Irene and {Garrido Cuadra}, Daniel and {Cunha}, Katia and {Damke}, Guillermo J. and {Darling}, Jeremy and {Davies}, Roger and {Dawson}, Kyle and {de la Macorra}, Axel and {Dell'Agli}, Flavia and {De Lee}, Nathan and {Delubac}, Timoth{\'e}e and {Di Mille}, Francesco and {Diamond-Stanic}, Aleks and {Cano-D{\'\i}az}, Mariana and {Donor}, John and {Downes}, Juan Jos{\'e} and {Drory}, Niv and {du Mas des Bourboux}, H{\'e}lion and {Duckworth}, Christopher J. and {Dwelly}, Tom and {Dyer}, Jamie and {Ebelke}, Garrett and {Eigenbrot}, Arthur D. and {Eisenstein}, Daniel J. and {Emsellem}, Eric and {Eracleous}, Mike and {Escoffier}, Stephanie and {Evans}, Michael L. and {Fan}, Xiaohui and {Fern{\'a}ndez-Alvar}, Emma and {Fernandez-Trincado}, J.~G. and {Feuillet}, Diane K. and {Finoguenov}, Alexis and {Fleming}, Scott W. and {Font-Ribera}, Andreu and {Fredrickson}, Alexander and {Freischlad}, Gordon and {Frinchaboy}, Peter M. and {Fuentes}, Carla E. and {Galbany}, Llu{\'\i}s and {Garcia-Dias}, R. and {Garc{\'\i}a-Hern{\'a}ndez}, D.~A. and {Gaulme}, Patrick and {Geisler}, Doug and {Gelfand}, Joseph D. and {Gil-Mar{\'\i}n}, H{\'e}ctor and {Gillespie}, Bruce A. and {Goddard}, Daniel and {Gonzalez-Perez}, Violeta and {Grabowski}, Kathleen and {Green}, Paul J. and {Grier}, Catherine J. and {Gunn}, James E. and {Guo}, Hong and {Guy}, Julien and {Hagen}, Alex and {Hahn}, ChangHoon and {Hall}, Matthew and {Harding}, Paul and {Hasselquist}, Sten and {Hawley}, Suzanne L. and {Hearty}, Fred and {Gonzalez Hern{\'a}ndez}, Jonay I. and {Ho}, Shirley and {Hogg}, David W. and {Holley-Bockelmann}, Kelly and {Holtzman}, Jon A. and {Holzer}, Parker H. and {Huehnerhoff}, Joseph and {Hutchinson}, Timothy A. and {Hwang}, Ho Seong and {Ibarra-Medel}, H{\'e}ctor J. and {da Silva Ilha}, Gabriele and {Ivans}, Inese I. and {Ivory}, KeShawn and {Jackson}, Kelly and {Jensen}, Trey W. and {Johnson}, Jennifer A. and {Jones}, Amy and {J{\"o}nsson}, Henrik and {Jullo}, Eric and {Kamble}, Vikrant and {Kinemuchi}, Karen and {Kirkby}, David and {Kitaura}, Francisco-Shu and {Klaene}, Mark and {Knapp}, Gillian R. and {Kneib}, Jean-Paul and {Kollmeier}, Juna A. and {Lacerna}, Ivan and {Lane}, Richard R. and {Lang}, Dustin and {Law}, David R. and {Lazarz}, Daniel and {Lee}, Youngbae and {Le Goff}, Jean-Marc and {Liang}, Fu-Heng and {Li}, Cheng and {Li}, Hongyu and {Lian}, Jianhui and {Lima}, Marcos and {Lin}, Lihwai and {Lin}, Yen-Ting and {Bertran de Lis}, Sara and {Liu}, Chao and {de Icaza Lizaola}, Miguel Angel C. and {Long}, Dan and {Lucatello}, Sara and {Lundgren}, Britt and {MacDonald}, Nicholas K. and {Deconto Machado}, Alice and {MacLeod}, Chelsea L. and {Mahadevan}, Suvrath and {Geimba Maia}, Marcio Antonio and {Maiolino}, Roberto and {Majewski}, Steven R. and {Malanushenko}, Elena and {Malanushenko}, Viktor and {Manchado}, Arturo and {Mao}, Shude and {Maraston}, Claudia and {Marques-Chaves}, Rui and {Masseron}, Thomas and {Masters}, Karen L. and {McBride}, Cameron K. and {McDermid}, Richard M. and {McGrath}, Brianne and {McGreer}, Ian D. and {Medina Pe{\~n}a}, Nicol{\'a}s and {Melendez}, Matthew},
        title = "{Sloan Digital Sky Survey IV: Mapping the Milky Way, Nearby Galaxies, and the Distant Universe}",
      journal = {\aj},
         year = 2017,
        month = jul,
       volume = {154},
       number = {1},
          eid = {28},
        pages = {28},
          doi = {10.3847/1538-3881/aa7567},
archivePrefix = {arXiv},
       eprint = {1703.00052},
 primaryClass = {astro-ph.GA},
       adsurl = {https://ui.adsabs.harvard.edu/abs/2017AJ....154...28B}
}

@ARTICLE{Abdurro_2022,
       author = {{Abdurro'uf} and {Accetta}, Katherine and {Aerts}, Conny and {Silva Aguirre}, V{\'\i}ctor and {Ahumada}, Romina and {Ajgaonkar}, Nikhil and {Filiz Ak}, N. and {Alam}, Shadab and {Allende Prieto}, Carlos and {Almeida}, Andr{\'e}s and {Anders}, Friedrich and {Anderson}, Scott F. and {Andrews}, Brett H. and {Anguiano}, Borja and {Aquino-Ort{\'\i}z}, Erik and {Arag{\'o}n-Salamanca}, Alfonso and {Argudo-Fern{\'a}ndez}, Maria and {Ata}, Metin and {Aubert}, Marie and {Avila-Reese}, Vladimir and {Badenes}, Carles and {Barb{\'a}}, Rodolfo H. and {Barger}, Kat and {Barrera-Ballesteros}, Jorge K. and {Beaton}, Rachael L. and {Beers}, Timothy C. and {Belfiore}, Francesco and {Bender}, Chad F. and {Bernardi}, Mariangela and {Bershady}, Matthew A. and {Beutler}, Florian and {Bidin}, Christian Moni and {Bird}, Jonathan C. and {Bizyaev}, Dmitry and {Blanc}, Guillermo A. and {Blanton}, Michael R. and {Boardman}, Nicholas Fraser and {Bolton}, Adam S. and {Boquien}, M{\'e}d{\'e}ric and {Borissova}, Jura and {Bovy}, Jo and {Brandt}, W.~N. and {Brown}, Jordan and {Brownstein}, Joel R. and {Brusa}, Marcella and {Buchner}, Johannes and {Bundy}, Kevin and {Burchett}, Joseph N. and {Bureau}, Martin and {Burgasser}, Adam and {Cabang}, Tuesday K. and {Campbell}, Stephanie and {Cappellari}, Michele and {Carlberg}, Joleen K. and {Wanderley}, F{\'a}bio Carneiro and {Carrera}, Ricardo and {Cash}, Jennifer and {Chen}, Yan-Ping and {Chen}, Wei-Huai and {Cherinka}, Brian and {Chiappini}, Cristina and {Choi}, Peter Doohyun and {Chojnowski}, S. Drew and {Chung}, Haeun and {Clerc}, Nicolas and {Cohen}, Roger E. and {Comerford}, Julia M. and {Comparat}, Johan and {da Costa}, Luiz and {Covey}, Kevin and {Crane}, Jeffrey D. and {Cruz-Gonzalez}, Irene and {Culhane}, Connor and {Cunha}, Katia and {Dai}, Y. Sophia and {Damke}, Guillermo and {Darling}, Jeremy and {Davidson}, Jr., James W. and {Davies}, Roger and {Dawson}, Kyle and {De Lee}, Nathan and {Diamond-Stanic}, Aleksandar M. and {Cano-D{\'\i}az}, Mariana and {S{\'a}nchez}, Helena Dom{\'\i}nguez and {Donor}, John and {Duckworth}, Chris and {Dwelly}, Tom and {Eisenstein}, Daniel J. and {Elsworth}, Yvonne P. and {Emsellem}, Eric and {Eracleous}, Mike and {Escoffier}, Stephanie and {Fan}, Xiaohui and {Farr}, Emily and {Feng}, Shuai and {Fern{\'a}ndez-Trincado}, Jos{\'e} G. and {Feuillet}, Diane and {Filipp}, Andreas and {Fillingham}, Sean P. and {Frinchaboy}, Peter M. and {Fromenteau}, Sebastien and {Galbany}, Llu{\'\i}s and {Garc{\'\i}a}, Rafael A. and {Garc{\'\i}a-Hern{\'a}ndez}, D.~A. and {Ge}, Junqiang and {Geisler}, Doug and {Gelfand}, Joseph and {G{\'e}ron}, Tobias and {Gibson}, Benjamin J. and {Goddy}, Julian and {Godoy-Rivera}, Diego and {Grabowski}, Kathleen and {Green}, Paul J. and {Greener}, Michael and {Grier}, Catherine J. and {Griffith}, Emily and {Guo}, Hong and {Guy}, Julien and {Hadjara}, Massinissa and {Harding}, Paul and {Hasselquist}, Sten and {Hayes}, Christian R. and {Hearty}, Fred and {Hern{\'a}ndez}, Jes{\'u}s and {Hill}, Lewis and {Hogg}, David W. and {Holtzman}, Jon A. and {Horta}, Danny and {Hsieh}, Bau-Ching and {Hsu}, Chin-Hao and {Hsu}, Yun-Hsin and {Huber}, Daniel and {Huertas-Company}, Marc and {Hutchinson}, Brian and {Hwang}, Ho Seong and {Ibarra-Medel}, H{\'e}ctor J. and {Chitham}, Jacob Ider and {Ilha}, Gabriele S. and {Imig}, Julie and {Jaekle}, Will and {Jayasinghe}, Tharindu and {Ji}, Xihan and {Johnson}, Jennifer A. and {Jones}, Amy and {J{\"o}nsson}, Henrik and {Katkov}, Ivan and {Khalatyan}, Dr., Arman and {Kinemuchi}, Karen and {Kisku}, Shobhit and {Knapen}, Johan H. and {Kneib}, Jean-Paul and {Kollmeier}, Juna A. and {Kong}, Miranda and {Kounkel}, Marina and {Kreckel}, Kathryn and {Krishnarao}, Dhanesh and {Lacerna}, Ivan and {Lane}, Richard R. and {Langgin}, Rachel and {Lavender}, Ramon and {Law}, David R. and {Lazarz}, Daniel and {Leung}, Henry W. and {Leung}, Ho-Hin and {Lewis}, Hannah M. and {Li}, Cheng and {Li}, Ran and {Lian}, Jianhui and {Liang}, Fu-Heng and {Lin}, Lihwai and {Lin}, Yen-Ting and {Lin}, Sicheng and {Lintott}, Chris and {Long}, Dan and {Longa-Pe{\~n}a}, Pen{\'e}lope and {L{\'o}pez-Cob{\'a}}, Carlos and {Lu}, Shengdong and {Lundgren}, Britt F. and {Luo}, Yuanze and {Mackereth}, J. Ted and {de la Macorra}, Axel and {Mahadevan}, Suvrath and {Majewski}, Steven R. and {Manchado}, Arturo and {Mandeville}, Travis and {Maraston}, Claudia and {Margalef-Bentabol}, Berta and {Masseron}, Thomas and {Masters}, Karen L. and {Mathur}, Savita and {McDermid}, Richard M. and {Mckay}, Myles and {Merloni}, Andrea and {Merrifield}, Michael and {Meszaros}, Szabolcs and {Miglio}, Andrea and {Di Mille}, Francesco and {Minniti}, Dante and {Minsley}, Rebecca and {Monachesi}, Antonela},
        title = "{The Seventeenth Data Release of the Sloan Digital Sky Surveys: Complete Release of MaNGA, MaStar, and APOGEE-2 Data}",
      journal = {\apjs},
         year = 2022,
        month = apr,
       volume = {259},
       number = {2},
          eid = {35},
        pages = {35},
          doi = {10.3847/1538-4365/ac4414},
archivePrefix = {arXiv},
       eprint = {2112.02026},
 primaryClass = {astro-ph.GA},
       adsurl = {https://ui.adsabs.harvard.edu/abs/2022ApJS..259...35A}
}

@article{Smee_2013,
doi = {10.1088/0004-6256/146/2/32},
url = {https://doi.org/10.1088/0004-6256/146/2/32},
year = {2013},
month = {jul},
publisher = {The American Astronomical Society},
volume = {146},
number = {2},
pages = {32},
author = {Smee, Stephen A. and Gunn, James E. and Uomoto, Alan and Roe, Natalie and Schlegel, David and Rockosi, Constance M. and Carr, Michael A. and Leger, French and Dawson, Kyle S. and Olmstead, Matthew D. and Brinkmann, Jon and Owen, Russell and Barkhouser, Robert H. and Honscheid, Klaus and Harding, Paul and Long, Dan and Lupton, Robert H. and Loomis, Craig and Anderson, Lauren and Annis, James and Bernardi, Mariangela and Bhardwaj, Vaishali and Bizyaev, Dmitry and Bolton, Adam S. and Brewington, Howard and Briggs, John W. and Burles, Scott and Burns, James G. and Castander, Francisco Javier and Connolly, Andrew and Davenport, James R. A. and Ebelke, Garrett and Epps, Harland and Feldman, Paul D. and Friedman, Scott D. and Frieman, Joshua and Heckman, Timothy and Hull, Charles L. and Knapp, Gillian R. and Lawrence, David M. and Loveday, Jon and Mannery, Edward J. and Malanushenko, Elena and Malanushenko, Viktor and Merrelli, Aronne James and Muna, Demitri and Newman, Peter R. and Nichol, Robert C. and Oravetz, Daniel and Pan, Kaike and Pope, Adrian C. and Ricketts, Paul G. and Shelden, Alaina and Sandford, Dale and Siegmund, Walter and Simmons, Audrey and Smith, D. Shane and Snedden, Stephanie and Schneider, Donald P. and SubbaRao, Mark and Tremonti, Christy and Waddell, Patrick and York, Donald G.},
title = {THE MULTI-OBJECT, FIBER-FED SPECTROGRAPHS FOR THE SLOAN DIGITAL SKY SURVEY AND THE BARYON OSCILLATION SPECTROSCOPIC SURVEY},
journal = {AJ}
}

@article{Dawson_2016,
doi = {10.3847/0004-6256/151/2/44},
url = {https://doi.org/10.3847/0004-6256/151/2/44},
year = {2016},
month = {feb},
publisher = {The American Astronomical Society},
volume = {151},
number = {2},
pages = {44},
author = {Dawson, Kyle S. and Kneib, Jean-Paul and Percival, Will J. and Alam, Shadab and Albareti, Franco D. and Anderson, Scott F. and Armengaud, Eric and Aubourg, Éric and Bailey, Stephen and Bautista, Julian E. and Berlind, Andreas A. and Bershady, Matthew A. and Beutler, Florian and Bizyaev, Dmitry and Blanton, Michael R. and Blomqvist, Michael and Bolton, Adam S. and Bovy, Jo and Brandt, W. N. and Brinkmann, Jon and Brownstein, Joel R. and Burtin, Etienne and Busca, N. G. and Cai, Zheng and Chuang, Chia-Hsun and Clerc, Nicolas and Comparat, Johan and Cope, Frances and Croft, Rupert A. C. and Cruz-Gonzalez, Irene and Costa, Luiz N. da and Cousinou, Marie-Claude and Darling, Jeremy and Macorra, Axel de la and Torre, Sylvain de la and Delubac, Timothée and Bourboux, Hélion du Mas des and Dwelly, Tom and Ealet, Anne and Eisenstein, Daniel J. and Eracleous, Michael and Escoffier, S. and Fan, Xiaohui and Finoguenov, Alexis and Font-Ribera, Andreu and Frinchaboy, Peter and Gaulme, Patrick and Georgakakis, Antonis and Green, Paul and Guo, Hong and Guy, Julien and Ho, Shirley and Holder, Diana and Huehnerhoff, Joe and Hutchinson, Timothy and Jing, Yipeng and Jullo, Eric and Kamble, Vikrant and Kinemuchi, Karen and Kirkby, David and Kitaura, Francisco-Shu and Klaene, Mark A. and Laher, Russ R. and Lang, Dustin and Laurent, Pierre and Goff, Jean-Marc Le and Li, Cheng and Liang, Yu and Lima, Marcos and Lin, Qiufan and Lin, Weipeng and Lin, Yen-Ting and Long, Daniel C. and Lundgren, Britt and MacDonald, Nicholas and Maia, Marcio Antonio Geimba and Malanushenko, Elena and Malanushenko, Viktor and Mariappan, Vivek and McBride, Cameron K. and McGreer, Ian D. and Ménard, Brice and Merloni, Andrea and Meza, Andres and Montero-Dorta, Antonio D. and Muna, Demitri and Myers, Adam D. and Nandra, Kirpal and Naugle, Tracy and Newman, Jeffrey A. and Noterdaeme, Pasquier and Nugent, Peter and Ogando, Ricardo and Olmstead, Matthew D. and Oravetz, Audrey and Oravetz, Daniel J. and Padmanabhan, Nikhil and Palanque-Delabrouille, Nathalie and Pan, Kaike and Parejko, John K. and Pâris, Isabelle and Peacock, John A. and Petitjean, Patrick and Pieri, Matthew M. and Pisani, Alice and Prada, Francisco and Prakash, Abhishek and Raichoor, Anand and Reid, Beth and Rich, James and Ridl, Jethro and Rodriguez-Torres, Sergio and Rosell, Aurelio Carnero and Ross, Ashley J. and Rossi, Graziano and Ruan, John and Salvato, Mara and Sayres, Conor and Schneider, Donald P. and Schlegel, David J. and Seljak, Uros and Seo, Hee-Jong and Sesar, Branimir and Shandera, Sarah and Shu, Yiping and Slosar, Anže and Sobreira, Flavia and Streblyanska, Alina and Suzuki, Nao and Taylor, Donna and Tao, Charling and Tinker, Jeremy L. and Tojeiro, Rita and Vargas-Magaña, Mariana and Wang, Yuting and Weaver, Benjamin A. and Weinberg, David H. and White, Martin and Wood-Vasey, W. M. and Yeche, Christophe and Zhai, Zhongxu and Zhao, Cheng and Zhao, Gong-bo and Zheng, Zheng and Zhu, Guangtun Ben and Zou, Hu},
title = {THE SDSS-IV EXTENDED BARYON OSCILLATION SPECTROSCOPIC SURVEY: OVERVIEW AND EARLY DATA},
journal = {AJ}
}

@article{Gansicke_2009,
    author = {Gänsicke, B. T. and Dillon, M. and Southworth, J. and Thorstensen, J. R. and Rodríguez-Gil, P. and Aungwerojwit, A. and Marsh, T. R. and Szkody, P. and Barros, S. C. C. and Casares, J. and De Martino, D. and Groot, P. J. and Hakala, P. and Kolb, U. and Littlefair, S. P. and Martínez-Pais, I. G. and Nelemans, G. and Schreiber, M. R.},
    title = {SDSS unveils a population of intrinsically faint cataclysmic variables at the minimum orbital period},
    journal = {MNRAS},
    volume = {397},
    number = {4},
    pages = {2170-2188},
    year = {2009},
    month = {08},
    issn = {0035-8711},
    doi = {10.1111/j.1365-2966.2009.15126.x},
    url = {https://doi.org/10.1111/j.1365-2966.2009.15126.x},
    eprint = {https://academic.oup.com/mnras/article-pdf/397/4/2170/3207035/mnras0397-2170.pdf}
}

@article{Echevarria_1988,
    author = {Echevarría, J.},
    title = {A statistical analysis of the emission line ratios in cataclysmic variables},
    journal = {MNRAS},
    volume = {233},
    number = {3},
    pages = {513-527},
    year = {1988},
    month = {08},
    issn = {0035-8711},
    doi = {10.1093/mnras/233.3.513},
    url = {https://doi.org/10.1093/mnras/233.3.513},
    eprint = {https://academic.oup.com/mnras/article-pdf/233/3/513/3135610/mnras233-0513.pdf}
}

@article{Hoerl_1970a,
 ISSN = {00401706},
 URL = {http://www.jstor.org/stable/1267351},
 author = {Arthur E. Hoerl and Robert W. Kennard},
 journal = {Technometrics},
 number = {1},
 pages = {55--67},
 publisher = {[Taylor & Francis, Ltd., American Statistical Association, American Society for Quality]},
 title = {Ridge Regression: Biased Estimation for Nonorthogonal Problems},
 urldate = {2025-11-05},
 volume = {12},
 year = {1970a}
}

@article{Hoerl_1970b,
author = {Arthur E. Hoerl and Robert W. Kennard},
title = {Ridge Regression: Applications to Nonorthogonal Problems},
journal = {Technometrics},
volume = {12},
number = {1},
pages = {69--82},
year = {1970b},
publisher = {ASA Website},
doi = {10.1080/00401706.1970.10488635},
URL = {https://www.tandfonline.com/doi/abs/10.1080/00401706.1970.10488635},
eprint = {https://www.tandfonline.com/doi/pdf/10.1080/00401706.1970.10488635}
}

@article{Berkson_1944,
author = {Joseph Berkson},
title = {Application of the Logistic Function to Bio-Assay},
journal = {JASA},
volume = {39},
number = {227},
pages = {357--365},
year = {1944},
publisher = {ASA Website},
doi = {10.1080/01621459.1944.10500699},
URL = {https://doi.org/10.1080/01621459.1944.10500699},
eprint = {https://doi.org/10.1080/01621459.1944.10500699}
}

@article{Cox_1958,
 ISSN = {00359246},
 URL = {http://www.jstor.org/stable/2983890},
 author = {D. R. Cox},
 journal = {Journal of the Royal Statistical Society. Series B (Methodological)},
 number = {2},
 pages = {215--242},
 publisher = {[Royal Statistical Society, Oxford University Press]},
 title = {The Regression Analysis of Binary Sequences},
 urldate = {2025-11-05},
 volume = {20},
 year = {1958}
}

@inproceedings{Kohavi_1995,
author = {Kohavi, Ron},
title = {A study of cross-validation and bootstrap for accuracy estimation and model selection},
year = {1995},
isbn = {1558603638},
publisher = {Morgan Kaufmann Publishers Inc.},
address = {San Francisco, CA, USA},
booktitle = {Proceedings of the 14th International Joint Conference on Artificial Intelligence - Volume 2},
pages = {1137–1143},
numpages = {7},
location = {Montreal, Quebec, Canada},
series = {IJCAI'95}
}

@article{Kato_2015,
    author = {Kato, Taichi},
    title = {WZ Sge-type dwarf novae},
    journal = {PASJ},
    volume = {67},
    number = {6},
    pages = {108},
    year = {2015},
    month = {09},
    issn = {0004-6264},
    doi = {10.1093/pasj/psv077},
    url = {https://doi.org/10.1093/pasj/psv077},
    eprint = {https://academic.oup.com/pasj/article-pdf/67/6/108/54682604/pasj_67_6_108.pdf},
}

@article{Casagrande,
    author = {Casagrande, L. and VandenBerg, Don A.},
    title = {Synthetic stellar photometry – I. General considerations and new transformations for broad-band systems},
    journal = {MNRAS},
    volume = {444},
    number = {1},
    pages = {392-419},
    year = {2014},
    month = {08},
}

@article{predehl2021,
  title={The eROSITA X-ray telescope on SRG},
  author={Predehl, P and Andritschke, R and Arefiev, V and Babyshkin, V and Batanov, O and Becker, W and B{\"o}hringer, H and Bogomolov, A and Boller, T and Borm, K and others},
  journal={A\&A},
  volume={647},
  pages={A1},
  year={2021},
  publisher={EDP Sciences}
}

@article{sunyaev2021,
  title={SRG X-ray orbital observatory-Its telescopes and first scientific results},
  author={Sunyaev, R and Arefiev, V and Babyshkin, V and Bogomolov, A and Borisov, K and Buntov, M and Brunner, H and Burenin, R and Churazov, E and Coutinho, D and others},
  journal={A\&A},
  volume={656},
  pages={A132},
  year={2021},
  publisher={EDP Sciences}
}

@ARTICLE{Gansicke_2014,
       author = {{Manser}, C.~J. and {Gänsicke}, B.~T.},
        title = "{Spectroscopy of the enigmatic short-period cataclysmic variable IR Com in an extended low state.}",
      journal = {\mnras},
         year = 2014,
        month = jul,
       volume = {442},
        pages = {L23-L27},
          doi = {10.1093/mnrasl/slu049},
archivePrefix = {arXiv},
       eprint = {1401.5055},
 primaryClass = {astro-ph.SR},
       adsurl = {https://ui.adsabs.harvard.edu/abs/2014MNRAS.442L..23M}
}

@article{Kafka_2005,
doi = {10.1086/431793},
url = {https://doi.org/10.1086/431793},
year = {2005},
month = {aug},
publisher = {},
volume = {130},
number = {2},
pages = {742},
author = {Kafka, S. and Honeycutt, R. K.},
title = {High/Low States in Magnetic Cataclysmic Variables},
journal = {AJ}
}

@article{Chavez_2022,
    author = {Chavez, Carlos E and Georgakarakos, Nikolaos and Aviles, Andres and Aceves, Hector and Tovmassian, Gagik and Zharikov, Sergey and Perez–Leon, J E and Tamayo, Francisco},
    title = {Testing the third-body hypothesis in the cataclysmic variables LU Camelopardalis, QZ Serpentis, V1007 Herculis and BK Lyncis},
    journal = {MNRAS},
    volume = {514},
    number = {3},
    pages = {4629-4638},
    year = {2022},
    month = {07},
    issn = {0035-8711},
    doi = {10.1093/mnras/stac1112},
    url = {https://doi.org/10.1093/mnras/stac1112},
    eprint = {https://academic.oup.com/mnras/article-pdf/514/3/4629/44395804/stac1112.pdf},
}

@article{Thorstensen_2002,
doi = {10.1086/342513},
url = {https://doi.org/10.1086/342513},
year = {2002},
month = {oct},
publisher = {The University of Chicago Press},
volume = {114},
number = {800},
pages = {1117},
author = {Thorstensen, John R. and Fenton, William H. and Patterson, Joseph and Kemp, Jonathan and Halpern, Jules and Baraffe, Isabelle},
title = {QZ Serpentis: A Dwarf Nova with a 2 Hour Orbital Period and an Anomalously Hot, Bright Secondary Star1},
journal = {PASP}
}

@ARTICLE{Shakura_Sunyaev_1973,
       author = {{Shakura}, N.~I. and {Sunyaev}, R.~A.},
        title = "{Black holes in binary systems. Observational appearance.}",
      journal = {A\&A},
         year = 1973,
        month = jan,
       volume = {24},
        pages = {337-355},
       adsurl = {https://ui.adsabs.harvard.edu/abs/1973A&A....24..337S}
}

@article{Imbalanced_Classes,
  title={Learning from Imbalanced Data},
  author={Haibo He and Edwardo A. Garcia},
  journal={IEEE Transactions on Knowledge and Data Engineering},
  year={2009},
  volume={21},
  pages={1263-1284},
  url={https://api.semanticscholar.org/CorpusID:206742563}
}

@article{scikit-learn,
  title={Scikit-learn: Machine Learning in {P}ython},
  author={Pedregosa, F. and Varoquaux, G. and Gramfort, A. and Michel, V.
          and Thirion, B. and Grisel, O. and Blondel, M. and Prettenhofer, P.
          and Weiss, R. and Dubourg, V. and Vanderplas, J. and Passos, A. and
          Cournapeau, D. and Brucher, M. and Perrot, M. and Duchesnay, E.},
  journal={JMLR},
  volume={12},
  pages={2825--2830},
  year={2011}
}

@article{Blair_1979,
author = {Blair, David C.},
title = {Information Retrieval, 2nd ed. C.J. Van Rijsbergen. London: Butterworths; 1979: 208 pp. Price: \$32.50},
journal = {JASIS},
volume = {30},
number = {6},
pages = {374-375},
doi = {https://doi.org/10.1002/asi.4630300621},
url = {https://asistdl.onlinelibrary.wiley.com/doi/abs/10.1002/asi.4630300621},
eprint = {https://asistdl.onlinelibrary.wiley.com/doi/pdf/10.1002/asi.4630300621},
year = {1979}
}

@article{Kent_1955,
author = {Kent, Allen and Berry, Madeline M. and Luehrs Jr., Fred U. and Perry, J. W.},
title = {Machine literature searching VIII. Operational criteria for designing information retrieval systems},
journal = {American Documentation},
volume = {6},
number = {2},
pages = {93-101},
doi = {https://doi.org/10.1002/asi.5090060209},
url = {https://onlinelibrary.wiley.com/doi/abs/10.1002/asi.5090060209},
eprint = {https://onlinelibrary.wiley.com/doi/pdf/10.1002/asi.5090060209},
year = {1955}
}

@article{Apte_1994,
author = {Apt\'{e}, Chidanand and Damerau, Fred and Weiss, Sholom M.},
title = {Automated learning of decision rules for text categorization},
year = {1994},
issue_date = {July 1994},
publisher = {Association for Computing Machinery},
address = {New York, NY, USA},
volume = {12},
number = {3},
issn = {1046-8188},
url = {https://doi.org/10.1145/183422.183423},
doi = {10.1145/183422.183423},
journal = {ACM Trans. Inf. Syst.},
month = jul,
pages = {233–251},
numpages = {19}
}

@article{Hanley_1982,
  title={The meaning and use of the area under a receiver operating characteristic (ROC) curve.},
  author={James A. Hanley and Barbara J. McNeil},
  journal={Radiology},
  year={1982},
  volume={143 1},
  pages={
          29-36
        },
  url={https://api.semanticscholar.org/CorpusID:10511727}
}

@article{Swets_1988,
author = {John A. Swets },
title = {Measuring the Accuracy of Diagnostic Systems},
journal = {Science},
volume = {240},
number = {4857},
pages = {1285-1293},
year = {1988},
doi = {10.1126/science.3287615},
URL = {https://www.science.org/doi/abs/10.1126/science.3287615},
eprint = {https://www.science.org/doi/pdf/10.1126/science.3287615}}

@article{Spearman_1904,
 ISSN = {00029556},
 URL = {http://www.jstor.org/stable/1412159},
 author = {C. Spearman},
 journal = {AJP},
 number = {1},
 pages = {72--101},
 publisher = {University of Illinois Press},
 title = {The Proof and Measurement of Association between Two Things},
 urldate = {2025-11-19},
 volume = {15},
 year = {1904}
}

@ARTICLE{Smak_1969,
       author = {{Smak}, J.},
        title = "{On the Rotational Velocities of Gaseous Rings in Close Binary Systems}",
      journal = {\actaa},
         year = 1969,
        month = jan,
       volume = {19},
        pages = {155},
       adsurl = {https://ui.adsabs.harvard.edu/abs/1969AcA....19..155S}
}

@ARTICLE{Smak_1982,
       author = {{Smak}, J.},
        title = "{On the emission lines from rotating gaseous disks.}",
      journal = {\actaa},
         year = 1982,
        month = jan,
       volume = {31},
        pages = {395-408},
       adsurl = {https://ui.adsabs.harvard.edu/abs/1982AcA....31..395S}
}

@ARTICLE{Harlaftis_1996,
       author = {{Harlaftis}, E.~T. and {Marsh}, T.~R.},
        title = "{OY Carinae in outburst: Balmer emission from the red star and the gas stream.}",
      journal = {\aap},
         year = 1996,
        month = apr,
       volume = {308},
        pages = {97-106},
          doi = {10.48550/arXiv.astro-ph/9510025},
archivePrefix = {arXiv},
       eprint = {astro-ph/9510025},
 primaryClass = {astro-ph},
       adsurl = {https://ui.adsabs.harvard.edu/abs/1996A&A...308...97H}
}

@ARTICLE{Stephen_1992,
       author = {{Davey}, Stephen and {Smith}, Robert C.},
        title = "{Irradiation of the secondary star in cataclysmic variables.}",
      journal = {\mnras},
         year = 1992,
        month = aug,
       volume = {257},
        pages = {476-484},
          doi = {10.1093/mnras/257.3.476},
       adsurl = {https://ui.adsabs.harvard.edu/abs/1992MNRAS.257..476D}
}

@BOOK{Warner_Book,
       author = {{Warner}, Brian},
        title = "{Cataclysmic variable stars}",
         year = 1995,
       volume = {28},
       adsurl = {https://ui.adsabs.harvard.edu/abs/1995cvs..book.....W}
}

@article{Smith_2006,
    author = {Smith, Amanda J. and Haswell, Carole A. and Hynes, Robert I.},
    title = {VW Hyi: optical spectroscopy and Doppler tomography},
    journal = {MNRAS},
    volume = {369},
    number = {4},
    pages = {1537-1546},
    year = {2006},
    month = {07},
    issn = {0035-8711},
    doi = {10.1111/j.1365-2966.2006.10409.x},
    url = {https://doi.org/10.1111/j.1365-2966.2006.10409.x},
    eprint = {https://academic.oup.com/mnras/article-pdf/369/4/1537/3793101/mnras0369-1537.pdf},
}

@ARTICLE{Cropper_Polars,
       author = {{Cropper}, Mark},
        title = "{The Polars}",
      journal = {\ssr},
         year = 1990,
        month = dec,
       volume = {54},
       number = {3-4},
        pages = {195-295},
          doi = {10.1007/BF00177799},
       adsurl = {https://ui.adsabs.harvard.edu/abs/1990SSRv...54..195C}
}

@article{Patterson_1994,
doi = {10.1086/133375},
url = {https://dx.doi.org/10.1086/133375},
year = {1994},
month = {mar},
publisher = {The Astronomical Society of the Pacific},
volume = {106},
number = {697},
pages = {209},
author = {Patterson, Joseph},
title = {THE DQ HERCULIS STARS},
journal = {PASP}
}

@ARTICLE{Popham_1995,
       author = {{Popham}, Robert and {Narayan}, Ramesh},
        title = "{Accretion Disk Boundary Layers in Cataclysmic Variables. I. Optically Thick Boundary Layers}",
      journal = {\apj},
         year = 1995,
        month = mar,
       volume = {442},
        pages = {337},
          doi = {10.1086/175444},
       adsurl = {https://ui.adsabs.harvard.edu/abs/1995ApJ...442..337P}
}

@article{Kalomeni_2016,
doi = {10.3847/1538-4357/833/1/83},
url = {https://doi.org/10.3847/1538-4357/833/1/83},
year = {2016},
month = {dec},
publisher = {The American Astronomical Society},
volume = {833},
number = {1},
pages = {83},
author = {Kalomeni, B. and Nelson, L. and Rappaport, S. and Molnar, M. and Quintin, J. and Yakut, K.},
title = {EVOLUTION OF CATACLYSMIC VARIABLES AND RELATED BINARIES CONTAINING A WHITE DWARF},
journal = {ApJ}
}

@article{Kolb_1993,
  title={A model for the intrinsic population of cataclysmic variables},
  author={Kolb, Ulrich},
  journal={A\&A},
  volume={271},
  pages={149},
  year={1993}
}

@article{Kolb_1998,
  title={The cataclysmic variable period gap: still there},
  author={Kolb, U and King, AR and Ritter, H},
  journal={MNRAS},
  volume={298},
  number={2},
  pages={L29--L33},
  year={1998},
  publisher={Blackwell Science Ltd Oxford, UK and Cambridge, USA}
}

@article{Mestel_1968,
  title={Magnetic braking by a stellar wind-I},
  author={Mestel, L},
  journal={MNRAS},
  volume={138},
  pages={359},
  year={1968}
}

@article{Mestel_1987,
  title={On magnetic braking of late-type stars},
  author={Mestel, L and Spruit, HC},
  journal={MNRAS},
  volume={226},
  number={1},
  pages={57--66},
  year={1987},
  publisher={The Royal Astronomical Society}
}

@article{Patterson_2011,
  title={Distances and absolute magnitudes of dwarf novae: murmurs of period bounce},
  author={Patterson, Joseph},
  journal={MNRAS},
  volume={411},
  number={4},
  pages={2695--2716},
  year={2011},
  publisher={Blackwell Publishing Ltd Oxford, UK}
}

@article{Goliasch_2015,
  title={Population synthesis of cataclysmic variables. I. Inclusion of detailed nuclear evolution},
  author={Goliasch, Jonas and Nelson, Lorne},
  journal={ApJ},
  volume={809},
  number={1},
  pages={80},
  year={2015},
  publisher={IOP Publishing}
}

@article{Howell_2001,
  title={An exploration of the paradigm for the 2-3 hour period gap in cataclysmic variables},
  author={Howell, Steve B and Nelson, Lorne A and Rappaport, Saul},
  journal={ApJ},
  volume={550},
  number={2},
  pages={897},
  year={2001},
  publisher={IOP Publishing}
}

@ARTICLE{Paczynski_1983,
       author = {{Paczynski}, B. and {Sienkiewicz}, R.},
        title = "{The minimum period and the gap in periods of cataclysmic binaries}",
      journal = {\apj},
         year = 1983,
        month = may,
       volume = {268},
        pages = {825-831},
          doi = {10.1086/161004},
       adsurl = {https://ui.adsabs.harvard.edu/abs/1983ApJ...268..825P}
}

@ARTICLE{Rappaport_1982,
       author = {{Rappaport}, S. and {Joss}, P.~C. and {Webbink}, R.~F.},
        title = "{The evolution of highly compact binary stellar systems.}",
      journal = {\apj},
         year = 1982,
        month = mar,
       volume = {254},
        pages = {616-640},
          doi = {10.1086/159772},
       adsurl = {https://ui.adsabs.harvard.edu/abs/1982ApJ...254..616R}
}

@ARTICLE{Knigge_2006,
       author = {{Knigge}, Christian},
        title = "{The donor stars of cataclysmic variables}",
      journal = {\mnras},
         year = 2006,
        month = dec,
       volume = {373},
       number = {2},
        pages = {484-502},
          doi = {10.1111/j.1365-2966.2006.11096.x},
archivePrefix = {arXiv},
       eprint = {astro-ph/0609671},
 primaryClass = {astro-ph},
       adsurl = {https://ui.adsabs.harvard.edu/abs/2006MNRAS.373..484K}
}

@ARTICLE{Faulkner_1971,
       author = {{Faulkner}, John},
        title = "{Ultrashort-Period Binaries, Gravitational Radiation, and Mass Transfer. I. The Standard Model, with Applications to WZ Sagittae and Z Camelopardalis}",
      journal = {\apjl},
         year = 1971,
        month = dec,
       volume = {170},
        pages = {L99},
          doi = {10.1086/180848},
       adsurl = {https://ui.adsabs.harvard.edu/abs/1971ApJ...170L..99F}
}

@ARTICLE{Paczynski_1981,
       author = {{Paczynski}, B. and {Sienkiewicz}, R.},
        title = "{Gravitational radiation and the evolution of cataclysmic binaries}",
      journal = {\apjl},
         year = 1981,
        month = aug,
       volume = {248},
        pages = {L27-L30},
          doi = {10.1086/183616},
       adsurl = {https://ui.adsabs.harvard.edu/abs/1981ApJ...248L..27P}
}

@article{Patterson_1998,
  title={Late evolution of cataclysmic variables},
  author={Patterson, Joseph},
  journal={PASP},
  volume={110},
  number={752},
  pages={1132},
  year={1998},
  publisher={IOP Publishing}
}

@article{Belloni_2020,
  title={Evidence for reduced magnetic braking in polars from binary population models},
  author={Belloni, Diogo and Schreiber, Matthias R and Pala, Anna F and G{\"a}nsicke, Boris T and Zorotovic, M{\'o}nica and Rodrigues, Claudia V},
  journal={MNRAS},
  volume={491},
  number={4},
  pages={5717--5731},
  year={2020},
  publisher={Oxford University Press}
}

@ARTICLE{Inight_2023a,
       author = {{Inight}, K. and {G{\"a}nsicke}, Boris T. and {Schwope}, A. and {Anderson}, S. F. and {Badenes}, C. and {Breedt}, E. and {Chandra}, V. and {Davies}, B. D. R. and {Gentile Fusillo}, N. P. and {Green}, M. J. and {Hermes}, J. J. and {Huamani}, I. Achaica and {Hwang}, H. and {Knauff}, K. and {Kurpas}, J. and {Long}, K. S. and {Malanushenko}, V. and {Morrison}, S. and {Quiroz C.}, I. J. and {Ramos}, G. N. Aichele and {Roman-Lopes}, A. and {Schreiber}, M. R.  and {Standke}, A. and {Stutz}, L. and {Thorstensen}, J. R. and {Toloza}, O. and {Tovmassian}, G. and {Zakamska}, N. L.},
        title = "{Cataclysmic Variables from Sloan Digital Sky Survey - V. The search for period bouncers continues}",
      journal = {MNRAS},
         year = 2023,
        month = nov,
       volume = {525},
       number = {3},
        pages = {3597-3625},
          doi = {10.1093/mnras/stad2409},
archivePrefix = {arXiv},
       eprint = {2305.13371},
 primaryClass = {astro-ph.SR},
       adsurl = {https://ui.adsabs.harvard.edu/abs/2023MNRAS.525.3597I}
}

@article{Pala_2020,
  title={A volume-limited sample of cataclysmic variables from Gaia DR2: space density and population properties},
  author={Pala, AF and G{\"a}nsicke, BT and Breedt, E and Knigge, C and Hermes, JJ and Gentile Fusillo, NP and Hollands, MA and Naylor, T and Pelisoli, I and Schreiber, MR and others},
  journal={MNRAS},
  volume={494},
  number={3},
  pages={3799--3827},
  year={2020},
  publisher={Oxford University Press}
}

@article{Rodriguez_2025,
  title={Cataclysmic Variables and AM CVn Binaries in SRG/eROSITA+ Gaia: Volume Limited Samples, X-Ray Luminosity Functions, and Space Densities},
  author={Rodriguez, Antonio C and El-Badry, Kareem and Suleimanov, Valery and Pala, Anna F and Kulkarni, Shrinivas R and Gaensicke, Boris and Mori, Kaya and Rich, R Michael and Sarkar, Arnab and Bao, Tong and others},
  journal={PASP},
  volume={137},
  number={1},
  pages={014201},
  year={2025},
  publisher={IOP Publishing}
}

@article{Czerny_1989,
    author = {Czerny, M. and King, A. R.},
    title = {Accretion disc winds and coronae},
    journal = {MNRAS},
    volume = {236},
    number = {4},
    pages = {843-850},
    year = {1989},
    month = {02},
    issn = {0035-8711},
    doi = {10.1093/mnras/236.4.843},
    url = {https://doi.org/10.1093/mnras/236.4.843},
    eprint = {https://academic.oup.com/mnras/article-pdf/236/4/843/3924624/mnras236-0843.pdf},
}

@article{Kool_1999,
    author = {De Kool, Martijn and Wickramasinghe, Dayal},
    title = {Thermal instability and evaporation of accretion disc atmospheres},
    journal = {MNRAS},
    volume = {307},
    number = {2},
    pages = {449-458},
    year = {1999},
    month = {08},
    issn = {0035-8711},
    doi = {10.1046/j.1365-8711.1999.02641.x},
    url = {https://doi.org/10.1046/j.1365-8711.1999.02641.x},
    eprint = {https://academic.oup.com/mnras/article-pdf/307/2/449/3863527/307-2-449.pdf},
}

@ARTICLE{Stone_1996,
       author = {{Stone}, James M. and {Hawley}, John F. and {Gammie}, Charles F. and {Balbus}, Steven A.},
        title = "{Three-dimensional Magnetohydrodynamical Simulations of Vertically Stratified Accretion Disks}",
      journal = {\apj},
         year = 1996,
        month = jun,
       volume = {463},
        pages = {656},
          doi = {10.1086/177280},
       adsurl = {https://ui.adsabs.harvard.edu/abs/1996ApJ...463..656S}
}

@article{Tout_1992,
    author = {Tout, C. A. and Pringle, J. E.},
    title = {Accretion disc viscosity: a simple model for a magnetic dynamo},
    journal = {MNRAS},
    volume = {259},
    number = {4},
    pages = {604-612},
    year = {1992},
    month = {12},
    issn = {0035-8711},
    doi = {10.1093/mnras/259.4.604},
    url = {https://doi.org/10.1093/mnras/259.4.604},
    eprint = {https://academic.oup.com/mnras/article-pdf/259/4/604/3178586/mnras259-0604.pdf},
}

@article{Daniela_2026,
	author = {{Mu\~noz-Giraldo}, Daniela and {Stelzer}, Beate and {Schwope}, Axel and {Hern\'andez-D\'{\i}az}, Santiago and {Anderson}, Scott F. and {Demasi}, Sebastian},
	title = {eROSITA selection of new period-bounce cataclysmic variables - First follow-up confirmation using TESS and SDSS},
	DOI= "10.1051/0004-6361/202557266",
	url= "https://doi.org/10.1051/0004-6361/202557266",
	journal = {A\&A},
	year = 2026,
	volume = 707,
	pages = "A62",
}

@ARTICLE{Murray_1992,
       author = {{Murray}, Stephen D. and {Lin}, Douglas N.~C.},
        title = "{The Formation of Coronal Regions in Accretion Disks}",
      journal = {\apj},
         year = 1992,
        month = jan,
       volume = {384},
        pages = {177},
          doi = {10.1086/170861},
       adsurl = {https://ui.adsabs.harvard.edu/abs/1992ApJ...384..177M}
}

@ARTICLE{Balbus_1991,
       author = {{Balbus}, Steven A. and {Hawley}, John F.},
        title = "{A Powerful Local Shear Instability in Weakly Magnetized Disks. I. Linear Analysis}",
      journal = {\apj},
         year = 1991,
        month = jul,
       volume = {376},
        pages = {214},
          doi = {10.1086/170270},
       adsurl = {https://ui.adsabs.harvard.edu/abs/1991ApJ...376..214B}
}

@ARTICLE{Gammie_1996,
       author = {{Gammie}, Charles F.},
        title = "{Layered Accretion in T Tauri Disks}",
      journal = {\apj},
         year = 1996,
        month = jan,
       volume = {457},
        pages = {355},
          doi = {10.1086/176735},
       adsurl = {https://ui.adsabs.harvard.edu/abs/1996ApJ...457..355G}
}

@ARTICLE{Martin_2005,
       author = {{Martin}, D. Christopher and {Fanson}, James and {Schiminovich}, David and {Morrissey}, Patrick and {Friedman}, Peter G. and {Barlow}, Tom A. and {Conrow}, Tim and {Grange}, Robert and {Jelinsky}, Patrick N. and {Milliard}, Bruno and {Siegmund}, Oswald H.~W. and {Bianchi}, Luciana and {Byun}, Yong-Ik and {Donas}, Jose and {Forster}, Karl and {Heckman}, Timothy M. and {Lee}, Young-Wook and {Madore}, Barry F. and {Malina}, Roger F. and {Neff}, Susan G. and {Rich}, R. Michael and {Small}, Todd and {Surber}, Frank and {Szalay}, Alex S. and {Welsh}, Barry and {Wyder}, Ted K.},
        title = "{The Galaxy Evolution Explorer: A Space Ultraviolet Survey Mission}",
      journal = {\apjl},
         year = 2005,
        month = jan,
       volume = {619},
       number = {1},
        pages = {L1-L6},
          doi = {10.1086/426387},
archivePrefix = {arXiv},
       eprint = {astro-ph/0411302},
 primaryClass = {astro-ph},
       adsurl = {https://ui.adsabs.harvard.edu/abs/2005ApJ...619L...1M}
}

@ARTICLE{Morrissey_2007,
       author = {{Morrissey}, Patrick and {Conrow}, Tim and {Barlow}, Tom A. and {Small}, Todd and {Seibert}, Mark and {Wyder}, Ted K. and {Budav{\'a}ri}, Tam{\'a}s and {Arnouts}, Stephane and {Friedman}, Peter G. and {Forster}, Karl and {Martin}, D. Christopher and {Neff}, Susan G. and {Schiminovich}, David and {Bianchi}, Luciana and {Donas}, Jos{\'e} and {Heckman}, Timothy M. and {Lee}, Young-Wook and {Madore}, Barry F. and {Milliard}, Bruno and {Rich}, R. Michael and {Szalay}, Alex S. and {Welsh}, Barry Y. and {Yi}, Sukyoung K.},
        title = "{The Calibration and Data Products of GALEX}",
      journal = {\apjs},
         year = 2007,
        month = dec,
       volume = {173},
       number = {2},
        pages = {682-697},
          doi = {10.1086/520512},
archivePrefix = {arXiv},
       eprint = {0706.0755},
 primaryClass = {astro-ph},
       adsurl = {https://ui.adsabs.harvard.edu/abs/2007ApJS..173..682M}
}

@ARTICLE{Bianchi_2017,
       author = {{Bianchi}, Luciana and {Shiao}, Bernie and {Thilker}, David},
        title = "{Revised Catalog of GALEX Ultraviolet Sources. I. The All-Sky Survey: GUVcat\_AIS}",
      journal = {\apjs},
         year = 2017,
        month = jun,
       volume = {230},
       number = {2},
          eid = {24},
        pages = {24},
          doi = {10.3847/1538-4365/aa7053},
archivePrefix = {arXiv},
       eprint = {1704.05903},
 primaryClass = {astro-ph.GA},
       adsurl = {https://ui.adsabs.harvard.edu/abs/2017ApJS..230...24B}
}

@article{King_2007,
    author = {King, A. R. and Pringle, J. E. and Livio, M.},
    title = {Accretion disc viscosity: how big is alpha?},
    journal = {MNRAS},
    volume = {376},
    number = {4},
    pages = {1740-1746},
    year = {2007},
    month = {03},
    issn = {0035-8711},
    doi = {10.1111/j.1365-2966.2007.11556.x},
    url = {https://doi.org/10.1111/j.1365-2966.2007.11556.x},
    eprint = {https://academic.oup.com/mnras/article-pdf/376/4/1740/18670094/mnras0376-1740.pdf},
}

@article{Kromer_2007,
	author = {{Kromer, M.} and {Nagel, T.} and {Werner, K.}},
	title = {Synthetic NLTE accretion disc spectra for the dwarf nova SS Cygni during an outburst cycle},
	DOI= "10.1051/0004-6361:20077898",
	url= "https://doi.org/10.1051/0004-6361:20077898",
	journal = {A\&A},
	year = 2007,
	volume = 475,
	number = 1,
	pages = "301-308",
}

@article{Mennickent_2006,
	author = {{Mennickent, R. E.} and {Unda-Sanzana, E.} and {Tappert, C.}},
	title = {Doppler tomography of the dwarf novae VYari and WXi},
	DOI= "10.1051/0004-6361:20064850",
	url= "https://doi.org/10.1051/0004-6361:20064850",
	journal = {A\&A},
	year = 2006,
	volume = 451,
	number = 2,
	pages = "613-619",
}

@ARTICLE{Neustroev_2002,
       author = {{Neustroev}, V.~V. and {Borisov}, N.~V. and {Barwig}, H. and {Bobinger}, A. and {Mantel}, K.~H. and {{\v{S}}imi{\'c}}, D. and {Wolf}, S.},
        title = "{IP Pegasi: Investigation of the accretion disk structure. Searching evidences for spiral shocks in the quiescent accretion disk}",
      journal = {\aap},
         year = 2002,
        month = oct,
       volume = {393},
        pages = {239-250},
          doi = {10.1051/0004-6361:20021008},
archivePrefix = {arXiv},
       eprint = {astro-ph/0207418},
 primaryClass = {astro-ph},
       adsurl = {https://ui.adsabs.harvard.edu/abs/2002A&A...393..239N}
}

@ARTICLE{Marsh_2005,
       author = {{Marsh}, T.~R.},
        title = "{Doppler Tomography}",
      journal = {\apss},
         year = 2005,
        month = apr,
       volume = {296},
       number = {1-4},
        pages = {403-415},
          doi = {10.1007/s10509-005-4859-3},
       adsurl = {https://ui.adsabs.harvard.edu/abs/2005Ap&SS.296..403M}
}

@article{Belloni_2018,
    author = {Belloni, Diogo and Schreiber, Matthias R and Zorotovic, Mónica and Iłkiewicz, Krystian and Hurley, Jarrod R and Giersz, Mirek and Lagos, Felipe},
    title = {No cataclysmic variables missing: higher merger rate brings into agreement observed and predicted space densities},
    journal = {MNRAS},
    volume = {478},
    number = {4},
    pages = {5626-5637},
    year = {2018},
    month = {06},
    issn = {0035-8711},
    doi = {10.1093/mnras/sty1421},
    url = {https://doi.org/10.1093/mnras/sty1421},
    eprint = {https://academic.oup.com/mnras/article-pdf/478/4/5626/25204431/sty1421.pdf},
}

@article{Bolton_2012,
doi = {10.1088/0004-6256/144/5/144},
url = {https://doi.org/10.1088/0004-6256/144/5/144},
year = {2012},
month = {oct},
publisher = {The American Astronomical Society},
volume = {144},
number = {5},
pages = {144},
author = {Bolton, Adam S. and Schlegel, David J. and Aubourg, {\'E}ric and Bailey, Stephen and Bhardwaj, Vaishali and Brownstein, Joel R. and Burles, Scott and Chen, Yan-Mei and Dawson, Kyle and Eisenstein, Daniel J. and Gunn, James E. and Knapp, G. R. and Loomis, Craig P. and Lupton, Robert H. and Maraston, Claudia and Muna, Demitri and Myers, Adam D. and Olmstead, Matthew D. and Padmanabhan, Nikhil and P{\^a}ris, Isabelle and Percival, Will J. and Petitjean, Patrick and Rockosi, Constance M. and Ross, Nicholas P. and Schneider, Donald P. and Shu, Yiping and Strauss, Michael A. and Thomas, Daniel and Tremonti, Christy A. and Wake, David A. and Weaver, Benjamin A. and Wood-Vasey, W. Michael},
title = {SPECTRAL CLASSIFICATION AND REDSHIFT MEASUREMENT FOR THE SDSS-III BARYON OSCILLATION SPECTROSCOPIC SURVEY},
journal = {\aj}
}

@article{Schreiber_2023,
	author = {{Schreiber}, Matthias R. and {Belloni}, Diogo and {van Roestel}, Jan},
	title = {Period bouncers as detached magnetic cataclysmic variables},
	DOI= "10.1051/0004-6361/202347766",
	url= "https://doi.org/10.1051/0004-6361/202347766",
	journal = {A\&A},
	year = 2023,
	volume = 679,
	pages = "L8",
}

@ARTICLE{Schreiber_2016,
       author = {{Schreiber}, M.~R. and {Zorotovic}, M. and {Wijnen}, T.~P.~G.},
        title = "{Three in one go: consequential angular momentum loss can solve major problems of CV evolution}",
      journal = {\mnras},
         year = 2016,
        month = jan,
       volume = {455},
       number = {1},
        pages = {L16-L20},
          doi = {10.1093/mnrasl/slv144},
archivePrefix = {arXiv},
       eprint = {1510.04294},
 primaryClass = {astro-ph.SR},
       adsurl = {https://ui.adsabs.harvard.edu/abs/2016MNRAS.455L..16S}
}

@article{Rodriguez_2005,
  title={Detection of the white dwarf and the secondary star in the new SU UMa dwarf nova HS 2219+ 1824},
  author={Rodriguez-Gil, Pablo and G{\"a}nsicke, BT and Hagen, H-J and Marsh, TR and Harlaftis, ET and Kitsionas, S and Engels, Dieter},
  journal={A\&A},
  volume={431},
  number={1},
  pages={269--277},
  year={2005},
  publisher={EDP Sciences}
}

@article{Szegedi_2022,
  title={Transient behaviour of three SU UMa-type dwarf novae: AR Pic, QW Ser, and V521 Peg},
  author={Szegedi, H{\'e}l{\`e}ne and Charles, Philip A and Meintjes, Pieter J and Odendaal, Alida},
  journal={MNRAS},
  volume={513},
  number={4},
  pages={4682--4695},
  year={2022},
  publisher={Oxford University Press}
}

@article{Pala_2022,
    author = {Pala, A F and Gänsicke, B T and Belloni, D and Parsons, S G and Marsh, T R and Schreiber, M R and Breedt, E and Knigge, C and Sion, E M and Szkody, P and Townsley, D and Bildsten, L and Boyd, D and Cook, M J and De Martino, D and Godon, P and Kafka, S and Kouprianov, V and Long, K S and Monard, B and Myers, G and Nelson, P and Nogami, D and Oksanen, A and Pickard, R and Poyner, G and Reichart, D E and Rodriguez Perez, D and Shears, J and Stubbings, R and Toloza, O},
    title = {Constraining the evolution of cataclysmic variables via the masses and accretion rates of their underlying white dwarfs},
    journal = {MNRAS},
    volume = {510},
    number = {4},
    pages = {6110-6132},
    year = {2022},
    month = {11},
    issn = {0035-8711},
    doi = {10.1093/mnras/stab3449},
    url = {https://doi.org/10.1093/mnras/stab3449},
    eprint = {https://academic.oup.com/mnras/article-pdf/510/4/6110/42357931/stab3449.pdf},
}

\begin{appendix}





\onecolumn

\section{Discarded observations}
\label{sec:Discarded_Observations}

\begin{table*}[h!]
    \caption{SDSS spectra discarded from the analysis due to outburst contamination or poor quality.}
    \label{table:Discarded}
    \centering
    \begin{tabular}{l c c c l}
        \hline\hline
        Name & Plate & Fibre & MJD & Comment \\
        \hline
         
        QW\,Ser & 1721 & 21 & 53857 & Likely observed during or near an outburst \\ 
        V1240\,Her & 1688  &  232 & 53462 & Likely observed during or near an outburst \\ 
        AK\,Cnc & 5293 & 322 & 55953 & Likely observed during or near an outburst \\ 
        KS\,UMa & 904 & 147 & 52381 & Likely observed during or near an outburst \\
        DI\,UMa & 5730 & 518 & 56607 & Likely observed during or near an outburst \\
        EZ\,Lyn & 3699 & 734 & 55517 & Likely observed during or near an outburst \\
        EZ\,Lyn & 3700 & 450 & 55542 & Likely observed during or near an outburst \\
        EZ\,Lyn & 4528 & 442 & 55559 & Likely observed during or near an outburst \\
        EZ\,Lyn & 4527 & 728 & 55590 & Likely observed during or near an outburst \\
        CY\,UMa  & 6700 & 434 & 56384 & Gaps in wavelength coverage \\
        PM\,J11384+0619 & 5377 & 374 & 55957 & Poor continuum calibration \\
        V355\,UMa & 6744 & 345 & 56399 & Poor continuum calibration \\
        \hline
    \end{tabular}
\end{table*}

\section{Sample properties and results from Balmer line analysis}
\label{sec:Table}
\begin{table*}[h!]

\caption {Description of the 34 columns in the sample, including identifiers, coordinates, photometric data, distance, system's classification, $P_{\rm orb}$, Balmer line fluxes, Balmer line luminosities, and Balmer decrements.}
\label{table:Master} 
\centering
\small
\begin{tabular}{clll}
\hline\hline             
\# & Name & Unit & Description \\
\hline
1 & System &  & Object's common name. \\
2 & GaiaDR3 &  & \textit{Gaia} ID from data release 3. \\
3 & ra\_DR3 & deg & \textit{Gaia}\,DR3 Right Ascension. \\
4 & dec\_DR3 & deg & \textit{Gaia}\,DR3 Declination. \\
5 & SDSS\_Name &  & SDSS name. \\
6 & Plate &  & SDSS spectroscopic plate number. \\
7 & Fibre &  & SDSS fibre identification number. \\
8 & MJD &  & Modified Julian Date of the SDSS spectroscopic observation. \\
9 & bp\_phot & mag &  \textit{Gaia} $BP$ mean magnitude. \\
10 & rp\_phot & mag & \textit{Gaia} $RP$ mean magnitude. \\
11 & g\_phot & mag & \textit{Gaia} $G$ mean magnitude. \\
12 & Distance & pc & Distance to the object. \\
13 & lo\_e\_Distance & pc & Lower uncertainty in Distance. \\
14 & up\_e\_Distance & pc & Upper uncertainty in Distance. \\
13 & Type &  & Cataclysmic variable type \\
14 & Porb & min & Orbital period of the system. \\
15 & F\_Halpha & $10^{-17}\,$erg/cm$^{2}$/s & $\text{H}\alpha$ line flux. \\
16 & e\_F\_Halpha & $10^{-17}\,$erg/cm$^{2}$/s & Uncertainty in F\_Halpha \\
17 & F\_Hbeta & $10^{-17}\,$erg/cm$^{2}$/s & $\text{H}\beta$ line flux. \\
18 & e\_F\_Hbeta & $10^{-17}\,$erg/cm$^{2}$/s & Uncertainty in F\_Hbeta \\
19 & F\_Hgamma & $10^{-17}\,$erg/cm$^{2}$/s & $\text{H}\gamma$ line flux. \\
20 & e\_F\_Hgamma & $10^{-17}\,$erg/cm$^{2}$/s & Uncertainty in F\_Hgamma. \\
21 & L\_Halpha & $10^{27}\,$erg/s & $\text{H}\alpha$ line luminosity. \\
22 & lo\_e\_L\_Halpha & $10^{27}\,$erg/s & Lower uncertainty in L\_Halpha. \\
23 & up\_e\_L\_Halpha & $10^{27}\,$erg/s & Upper uncertainty in L\_Halpha. \\
24 & L\_Hbeta & $10^{27}\,$erg/s & $\text{H}\beta$ line luminosity. \\
25 & lo\_e\_L\_Hbeta & $10^{27}\,$erg/s & Lower uncertainty in L\_Hbeta. \\
26 & up\_e\_L\_Hbeta & $10^{27}\,$erg/s & Upper uncertainty in L\_Hbeta. \\
27 & L\_Hgamma & $10^{27}\,$erg/s & $\text{H}\gamma$ line luminosity. \\
28 & lo\_e\_L\_Hgamma & $10^{27}\,$erg/s & Lower uncertainty in L\_Hgamma. \\
29 & up\_e\_L\_Hgamma & $10^{27}\,$erg/s & Upper uncertainty in L\_Hgamma. \\
30 & Halpha/Hbeta &  & Ratio between H$\alpha$ and H$\beta$ lines. \\
31 & e\_Halpha/Hbeta &  & Uncertainty in Halpha/Hbeta. \\
32 & Hgamma/Hbeta &  & Ratio between H$\gamma$ and H$\beta$ lines. \\
33 & e\_Hgamma/Hbeta &  & Uncertainty in Hgamma/Hbeta. \\
34 & P\_PB &  & Period-Bouncer probability from the logistic regression model (expressed as a percentage). \\
\hline
\end{tabular}
\end{table*}

\clearpage

\section{White dwarf model fittings}
\label{sec:WD_fittings}

\begin{table*}[h!]
    \caption{Results from the WD model atmosphere fittings to the spectra corrected for the WD contribution, including observation identifiers, the effective temperature and surface gravity of the best-fitting model, the scaling factor ($K_{\rm WD}$), and a flag denoting the inclusion of GALEX photometry in the fit.} 
    \label{table:WD_fittings}
    \centering
    \begin{tabular}{l c c c c c c c}
        \hline\hline
        Name & Plate &  Fibre & MJD & $T_{\rm eff}$ [K] & $\log (g)$ & $K_{\rm WD}$ & GALEX \\
        \hline
        SDSS\,J102905.24+485515.2 & 6656 & 533 & 56624 & 10750 & 8.0 & $8.73\cdot10^{-25}$ &  N \\
       SDSS\,J100515.38+191107.9 & 2372 & 473 & 53768 & 14250 & 8.75 & $1.14\cdot10^{-24}$ &  Y \\
       SDSS\,J100515.38+191107.9 & 9571 & 520 & 57808 & 14750 & 9.25 & $9.52\cdot10^{-25}$ &  Y \\
       SDSS\,J100515.38+191107.9 & 5882 & 272 & 56029 & 15750 & 9.5 & $6.86\cdot10^{-25}$ &  Y \\
       V355\,UMa & 1283 & 591 & 52762 & 12000 & 8.5 & $3.29\cdot10^{-24}$ &  N \\
       LV\,Cnc & 1301 & 624 & 52976 & 13000 & 8.25 & $1.60\cdot10^{-24}$ &  Y \\
       IY\,UMa & 7088 & 960 & 56657 & 14750 & 6.75 & $2.49\cdot10^{-24}$ &  Y \\
       IY\,UMa & 949 & 358 & 52427 & 14500 & 6.5 & $2.72\cdot10^{-24}$ &  Y \\
      IR\,Com & 5985 & 232 & 56089 & 14750 & 9.5 & $7.93\cdot10^{-25}$ &  N \\
      GZ\,Cet & 662 & 541 & 52178 & 12250 & 9.5 & $1.04\cdot10^{-24}$ &  N \\
      GZ\,Cet & 662 & 552 & 52147 & 11500 & 9.5 & $1.25\cdot10^{-24}$ &  N \\
      V0493\,Ser & 344 & 315 & 51693 & 10500 & 8.75 & $1.83\cdot10^{-24}$ &  N \\
        AK\,Cnc & 2575 & 318 & 54085 & 14750 & 9.5 & $7.11\cdot10^{-25}$ &  Y \\
        BC\,UMa & 968 & 280 & 52412 & 15500 & 8.75 & $7.62\cdot10^{-25}$ &  Y \\
        BC\,UMa & 6677 & 612 & 56385 & 14250 & 8.5 & $1.16\cdot10^{-24}$ &  Y \\
        RZ\,Leo & 513 & 70 & 51989 & 14000 & 8.0 & $1.27\cdot10^{-24}$ &  Y \\
        RZ\,Leo & 4740 & 188 & 55651 & 15000 & 8.5 & $8.23\cdot10^{-25}$ &  Y \\
       EG\,Cnc & 1588 & 278 & 52965 & 11750 & 8.75 & $1.08\cdot10^{-24}$ &  Y \\
       MT\,Com & 5972 & 566 & 56334 & 10250 & 8.0 & $1.13\cdot10^{-24}$ &  N \\
       V406\,Vir & 335 & 85 & 52000 & 12250 & 8.0 & $2.70\cdot10^{-24}$ &  Y \\
        EZ\,Lyn  & 1780 & 431 & 53090 & 11500 & 8.75 & $2.87\cdot10^{-24}$ &  N \\
       PM\,J11384+0619 & 1620 & 290 & 53137 & 12500 & 8.75 & $8.57\cdot10^{-25}$ &  Y \\
       PM\,J12192+2049 & 5978 & 185 & 56073 & 11750 & 7.75 & $7.50\cdot10^{-25}$ &  Y \\
      PM\,J12192+2049  & 2611 & 376 & 54477 & 12000 & 8.0 & $6.85\cdot10^{-25}$ &  Y \\
        1RXS\,J101421.4+063855 & 4876 & 652 & 55679 & 11000 & 9.0 & $5.62\cdot10^{-25}$ &  N \\
      SDSS\,J075507.70+143547.6 & 4501 & 555 & 55590 & 17750 & 8.75 & $6.87\cdot10^{-25}$ &  Y \\
       SDSS\,J075507.70+143547.6 & 2264 & 416 & 53682 & 16500 & 8.75 & $1.00\cdot10^{-24}$ &  Y \\
       SDSS\,J103533.02+055158.4 & 999 & 55 & 52636 & 11000 & 8.25 & $1.59\cdot10^{-24}$ &  Y \\
       SDSS\,J103533.02+055158.4 & 4852 & 323 & 55689 & 11750 & 9.0 & $1.05\cdot10^{-24}$ &  Y \\
      SDSS\,J105754.25+275947.5  & 2359 & 105 & 53826 & 10250 & 6.5 & $6.97\cdot10^{-25}$ &  Y \\
      SDSS\,J105754.25+275947.5 & 2359 & 102 & 53800 & 10250 & 6.5 & $6.66\cdot10^{-25}$ &  Y \\
       SDSS\,J121607.03+052013.9 & 844 & 423 & 52378 & 10250 & 6.75 & $4.90\cdot10^{-25}$ &  Y \\
      SDSS\,J143317.78+101122.8  & 1709 & 153 & 53533 & 12500 & 7.0 & $1.17\cdot10^{-24}$ &  Y \\
      SDSS\,J143317.78+101122.8  & 5465 & 138 & 55988 & 12250 & 7.0 & $1.31\cdot10^{-24}$ &  Y \\
        \hline
    \end{tabular}
    \tablefoot{The fitted parameters were derived for the purpose of subtracting the photospheric Balmer absorption from the WD and recover the full Balmer emission-line fluxes. They should be regarded only as approximate physical parameters, especially for pre-bounce CVs, where the accretion disc is expected to contribute to the optical spectrum.}
\end{table*}

\clearpage
\twocolumn

\section{Uncertainties}
\label{sec:uncertainties}

As described in Sect.~\ref{Line_Flux_Measurement}, the line flux is obtained by subtracting a locally fitted linear continuum and integrating the residual profile over a selected wavelength window. The integrated line flux is then defined as,

\begin{equation}
F_{\text{line}} = \sum_i \left( f_i - c_i \right)\,\Delta\lambda_i
\end{equation}

\noindent where $f_i$ and $c_i$ are the flux density and local continuum estimate in the $i$-th spectral bin, respectively, and $\Delta\lambda_i$ is the bin width. The sum extends over all bins within the integration window including the line.

The uncertainties in the local continuum across the emission line region are obtained by propagating the uncertainties of the linear fit parameters,

\begin{equation}
\sigma_{\mathrm{c, i}} = \sqrt{\lambda_i^{\,2}\,\sigma_{m}^{2} + \sigma_{b}^{2}}
\end{equation}

\noindent where $\sigma_{\mathrm{c},i}$ is the uncertainty in the continuum estimate at the $i$-th bin, and $\sigma_m^2$ and $\sigma_b^2$ are the variances of the slope and intercept of the linear fit, respectively.

The uncertainty in the integrated flux is calculated as,

\begin{equation}
\sigma_{F_{\text{line}}} = \sqrt{3} \sqrt{ \sum_i \left[ (\sigma_{f,i}\,\Delta\lambda_i)^2 + (\sigma_{c,i}\,\Delta\lambda_i)^2 \right] }
\end{equation}

\noindent where $\sigma_{F_{\text{line}}}$ is the uncertainty in the integrated line flux and $\sigma_{f,i}$ is the uncertainty of the flux density in the $i$-th bin. The factor $\sqrt{3}$ accounts for short-range correlations between adjacent bins introduced during SDSS spectral processing (see \citealt{Bolton_2012} and \citealt{Dawson_2016}), assuming an effective correlation length of three bins.

Line luminosities, $L_{\text{line}}$, are computed from $F_{\text{line}}$ using the distance to the source from \citet{Bailer-Jones}. As the distances from \citet{Bailer-Jones} have asymmetric errors, we calculate upper and lower errors for the line luminosities,

\begin{equation}
\sigma_{L_\text{line}}^{+} = L_{\text{line}} \sqrt{\left(\frac{\sigma_{F_{\text{line}}}}{F_{\text{line}}}\right)^2 + \left(2\frac{\sigma_{d}^{+}}{d}\right)^2}
\end{equation}

\begin{equation}
\sigma_{L_\text{line}}^{-} = L_{\text{line}} \sqrt{\left(\frac{\sigma_{F_{\text{line}}}}{F_{\text{line}}}\right)^2 + \left(2\frac{\sigma_{d}^{-}}{d}\right)^2}
\end{equation}

\noindent where $\sigma_{L,\text{line}}^{+}$ and $\sigma_{L,\text{line}}^{-}$ are the upper and lower uncertainties on the line luminosity, $d$ is the distance to the source, and $\sigma_d^{+}$ and $\sigma_d^{-}$ denote the upper and lower uncertainties on the distance, respectively.

Finally, uncertainties in line ratios are propagated as,

\begin{equation}
\sigma_R = R \sqrt{\left(\frac{\sigma_{F_{\text{line}, 1}}}{F_{\text{line}, 1}}\right)^2 + \left(\frac{\sigma_{F_{\text{line}, 2}}}{F_{\text{line}, 2}}\right)^2}.
\end{equation}

\noindent where $F_{\text{line}, 1}$ and $F_{\text{line}, 2}$ are the fluxes of two emission lines, and $R = F_1/F_2$ (or equivalently $L_1/L_2$).

\section{Logistic regression model training and evaluation}
\label{sec:Logistic_Regression}

As explained in Sect.~\ref{subsect:diagnostic_diagram}, to classify CVs as pre-bouncers or period-bouncers based on their Balmer emission line ratios, we have trained a linear logistic regression model (see Fig.~\ref{Fig.Diagnostic_Diagram} and Eq.~\ref{Eq.LRModel} in Sect.~\ref{subsect:diagnostic_diagram}). Here, we explain the detailed training and evaluation procedures. The software routines used in this work are based on Python implementations available in the \textsl{scikit-learn} package (\citealt{scikit-learn}).

\subsection{Training}
\label{sec:Training_Model}

\begin{figure}
	\centering
	\includegraphics[width=0.99\linewidth]{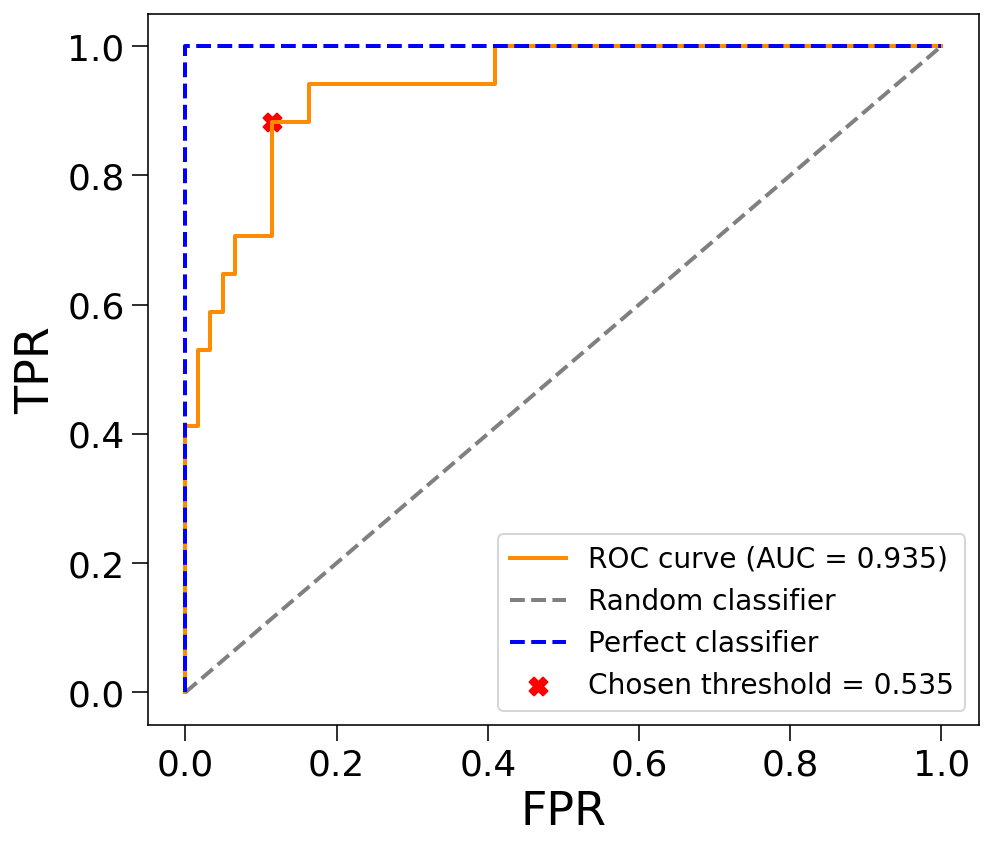}
	\caption{Receiver Operating Characteristic curve of the model, computed from cross-validated out-of-fold predictions and shown in comparison to a random classifier (grey dashed line) and a perfect classifier (blue dashed line). The Area Under Curve is 0.935. The red marker denotes the chosen decision threshold of $P_{\text{PB}}=0.535$.}
	\label{Fig.ROC}
\end{figure}

The two classes in our data set are not equally represented, with 61 SDSS spectra of pre-bouncers and 17 SDSS spectra of period-bouncers, corresponding to an approximate ratio of 3:1. To prevent the logistic regression from being biased towards the majority class, we used balanced class weights during training (see e.g. \citealt{Imbalanced_Classes}). This way, the loss function assigns a higher penalty to misclassifications of the minority class in proportion to the inverse of their class frequency. In consequence, the model becomes more sensitive to the underrepresented class, in our case the period-bouncers, resulting in a more balanced predictive performance across both classes. 

To prevent overfitting, L2 regularisation (ridge penalty; \citealt{Hoerl_1970a}, \citealt{Hoerl_1970b}) was applied during training, with the regularisation strength controlled by the inverse penalty parameter set to $\text{C}=0.4$.

\subsection{Evaluation}
\label{sec:Evaluation_Model}

Given the limited number of spectra for period-bouncers, we did not reserve a separate test data set for evaluating the performance of the model, instead, we implemented \textsl{k}-fold stratified cross-validation (see e.g. \citealt{Kohavi_1995}). In this method, the sample is split into \textsl{k} disjoint folds of approximately equal size. An untrained copy of the model is trained on \textsl{k}-1 folds and evaluated on the remaining fold, being this process repeated \textsl{k} times so that each object in the sample is used exactly once for testing the model. Stratification divides the data into distinct groups based on their class labels and then randomly selects examples from each group in proportion to their representation in the full data set. Thus, it ensures that the class proportions are preserved in each fold, which is particularly important in our case due to the class imbalance. We used \textsl{k} = 3 folds. 

The predictions from each test fold can be summarised in terms of four outcomes: number of true positives (TP), true negatives (TN), false positives (FP), and false negatives (FN) (see Fig.~\ref{Fig.Confusion_Matrix} in Sect.~\ref{subsect:diagnostic_diagram}). Based on these quantities, we compute both class-specific and global performance metrics. In each iteration, all evaluation metrics are computed on the test fold and the mean across the iterations is calculated to provide an estimate of the model’s generalisation performance. The variability across folds allows us to quantify an uncertainty for each metric, expressed as the standard deviation. We report in Table~\ref{table:ModelMetrics} the class-specific and global precision, recall, and F1-score (see Eqs.~\ref{eq:precision}–\ref{eq:f1}) (\citealt{Kent_1955}, \citealt{Blair_1979}, \citealt{Apte_1994}).

\begin{equation}
\text{Precision} = \dfrac{\text{TP}}{\text{TP}+\text{FP}},
\label{eq:precision}
\end{equation}

\begin{equation}
\text{Recall} = \dfrac{\text{TP}}{\text{TP}+\text{FN}},
\label{eq:recall}
\end{equation}

\begin{equation}
\text{F1} = \dfrac{2\cdot \text{Precision}\cdot\text{Recall}}{\text{Precision}+\text{Recall}},
\label{eq:f1}
\end{equation}

\begin{equation}
\text{Accuracy} = \dfrac{\text{TP}+\text{TN}}{\text{TP}+\text{TN}+\text{FP}+\text{FN}},
\label{eq:accuracy}
\end{equation}

\noindent Precision quantifies the reliability of positive classifications, recall measures the degree of completeness, and the F1-score, defined as the harmonic mean of precision and recall, summarises the trade-off between the two. Global metrics are obtained as the unweighted mean between the class-specific metrics (macro-averaging), thus ensuring that each class contributes equally, regardless of its frequency in the data set. Additionally, we report in Table~\ref{table:ModelMetrics} the accuracy of the model (see Eq.~\ref{eq:accuracy}), which measures the aggregate rate of correct predictions. We note that the accuracy of the model is defined only globally.

\subsection{Final model}
\label{sec:Final_Model}

The final model coefficients were determined by training on the complete data set, and coefficient uncertainties were quantified using a bootstrap approach. In this procedure, the data set of all Balmer decrement measurements was repeatedly resampled with replacement to generate new pseudo-samples, using stratification to preserve the class proportions in each pseudo-sample. Consequently, some measurements may appear multiple times within a pseudo-sample, while others may not be selected at all. In each iteration, an untrained copy of the model was trained on the resampled data. This process was iterated 1000 times, yielding distributions for each coefficient. The standard deviation of these distributions was adopted as the corresponding uncertainty estimate. The final model coefficients and their uncertainties are $\beta_{0}=-0.953\pm 0.085$,
$\beta_{1}=1.40\pm 0.11$, and 
$\beta_{2}=-1.01\pm 0.16$. 

\begin{table}[h!]
    \caption{Class-specific and global performance metrics of the logistic regression model.}
    \label{table:ModelMetrics}
    \centering
    \begin{tabular}{l c c c}
        \hline\hline
        Metric & Pre-bouncers & Period-bouncers & Global \\
        \hline
        Precision & $0.966 \pm 0.024$ & $0.701 \pm 0.114$ & $0.833 \pm 0.047$ \\
        Recall    & $0.887 \pm 0.058$ & $0.889 \pm 0.079$ & $0.888 \pm 0.016$ \\
        F1   & $0.923 \pm 0.023$ & $0.772 \pm 0.049$ & $0.847 \pm 0.036$ \\
        \hline
        Accuracy  & - & - & $0.885 \pm 0.031$ \\
        \hline
    \end{tabular}
\end{table}

The decision threshold, that is, the probability value above which a system is classified as a period-bouncer, was chosen to maximize the F1-score of the period-bouncer class, providing an optimal trade-off between precision and recall for period-bouncers. The optimal threshold value is $P_{\text{PB}}=0.535$, and it was determined using cross-validated out-of-fold predictions, that is, the prediction for each observation was generated by a model trained on the remaining folds during cross-validation, thus ensuring an unbiased estimate of model performance. The same threshold value was consistently applied during the cross-validation evaluation of the model (see Appendix~\ref{sec:Evaluation_Model}). In Fig.~\ref{Fig.ROC}, we additionally present the Receiver Operating Characteristic (ROC) curve (\citealt{Swets_1988}), which illustrates the true positive rate (TPR) against the false positive rate (FPR) across all possible threshold values. As a measure of the model’s discriminative capability, we report the ROC Area Under Curve (AUC) (\citealt{Hanley_1982}), which is $\text{ROC\,AUC}=0.935$, indicating strong discriminative performance. The ROC curve and ROC\,AUC metric were calculated using cross-validated out-of-fold predictions. The chosen optimal threshold value at $P_{\text{PB}}=0.535$ is indicated in Fig.~\ref{Fig.ROC}. As can be seen, the model’s performance lies close to that of a perfect classifier, and the chosen threshold value is also well supported from this perspective.

\end{appendix}
\end{document}